\documentclass[useAMS,usenatbib,fleqn]{mnras}

\usepackage{graphicx}
\usepackage{color}
\usepackage{times}
\usepackage{mathptmx}
\usepackage{natbib}
\usepackage{setspace}
\usepackage{amsmath}
\usepackage{amsbsy}
\usepackage{gensymb}

\newif\ifAMStwofonts
\AMStwofontstrue
\definecolor{red}{rgb}{1,0.,0.}

\newcommand{\eref}[1]{equation~(\ref{#1})}

\newcommand{\K}{{\rm K}}
\newcommand{\deltat}{\Delta t}

\newcommand{\arepo}{{\sc Arepo}}

\newcommand{\msun}{{\rm M}_\odot}
\newcommand{\Myr}{{\rm Myr}}
\newcommand{\Gyr}{{\rm Gyr}}
\newcommand{\yr}{{\rm yr}}
\newcommand{\pc}{{\rm pc}}
\newcommand{\kpc}{{\rm kpc}}
\newcommand{\erg}{{\rm erg}}
\newcommand{\kms}{{\rm km~s^{-1}}}
\newcommand{\cm}{{\rm cm}}
\newcommand{\sech}{{\rm sech}}

\newcommand{\gsim}{\,\lower.7ex\hbox{$\;\stackrel{\textstyle>}{\sim}\;$}}
\newcommand{\lsim}{\,\lower.7ex\hbox{$\;\stackrel{\textstyle<}{\sim}\;$}}
\newcommand{\namemodel}{\textit{SMUGGLE}}

\defcitealias{Planck2016}{Planck Collaboration XIII} 
\newcommand{\planckpappar}{\citetalias{Planck2016} \citeyear{Planck2016}}

\newcommand{\rev}[1]{#1}

\ifpdf
  \DeclareGraphicsExtensions{.pdf,.png}
\else
  \DeclareGraphicsExtensions{.eps}
\fi

\title[Simulating the ISM and stellar feedback on a moving mesh] 
{Simulating the interstellar medium and stellar feedback on a moving mesh:  Implementation and isolated galaxies}

\author[F. Marinacci et al.]
{Federico Marinacci$^{1,2,3}$\thanks{E-mail: federico.marinacci2@unibo.it}\href{https://orcid.org/0000-0003-3816-7028}{\includegraphics[scale=0.08]{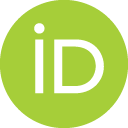}}, 
Laura V. Sales$^4$\thanks{Hellman Fellow}\href{https://orcid.org/0000-0002-3790-720X}{\includegraphics[scale=0.08]{Plots/orcid}},
Mark Vogelsberger$^3$\href{https://orcid.org/0000-0001-8593-7692}{\includegraphics[scale=0.08]{Plots/orcid}},
Paul Torrey$^{5}$\href{https://orcid.org/0000-0002-5653-0786}{\includegraphics[scale=0.08]{Plots/orcid}}, 
\newauthor Volker Springel$^6$\href{https://orcid.org/0000-0001-5976-4599}{\includegraphics[scale=0.08]{Plots/orcid}}\vspace*{0.2cm}\\
  $^1$Department of Physics \& Astronomy, University of Bologna, via Gobetti 93/2, 40129 Bologna, Italy\\
  $^2$Institute for Theory and Computation, Harvard-Smithsonian Center for Astrophysics, 
  60 Garden Street, Cambridge, MA 02138, USA\\
  $^3$Kavli Institute for Astrophysics and Space Research, 
  Massachusetts Institute of Technology, Cambridge, MA 02139, USA\\
  $^4$Department of Physics \& Astronomy, University of California, Riverside, 900 University Avenue, Riverside, CA 92521, USA\\
  $^5$Department of Astronomy, University of Florida, 211 Bryant Space Sciences 
  Center, Gainesville, FL 32611 USA\\
  $^6$Max-Planck-Institut f\"ur Astrophysik, Karl-Schwarzschild-Str. 1, D-85748, Garching, Germany\\
}

\date{Accepted 2019 August 25. Received 2019 August 06; in original form 2019 May 21}

\begin{document}

\pagerange{\pageref{firstpage}--\pageref{lastpage}}
\pubyear{2019}

\maketitle

\label{firstpage}

\begin{abstract} 
We introduce the {\bf S}tars and {\bf MU}ltiphase {\bf G}as in {\bf G}a{\bf L}axi{\bf E}s -- {\it \textbf{SMUGGLE}} model, an explicit and comprehensive stellar feedback model for the moving-mesh code \arepo. This novel sub-resolution model resolves the multiphase gas structure of the interstellar medium and self-consistently generates gaseous outflows. 
The model implements crucial aspects of stellar feedback including photoionization, radiation pressure, energy and momentum injection from stellar winds and from supernovae. 
We explore this model in high-resolution isolated simulations of Milky Way-like disc galaxies.
Stellar feedback regulates star formation to the observed level and naturally captures the establishment of a Kennicutt-Schmidt relation. 
This result is achieved independent of the numerical mass and spatial resolution of the simulations. 
Gaseous outflows are generated with average mass loading factors of the order of unity. 
Strong outflow activity is correlated with peaks in the star formation history of the galaxy with evidence that most of the ejected gas eventually rains down onto the disc in a galactic fountain flow that sustains late-time star formation.
Finally, the interstellar gas in the galaxy shows a distinct multiphase distribution with a coexistence of cold, warm and hot phases. 
\end{abstract}

\begin{keywords}
  galaxies: formation -- galaxies: evolution -- galaxies: ISM -- ISM: general
\end{keywords}

\section{Introduction} \label{sec:intro}

The $\Lambda$ Cold Dark Matter ($\Lambda$CDM) cosmological model
has emerged as the most successful theoretical framework to explain the formation and evolution of cosmic structures (\planckpappar). 
This model assumes that galaxies in the present-day Universe assembled from a nearly homogeneous initial state during almost 14 billion years of evolution. 
Invisible, non-baryonic, dark matter constitutes the backbone for this structure formation process.
Dark matter haloes assemble through gravitational collapse allowing baryons to cool and collapse effectively at their centres where the density is the highest, and the potential is deepest.
The potential wells associated with dark matter haloes host the formation of the first generation of stars, the formation of the first nascent galaxies, and eventually host the formation of more massive galaxies, as we are accustomed to seeing in the local Universe \citep{White1978}. 

Understanding and modelling the formation of galaxies within their cosmological context is a daunting task owing to the large dynamical range in space and time as well as the wide range of physical mechanisms involved. For instance, the physics of a single supernova (SN) explosion occurs on sub-parsec scales, but megaparsec scales are required to capture the gravitational collapse and assembly of a halo such as that of the Milky Way.  Even larger scales are needed to properly capture the galactic environment including large-scale tidal torques and the past merger and interaction history of galaxies.
In other words, accurately simulating galaxy formation simultaneously demands modelling large volumes, while still resolving small scales in order to capture the relevant galaxy formation physics.

All currently existing cosmological simulations, both full volume and zoom-ins, rely on sub-resolution (or sub-grid) models. 
Sub-grid models are a set of effective numerical prescriptions that are needed to capture physical processes that occur on scales at or below the resolution of the simulation (i.e. those processes that are unresolved). For example, cosmological galaxy formation simulations are not able to resolve the formation of individual stars, nor stellar evolution.
Therefore, models are required that describe both the formation process of stars and their subsequent stellar evolution.  
Nearly all physical processes involved in galaxy formation are implemented to some degree as sub-grid models including radiative cooling, star formation, stellar evolution, SN feedback, Active Galactic Nuclei (AGN) feedback, etc.
Ideal sub-grid models are independent of numerical resolution above a minimum threshold resolution such that the model numerically converges with increasing resolution. 
However, a well converged model will also yield no new information once the convergent resolution requirement is achieved.
It is therefore important to separately evaluate simulation  
limitations imposed by numerical resolution 
versus limitations imposed by inadequate modelling of physical processes. 
For galaxy formation, the most important physical processes beyond gravity and hydrodynamics
are the formation of stars and their interactions with the surrounding interstellar medium (ISM). 

It is observed that star formation is inefficient with less than $20-25$ per cent of a galaxy's predicted total baryonic budget turning into stars, and with fractions as small as $1$ per cent estimated for low mass dwarf galaxies \citep{Conroy2009, Moster2013, Behroozi2018}. 
The remaining baryons may be kept in the form of gas associated with the disc in the ISM, dispersed into the more extended circumgalactic medium, or remain outside of their dark matter haloes \citep[][and references therein]{Bregman2007, Putman2012}. 
Most galaxies must have an active mechanism to prevent gas from cooling and turning into stars, with the  exception of very low mass objects where reionization may prevent gas from accreting into the gravitational potential well of the dark haloes \citep{Benson2002, Simpson2013}.
Several decades of research in this area have identified {\it stellar feedback} as one of the leading physical processes to explain this inefficiency in $L_*$ galaxies and below \rev{\citep[e.g.][]{Stinson2006, Sales2010, Ostriker2011, Fire,  Kimm2014, Marinacci2014, Vogelsberger2014a, Murante2015, Wang2015, Schaye2015, Martizzi2016, Li2017, Kim2018}}. 
Resolving {\it the multiphase structure of the ISM} and a detailed modelling of {\it stellar feedback} are therefore key targets in modern studies of galaxy formation. 

In fact, the ISM structure in galaxies is complex, with hot, warm and cold phases of gas coexisting and interacting~\citep[e.g.][]{McKee1977}. 
Following this complex structure becomes computationally more expensive when the numerical resolution increases.
In particular, the modelling of the cold dense regions is challenging since the time-steps demanded for hydrodynamical calculations can become rather short.
A common solution to overcome this obstacle is to impose an effective equation of state for the dense gas instead of directly resolving the individual gas phases and processes within the ISM. 
Effective equation of state models have been implemented in several different ways.
A widely-used approach is to treat the ISM as two-fluid gas assumed to be composed of cold clouds
embedded into a hot and diffuse medium \citep{Springel2003, Agertz2011}. 
An imposed polytropic relation between density and temperature $T \propto \rho^\gamma$ \citep[see also][]{Agertz2011, DallaVecchia2012} acts as a pressurization of the medium to prevent unresolved numerical fragmentation. 
Models of this nature are usually accompanied by a restricted minimum gas temperature comparable to warm
ionized gas, $T \sim 10^4 \,{\rm K}$, below which the cooling and
further collapse of the cold gas phase is not considered.  

Such an approach for the ISM, although numerically stable and 
with desirable convergence properties \citep[e.g.][]{Springel2000, Marinacci2014} may have unwanted consequences for the resulting structure of the simulated galaxies beyond simply not resolving the ISM complexity.
For instance, \citet{Benitez-Llambay2018} demonstrated that the scale height of dark matter dominated discs in similar models is set purely by the combination of sound speed in the 
mid-plane and the circular velocity. In this case
the minimum scale height of the disc becomes  
independent of numerical resolution, or in other words, while
one might increase the numerical accuracy of the simulation, 
the vertical structure of the discs will nonetheless
remain unresolved, resulting in discs that are thicker and 
kinematically hotter than observed in low mass galaxies. 
The overall  structure of galaxies in such models is 
overly-smooth on scales of a few hundred parsecs 
-- the scales of molecular clouds -- as a result of 
the missing complexity in the gas phases, severely limiting the 
predictive power of these simulations on small scales. In this 
case the appearance of the ISM will not change even if the numerical 
resolution is increased dramatically, as it is intimately linked 
to the effective equation of state chosen to treat the ISM.

Efforts to build models of resolved multiphase ISM gas structure must go hand-in-hand with the development of a stellar feedback model that can regulate and prevent the runaway collapse of the cold ISM gas. 
Several energy and momentum injection channels have to be considered including contributions 
from SN explosions at the end of the lifetime of 
massive stars, but also the impact of their early ionizing 
radiation on the surrounding media
\citep{Stinson2013, Sales2014, Gupta2016, Emerick2018, Haid2018}. 
Detailed modelling suggests that the 
feedback energy budget is dominated by radiation 
while the direct momentum injection budget
is comparable between SN and radiation \citep{Agertz2013}. 
Radiation impacts the density and
temperature of gas through photoionization 
and  radiation pressure. 
Moreover, dust in the ISM can boost the impact of radiation pressure via photon trapping where a single photon will undergo multiple scatterings. 
All these processes should be included in models that aim to resolve the multiphase ISM.

What remains less well understood in this picture is the efficiency with which the available energy and momentum couples to the surrounding gas. 
This coupling efficiency determines how important each feedback channel is in regulating star formation and launching galactic outflows. 
Although analytic arguments and some numerical simulations suggest that radiation pressure might be responsible for the launching
of outflows of several hundred ${\rm km\,s^{-1}}$ observed in galaxies \citep{Veilleux2005, Heckman2000, Pettini2001}, other authors find the effect of radiation pressure to be more modest \citep{Krumholz2012,Krumholz2013} and the overall impact of radiation to be determined mostly by photoionization effects \citep{Sales2014, Rosdahl2015}. 

Idealized radiative transfer experiments in \citet{Sales2014} indicate that although radiation pressure has the ability to eventually push the gas to galactic-outflow speeds, the timescales required to propel the gas to these speeds are much longer than those of photoionization. 
As a result, stars will heat their surrounding gas via photoionization, driving gas expansion and lowering the
surrounding gas density.
This process further limits the number of photon absorption events  and therefore limits the radiation pressure momentum coupling. 
The results from these idealized experiments are in agreement with other works in the literature that follow star formation and feedback on cloud scales \citep{Rosdahl2015, Dale2017, Walch2012, Walch2015, Kim2018b} as well as on idealized disc patches \citep{Kannan2018b}. 
The relevance of radiation pressure is therefore questionable.
There is general consensus that radiation pressure can dissipate local cold clouds, but uncertainties remain about its contribution to launching galactic 
scale outflows \citep[see ][]{Krumholz2018}.  
Furthermore, since the impact of radiation pressure
might be considerably larger for very dusty and high density star
forming regions, such as those characteristic of Ultra Luminous
Infrared Galaxies \citep[ULIRGs,][]{Murray2010, Ochsendorf2014}, the actual environment is also important.

Other sources of stellar feedback remain still under-explored and their coupling to SN and radiation feedback is not well understood.
Examples include the impact of cosmic rays \citep[e.g.][]{Recchia2016, Pais2018, Jacob2018}, stellar jets \citep{Frank2014}, high-energy photons from X-ray binaries \citep{Kannan2016}, runaway stars \citep{Kimm2014}, among others. 
As such, caution should be exercised when interpreting the results of the models or the meaning of parameters currently used in numerical simulations since they are, by construction, only approximations to the real physics driving the structure and evolution of galaxies. 

Similarly, even when the physics is well understood, there are still ad-hoc choices made in the numerical implementation of 
these processes. For instance, one might consider the total 
energy input per SN event in $10^{51}\,{\rm erg}$ as 
a relatively well constrained quantity. However, the number of
neighboring resolution elements used to distribute that energy or the corresponding
momentum can significantly change the overall impact of that SN
event \citep[e.g. ][]{Sales2010, DallaVecchia2012}. It is therefore
important to benchmark new ISM models against both, numerical
resolution as well as the robustness of internal numerical assumptions.

Current state-of-the-art zoom-in simulations have reached a resolution where it becomes desirable to remove the effective ISM approach and instead model the multiphase gas structure and related stellar feedback in more detail. 
Even full-volume 
simulations have reached mass resolutions of the order of
$\sim 10^5\, \rm M_\odot$ and therefore are able to resolve ${L^\star}$ 
galaxies with millions of gas and stellar resolution elements. 
For example, the IllustrisTNG TNG50 simulation achieves a $8.5\times 10^{4}\,{\rm M_\odot}$ mass resolution and a minimum gas softening of $74\,\pc$ throughout a simulation box of $\simeq 50\,{\rm Mpc}$ on a side~\citep{Nelson2019, Pillepich2019}.
Similarly, the galaxies in the zoom-in simulations of the Auriga project
\citep{Grand2017} contain several million of resolution elements to describe
the baryonic component of Milky Way (MW)-like galaxies, and tens of thousands for
the surrounding low mass dwarfs. These resolutions are therefore, in principle, sufficient to resolve more details within the ISM phase, and 
hence allow for physically improved models of stellar feedback and galactic outflows. 

Nevertheless, for the moment these simulations still rely on an effective equation of state model combined with a hydrodynamically decoupled wind-model, where winds are launched using phenomenological prescriptions.
These limitations motivate our work here, where we aim to construct a new ISM model that captures the multiphase gas structure alongside a more explicit local stellar feedback model.
Recent successful attempts in this direction 
have been presented \citep[e.g. ][]{Hopkins2011,Agertz2011,Agertz2013,Fire, Fire2}, 
demonstrating the possibility to achieve a more detailed treatment
of the gas and stars in galaxies, within the cosmological context 
and with high numerical resolution, on scales of MW galaxies and below
\citep{Wetzel2016, Onorbe2015,Fitts2017}.
Improved and refined ISM models should ideally be numerically well posed (this, for example, also means that they should be implementable both in particle and mesh codes) and be feasible to execute at the resolution typically achieved  by current state-of-the-art simulations. 
The goal of these novel ISM and stellar feedback models is to move the scales of numerical closure to smaller scales within the ISM. 
In contrast, effective models do not resolve structure within the ISM even if the numerical resolution is increased. 
The numerical closure scale is therefore essentially set by the ISM. Our goal is to go below this scale, to provide a model with a smaller scale for numerical closure that is better matched to the resolution of existing and upcoming simulations. 

Here we therefore present the \namemodel\ model, a novel ISM and stellar feedback model for the moving mesh code 
\arepo. 
The resolution requirements of this new model are such that it 
can yield converged results at the resolution of current state-of-the-art cosmological simulations. 
Our paper is organized as follows. Section~\ref{sec:models} gives a description of the
implementation of the most important physical processes occurring 
in the ISM that are considered in our model. Section~\ref{sec:results}
describes the initial conditions of the simulations carried out in 
this work, and presents their main features such as star formation 
rate and histories (and their sensitivity to resolution) and the ISM multiphase structure. Section~\ref{sec:outflows} shows the properties of the gaseous outflows generated by the model and the salient features of the gas in the simulations, while Section~\ref{sec:feedchan} studies the relative contribution of the different channels to the global momentum output of feedback. Finally, Section~\ref{sec:discussion} compares the numerical implementation of the physical processes considered in \namemodel\ to ISM and feedback models existing in the literature, while Section~\ref{sec:conclusion} gives a brief summary of our results. 

\begin{figure*}
\includegraphics[width=0.91\textwidth]{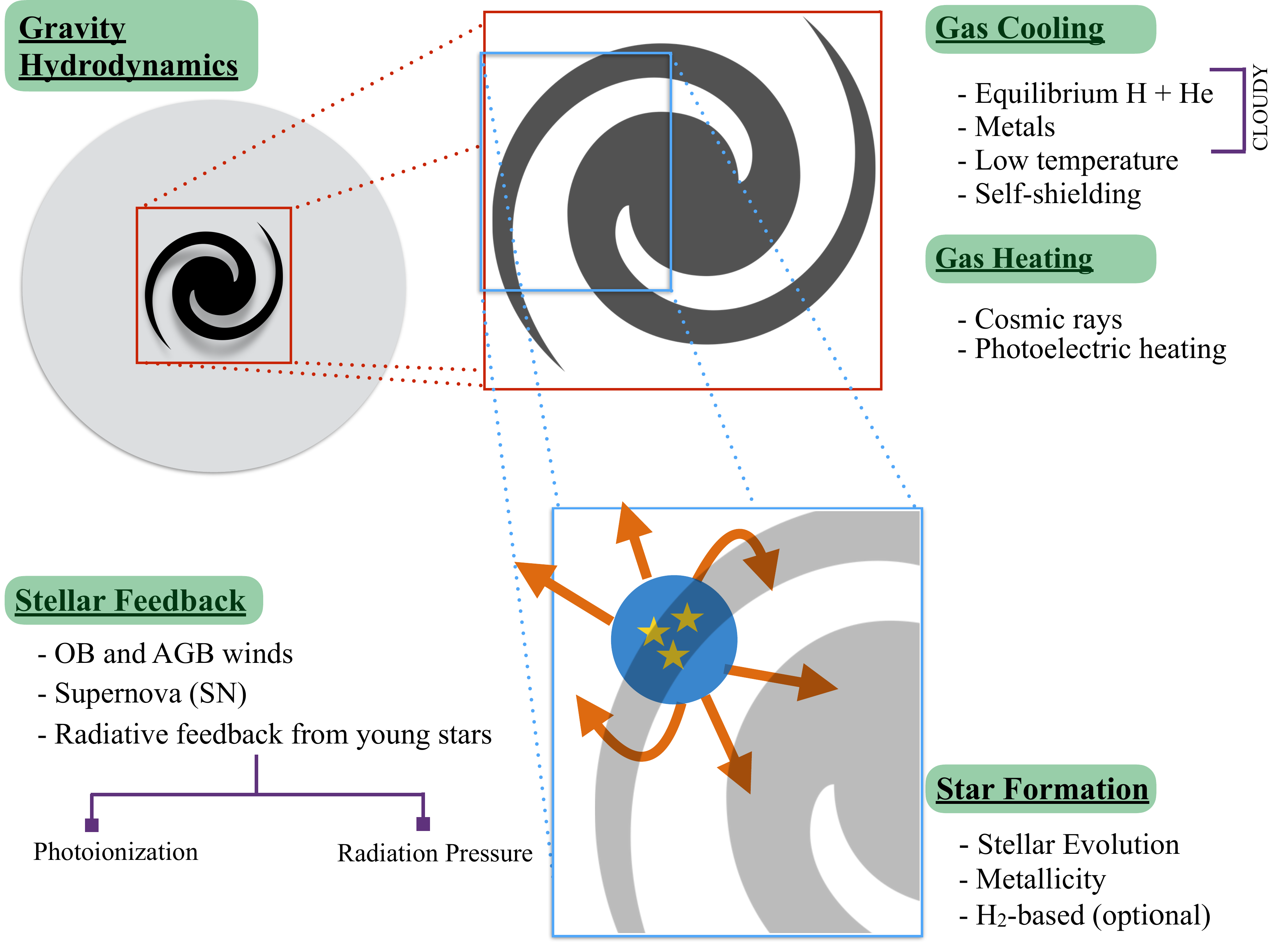}
\caption{Schematic overview of the main physical processes included in our model. These are: gravity and gas dynamics, radiative cooling of the gas -- including metal line cooling and low-temperature atomic and molecular cooling with a prescription for self-shielding from the UV radiation -- gas heating caused by the interaction with cosmic rays and photoelectric heating, a stochastic prescription for star formation, stellar evolution with the associated mass and metal return, and three main channels for stellar feedback, namely stellar winds, supernova feedback and radiation feedback from massive stars. See main text for a more detailed description of these processes.}
\label{fig:modeloverview}
\end{figure*}

\section{Methods} \label{sec:models}

In this Section we describe the main ingredients of our ISM and stellar feedback model. 
Figure \ref{fig:modeloverview} provides a high-level schematic overview of the physical processes that we consider in our model. 
These main processes are: 
$i)$ gravity and hydrodynamics -- modelled in \arepo\ with a standard oct-tree algorithm \citep{Barnes1986} for gravity and a finite volume solver over an unstructured Voronoi mesh that is allowed to move freely with the flow for hydrodynamics \citep{Springel2010,Pakmor2016}-- 
$ii)$ gas heating and cooling mechanisms (see Sec.~\ref{sec:coolheat}), which are needed to describe the emerging multiphase structure of the ISM, 
$iii)$ a stochastic implementation for the formation of stellar particles (the sites from which feedback, metal and mass enrichment are originating) from the gas phase (see Sec.~\ref{sec:stars}), and 
$iv)$ stellar feedback processes from three main channels, SNe (Sec.~\ref{sec:SN}), radiation (Sec.~\ref{sec:radiation}) and stellar winds (Sec.~\ref{sec:winds}), which are important for regulating star formation and shaping the properties of the ISM. 

\begin{figure*}
\includegraphics[width=0.98\textwidth]{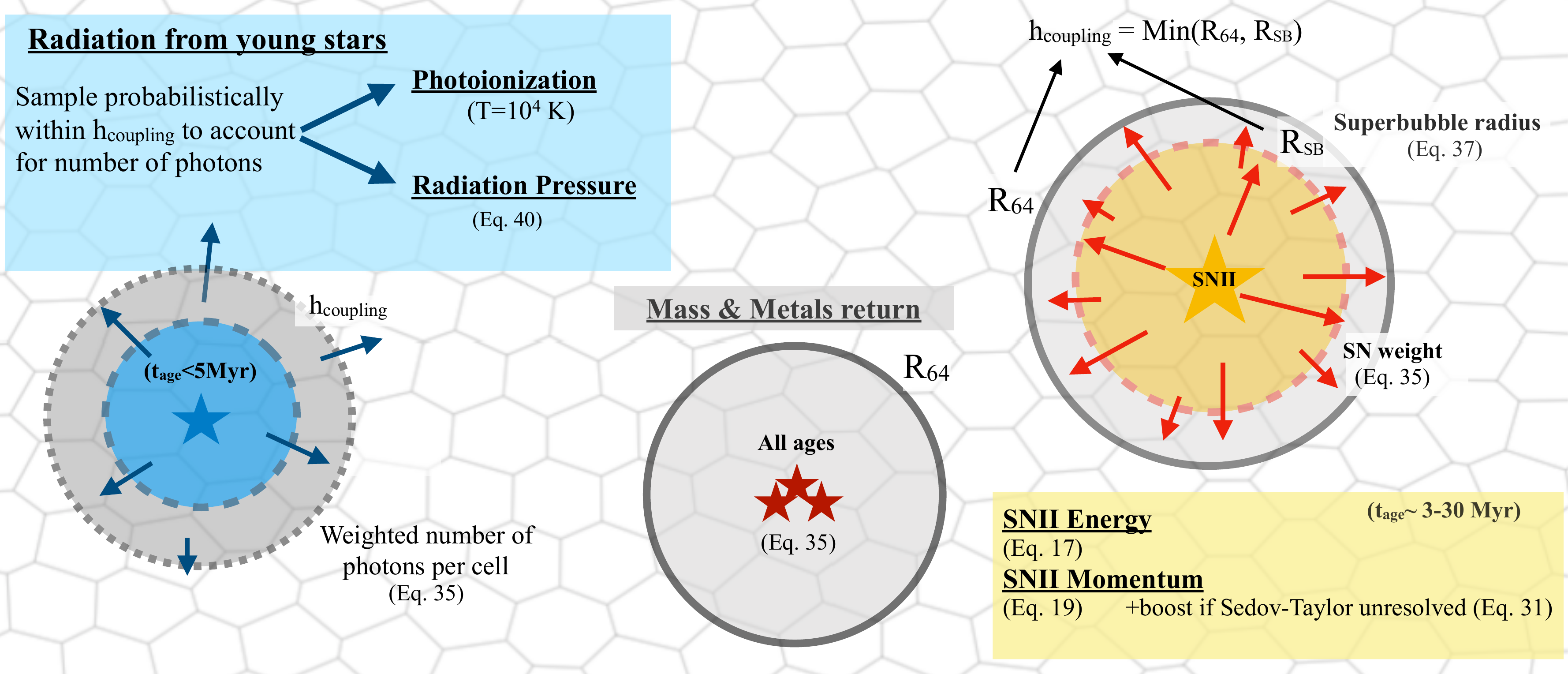}
\caption{Schematic picture illustrating the injection of feedback energy and momentum in different channels and mass and metal return from stellar particles to the interstellar medium due to stellar evolution. The background of the figure shows an illustrative Voronoi tessellation that partitions the gas distribution. These cells are affected by feedback processes and mass and metal return from stars if they fall within a predefined coupling radius, dependent on the average thermodynamic state of the gas, centred on any given stellar particle. For each process the figure indicates the key features and main equations used in the implementation. See the main text for a more complete description of the implementation of each process.}
\label{fig:model}
\end{figure*}

Figure \ref{fig:model} shows further details of the modelling of the feedback channels and mass and metal return illustrating their salient features -- in particular how feedback energy and momenta, as well as the mass returned and the metals synthesized from stars are coupled to the gas -- and referring to the key equations that we used for their implementation. In the subsections below we give a more detailed description of the physical processes that we consider in this model and their numerical implementation. 

\subsection{Cooling and heating}\label{sec:coolheat}
We first describe the gas cooling mechanisms implemented in our model. 
The thermal state of gas is modified by a cooling and heating network that models a primordial mix of hydrogen and helium undergoing two-body processes, such as collisional  excitation,  collisional  ionization, recombination,  dielectric  recombination  and  free–free  emission \citep[e.g.][]{Katz1996}, Compton cooling off CMB photons \citep{Ikeuchi1986}, and photoionization from a spatial-uniform UV background \citep{FaucherGiguere2009}. 
This primordial network is complemented by high-temperature ($T\gsim 10^4\,\K$) metal line cooling based on a self-consistently updated gas metallicity field.

Net metal cooling rates are tabulated as a function of temperature, gas density and redshift based on {\sc cloudy} calculations \citep{Ferland1998} in the presence of UV background radiation while assuming solar metallicity.
The net metal cooling rates are added to the primordial network, scaled by the gas total metallicity relative to solar. 
The details of these procedures are described in \citet{Vogelsberger2013}.
In the remaining part of this Section we only describe the implementation of the additional processes that we consider specifically for the \namemodel\ model.

An important aspect of our model is the creation of low-temperature ($T \lsim 10^4\,\K$) gas. 
Additional cooling processes must be considered to reach these low temperatures.  
We allow gas to cool via low-temperature metal line, fine-structure and molecular cooling processes, which enables the gas to reach $\sim10\,{\rm K}$ under a chemical and ionization equilibrium assumption. 
Cooling gas to low temperatures is particularly important for reaching the high densities necessary for the formation of a molecular gas phase and, ultimately, for star formation in our model (see Section~\ref{sec:stars}).  
We account for these cooling processes by implementing the fit to the {\sc cloudy} cooling tables presented in \citet[][]{Fire2}
\begin{equation} 
\begin{split}
\Lambda_{\rm mol} = & 2.896\times10^{-26}\left\{\left(\frac{T}{125.215\; \K}\right)^{-4.9202} + \right. \\
          & \left.\left(\frac{T}{1349.86 \;\K} \right)^{-1.7288}  +  \left( \frac{T}{6450.06 \;\K} \right)^{-0.3075}\right\}^{-1} \times \\
          & \left(0.001 + \frac{0.10\,n_{\rm H}}{1+n_{\rm H}} + \frac{0.09\,n_{\rm H}}{1+0.1\,n_{\rm H}} + \frac{(Z/Z_{\sun})^{2}}{1+n_{\rm H}} \right) \times \\ 
          & \left(\frac{1 + (Z/Z_{\sun})}{1+0.00143\,n_{\rm H}}\right)\, \times \, \exp{\left(-\left[ \frac{T}{158000 \;\K} \right]^{2} \right)} \;\;
          \erg\,{\rm s^{-1}}\,\cm^3.
\end{split}
\label{eq:molcooling}
\end{equation}
here $T$ is the gas temperature, $n_{\rm H}$ (in $\cm^{-3}$) is the hydrogen number density, and $Z$ and $Z_{\sun}$ are the gas and the solar metallicities. To model self-shielding of the gas at high densities ($n \gsim 10^{-3}\,\cm^{-3}$), we multiply equation (\ref{eq:molcooling}) by $1-f_{\rm ss}$, where $f_{\rm ss}$ is the self-shielding factor computed using the parameterization presented in \citet{Rahmati2013}. More specifically, the factor varies as a function of redshift and gas density as 
\begin{equation}
    f_{\rm ssh} = (1 - f) \left[1 + \left(\frac{n_{\rm H}}{n_0}\right)^\beta\right]^{\alpha_1} + f \left[1 + \left(\frac{n_{\rm H}}{n_0}\right)\right]^{\alpha_2},
    \label{eq:fss}
\end{equation}
where $n_{\rm H}$ is the hydrogen number density (in $\cm^{-3}$) of each cell and the parameters ($n_0$, $\alpha_1$, $\alpha_2$, $\beta$, $f$) encode the redshift dependence \citep[see][their Table A1]{Rahmati2013}. The self-shielding factor is also used to suppress the ionization and heating rates entering the primordial network and the normalization of the UV spectrum entering the determination of the contribution to the cooling of metal lines. Above $z = 6$ no self-shielding correction is considered. More details about self-shielding can be found in \citet{Vogelsberger2013}. Since the test simulations presented in Sec. \ref{sec:results} follow the evolution of an isolated, Milky Way-type galaxy (i.e. they are not cosmological), equation~(\ref{eq:fss}) adopts the redshift zero parameterization.  

We also take into account cosmic ray and photo-electric heating. 
Both processes are thought to be important for the thermal balance and stability of the cold ($T\sim 50\,\K$) and warm ($T\sim 8000\,\K$) phases of the ISM \citep[see e.g.][]{Field1969,Wolfire1995}. 
Cosmic ray heating is implemented using the form \citep[see][]{Guo2008}
\begin{equation}
\Lambda_{\rm CR} = -10^{-16}(0.98 + 1.65) \tilde{n}_e e_{\rm CR}n_{\rm H}^{-1} \,\,\,\,{\rm erg\,s^{-1} cm^3},\label{eq:CR}
\end{equation}
where $\tilde{n}_e$ is the electron number density in units of the hydrogen atom number density $n_{\rm H}$ (in ${\rm cm^{-3}}$) and $e_{\rm CR}$ is the cosmic ray energy density that
is parametrized by
\begin{equation}
 e_{\rm CR} =
 \begin{cases}
 \displaystyle 9\times 10^{-12} & \text{${\rm erg\,cm^{-3}}$} \,\,\,\, \text{for $n_{\rm H} > 0.01\,\cm^{-3}$}\\\\
 \displaystyle 9\times 10^{-12} \left(\frac{n_{\rm H}}{0.01\,\cm^{-3}}\right) & \text{${\rm erg\,cm^{-3}}$}\,\,\,\,\text{for $n_{\rm H} \leq 0.01 \,\cm^{-3}$},
 \end{cases}
\end{equation}
in such a way that it becomes progressively less important for low density gas \citep[see also][]{Fire2}. In equation (\ref{eq:CR}) the first term represents the heating through hadronic losses while the second term is related to Coulomb interactions between cosmic rays and the gas. By putting together the last two equations the heating rate in dense gas is $\Lambda_{\rm CR} \simeq  -2.37\times10^{-27}n_{\rm H}^{-1}\,{\rm erg\,s^{-1}\,cm^{3}}$, a typical value used in ISM studies \citep[see][equation 3.31]{Tielens2010}. 

For the photoelectric heating -- i.e. the emission of electrons from dust grains, in particular polycyclic aromatic hydrocarbons (PAHs), caused by the photoelectric effect due to the interstellar radiation field -- we employ the rate from \citet[][equation~19]{Wolfire2003} 
\begin{multline}
\Lambda_{\rm phot} = -1.3\times10^{-24}\,\tilde{e}_{\nu}^{\rm pe}\,n_{\rm H}^{-1}\,\left(\frac{Z}{Z_{\sun}} \right) \times \\ 
\left(\frac{0.049}{1 + (x_{\rm pe}/1925)^{0.73}} + \frac{0.037\,(T/10^{4}\;\K)^{0.7}}{1 + (x_{\rm pe}/5000)} \right)  \;\;
          \erg\,{\rm s^{-1}}\,\cm^3, 
\end{multline}
\begin{equation}
\begin{split}
x_{\rm pe} &\equiv \frac{\tilde{e}_{\nu}^{\rm pe}\,T^{0.5}}{\Phi_{\rm PAH}\,\tilde{n}_{e}\,n_{\rm H}},
\end{split}
\end{equation}
where $\tilde{e}_{\nu}^{\rm pe}$ is the photon energy density, normalized to the Milky Way units, $\tilde{e}_{\nu}^{\rm pe} \equiv e_{\nu}^{\rm pe} / (3.9\times10^{-14}\,{\rm erg\,cm^{-3}})$, which we set to one in our calculations, and $\Phi_{\rm PAH}$ is a factor that incorporates the uncertainties in the interaction rates between atoms and dust grains in the molecular regime. Observations of carbon density ratios in diffuse clouds suggest that $\Phi_{\rm PAH} = 0.5$ \citep{Jenkins2001}. Therefore, we fix this parameter to that value, which is also the fiducial value adopted in \citet{Wolfire2003}. The factor $(Z/Z_\odot)$ links the dust abundance in the gas to the metal abundance, which is a common choice in this type of modelling.

\subsection{Star formation implementation}\label{sec:stars}
Cold dense gas will eventually be converted to stars. To convert gas cells into star particles we follow a probabilistic approach~\citep[e.g.,][]{Springel2003}.
Star particles, representing a coeval simple stellar population following a \citet{Chabrier2001} initial mass function, are created stochastically according to the probability derived
from the star formation rate ($\dot{M}_{\star}$) of each gas cell computed as
follows
\begin{equation}
 \dot{M}_{\star} =\left\{
  \begin{array}{ll}
    \,\,\,\,\,\,0 & \rho < \rho_{\rm th}\\
    \\
  \displaystyle\epsilon\frac{M_{\rm gas}}{t_{\rm dyn}} & \rho \geq \rho_{\rm th},
  \end{array}\right.
 \label{eq:SFRStinson}
\end{equation}
\noindent where $\epsilon$ is an efficiency factor that we set to $0.01$, in line with observational determinations \citep[see e.g.][]{Krumholz2007}, and $t_{\rm
dyn}$ is the gravitational dynamical time of any given gas cell defined as 
\begin{equation}
 t_{\rm dyn} = \sqrt{\frac{3\pi}{32 G \rho_{\rm gas}}},
 \label{eq:tdyn}
\end{equation}
with $\rho_{\rm gas}$ and $M_{\rm gas}$ indicating the gas density and mass,
respectively. We note that other choices for the star formation efficiency parameter $\epsilon$ are possible. For instance, in \citet{Fire2} it is shown that the exact level of star formation is independent of the value of $\epsilon$, provided that star formation is indeed feedback regulated. We have not explicitly checked this in our model and instead we have chosen a value more in line with observational determinations \citep[see also][]{Smith2018}. Also, \citet{Agertz2013} report that variations of a factor of two in the normalization of the Kennicutt-Schmidt relation \citet{Kennicutt1998} are possible when passing from $\epsilon = 0.01$ to $\epsilon = 0.1$. We defer to future work for a detailed exploration of the sensitivity of the model to the values of the parameters (see Table~\ref{tab:parameters} for a list of the key parameters of the model).

A first criterion for a gas cell to be eligible for star formation
is to have a density above a specific density threshold $\rho_{\rm th}$. We  set this parameter to $100\,{\rm cm^{-3}}$, which is in the range of average densities ($10^{2}-10^{3}\,\cm^{-3}$) of giant molecular clouds \citep{Ferriere2001}. Other choices of the density threshold above this value \rev{do not significantly impact the star formation history and the establishment of the Kennicutt-Schmidt relation in the simulations that we are going to present in Section~\ref{sec:results}. As for the star formation efficiency value, we postpone a more complete analysis of this aspect to future work}. 

We also add an additional criterion based on the virial parameter $\alpha$ to restrict star formation only to gravitationally bound regions, i.e. those that are prone to the onset of gravitational collapse \citep{Semenov2017}. Namely we compute for each cell $i$
\citep[][]{Fire2}
\begin{equation}
 \alpha_i = \frac{||\nabla\otimes\boldsymbol{v}_i||^2 + (c_{s,i} / \Delta x_i)^2}{8\pi G \rho_i},
 \label{eq:alpha}
\end{equation}
and allow star formation only for cells in which $\alpha < 1$, which indicates that the gas is not able to overcome gravitational collapse via gas motion and thermal support. In
\eref{eq:alpha}, $\boldsymbol{v}$ is the gas velocity, $c_s$ the sound speed,
$\Delta x$ the cell size (defined as the radius of a sphere having the same
volume as the cell), $\rho$ is the gas density and $G$ the gravitational
constant, and $||\nabla\otimes\boldsymbol{v}_i||^2$ is defined as
\begin{equation}
||\nabla\otimes\boldsymbol{v}_i||^2 \equiv \sum_{i,j} \left(\frac{\partial v_i}{\partial x_j}\right)^2.
\end{equation}

Once the star formation rate is determined, the stellar mass that would 
be formed in the current time-step $\deltat$ by the cell $i$ is
\begin{equation}
 M_{\star,i} = M_i \left[ 1 - \exp\left(-\frac{\dot{M}_{\star}\Delta t}{M_i}\right)\right],
 \label{eq:starmass}
\end{equation}
where $M_i$ is the mass of gas in the cell and $\dot{M}_{\star}$ the star
formation rate determined in equation~(\ref{eq:SFRStinson}). Instead of forming
stellar mass in a continuous way as prescribed by the previous equation, a
stochastic approach, in which a gas cell is turned into a collisionless star
particle with a probability consistent with the local SFR, is adopted. This
probability can be easily found in \eref{eq:starmass} by noting that in a time
$\deltat$ a fraction $p\equiv 1 - \exp(-\dot{M}_{\star}\Delta
t/M_i)$ of the cell mass will be converted to stars. Hence, to decide
whether the conversion occurs, a uniformly distributed random variable $p^{\star}$ in the
interval $[0, 1]$ is extracted and compared with $p$. A cell is converted to
a star particle if $p^{\star} < p$.  The phase-space variables of the newly
created star particle are inherited from those of the converted gas cell. 

We note that with the procedure just described all the created star particles
would have the mass of the parent gas cell. It is sometimes however
desirable to change this behaviour. For instance, an equal mass for all stellar
particles can be desirable. This can easily be accommodated in this scheme by
multiplying $p$ with the ratio between the mass of the cell and the target mass
for the star particle. Another typical case is to enforce that the stellar mass is within a predetermined factor from a target value, in analogy with the standard Lagrangian refinement criterion for gas cells adopted in \arepo\ \citep[see][]{Vogelsberger2012}. As in these cases usually not all the gas cell mass is turned into a star, the conservative variables in the gas cell must be changed accordingly. We note that the scheme to generate stellar particles just described is also the standard approach in \arepo\ \citep[see][]{Vogelsberger2013}.

Finally, our model can also base the computation of the star formation rate on molecular gas only, given the strong observational correlation between these two quantities \citep{Bigiel2008,Leroy2013}. This is achieved by multiplying $M_{\rm gas}$, appearing 
in equation (\ref{eq:SFRStinson}), by the factor \citep[see][]{McKee2010, Krumholz2011}:
\begin{equation}
 f_{\rm H_2} = 1 - \frac{3s}{4 + s},
 \label{eq:H2}
\end{equation}
with 
\begin{equation}
s = \frac{\log[1 + \chi (0.6 + 0.01 \chi)]}{0.6 \tau},
\end{equation}
\begin{equation}
\chi = 2.3  \frac{1 + 3.1 (Z/Z_\odot)^{0.365}}{3}, 
\end{equation}
\begin{equation}
\tau = 0.067 \left(\frac{Z}{Z_\odot}\right) \left(\frac{\Sigma}{{\rm M_\odot\,pc^{-2}}}\right),
\end{equation}
\begin{equation}
\Sigma = \rho\left(\frac{\rho}{||\nabla \rho||}\right).
\end{equation}
However, for sufficiently dense gas, as it is the case for the simulations described below, $f_{\rm H_2} \sim 1$. $f_{\rm H_2}$ is enforced to be positive, meaning that if its value goes below zero then it is set to zero and no star formation occurs. This can represent a problem in the case of zero metallicity gas -- a situation that occurs in cosmological simulations, in which the gas starts from a metal free condition -- then no star will ever be formed. To circumvent this problem, a minimum metallicity floor of $Z = 10^{-5}Z_\odot$ is imposed, thus allowing star formation to occur in gas with zero or very low metallicity values.

\subsection{Feedback from supernovae}\label{sec:SN}

SN feedback is thought to play a crucial role in regulating star formation and the resulting ISM structure through the injection of momentum in the gas \citep[e.g.][]{Agertz2011, Agertz2013, Aumer2013b, Guedes2011, Hopkins2011, Fire, Kimm2014, Stinson2006, Stinson2013}.
SN momentum injection is responsible for driving turbulence \citep{Martizzi2016} and giving rise to galactic-scale outflows \citep{Li2017}. 
However, implementing SN feedback effectively is not trivial, especially in simulations with limited resolution \citep[see e.g.][]{Scannapieco2012}. 
The primary difficulty associated with modelling SN feedback is to properly capture the early energy-conserving expansion phase of a SN remnant, the so-called Sedov-Taylor phase.
During this phase, the momentum imparted to the ISM gas is generated by an overpressurized central gas bubble that expands into, sweeps up, and accelerates ambient material.
The radial momentum of an expanding SN shell can reach a factor of $\sim 10$ above the initially injected momentum \citep[see e.g.][]{Martizzi2015}.
It is this boosted momentum injection that is largely responsible for the regulation of star formation \citep{Ostriker2011} or the driving of galactic winds \rev{\citep{Kim2018}}. 
Therefore, an effective SN implementation that appropriately captures the total momentum and energy injection is important, especially if the early Sedov-Taylor blast phase expansion is not explicitly resolved.

Many approaches have been adopted to increase the efficiency of SN feedback. 
\rev{Common approaches are temporarily turning off the cooling of gas \citep{Stinson2006} and selectively \citep{Murante2010} or stochastically \citep{DallaVecchia2012} heat the gas affected by feedback.}
Both approaches allow sufficient momentum to build up to enable the launching of outflows. 
Additionally, some schemes inject momentum while hydrodynamically decoupling gas elements for a short period of time to facilitate the launching of a galactic wind \rev{\citep{Springel2003, oppenheimer2010,Vogelsberger2013, Dave2017, Valentini2017, Pillepich2018b}}. 
In this Section, we describe our implementation of SN feedback that aims for a local treatment of energy and momentum injection into the ISM, regulating the buildup of cool/cold gas and generating gas outflows self-consistently.
 
\subsubsection{SN energy and momentum budget at injection} 

We assume that the total energy injected in a single SN event is given by
\begin{equation}
E_{\mathrm{SN}} = f_{\mathrm{SN}} E_{51},
\end{equation}
where $E_{51} = 10^{51} \mathrm{ergs}$ and $f_{\mathrm{SN}}$ is a model
parameter that encodes the SN feedback efficiency ($f_{\mathrm{SN}}\equiv1$ in all our test runs). We assume that the blast wave velocity at explosion is
\begin{equation}
v_{\mathrm{SN}} = \sqrt{\frac{2 E_{\mathrm{SN}}}{M_{\mathrm{SN}}}},
\end{equation}
such that the momentum carried by the blast wave at explosion is given by
\begin{equation}
p_{\mathrm{SN}} = M_{\mathrm{SN}} v_{\mathrm{SN}} = \sqrt{2 E_{\mathrm{SN}}M_{\mathrm{SN}}}.
\end{equation}
and $M_{\mathrm{SN}}$ is the ejecta mass per SN.

The stellar particles in our simulations do not represent single stars, but rather a stellar population. 
As such, for any given time step, $\Delta t$, multiple SN events may occur.
To inject the proper amount of energy and momentum per timestep we must first compute the number of SN events and the total associated ejecta masses. 
We thus define for each stellar particle and for each time step $\Delta t$
\begin{equation}
E_{\mathrm{SN,tot}} = f_{\mathrm{SN}} E_{51} (N_{\mathrm{SNII}} + N_{\mathrm{SNIa}}),
\label{eq:SNenergy}
\end{equation}
and
\begin{equation}
\begin{split}
&p_{\mathrm{SN,tot}} = p_{\mathrm{SNII,tot}} + p_{\mathrm{SNIa,tot}} = \\
&\sqrt{2 N_{\mathrm{SNII}} E_{\mathrm{SN}}M_{\mathrm{SNII,tot}}} + \sqrt{2 N_{\mathrm{SNIa}} E_{\mathrm{SN}}M_{\mathrm{SNIa,tot}}}
\label{eq:SNmom}
\end{split}
\end{equation}
\noindent
where $N_{\mathrm{SNII}}$ and $N_{\mathrm{SNIa}}$ are the number of type II and
type Ia SN in the time step and $M_{\mathrm{SNII,tot}}$ and
$M_{\mathrm{SNIa,tot}}$ are the total ejecta mass associated with these events,
respectively.  The total number of SNII events and their associated ejecta mass
are found by integrating over the initial mass function of the stellar particle
as \citep{Vogelsberger2013}
\begin{equation}
 N_{\mathrm{SNII}} = M_{\star}\int_{M(t+\Delta t)}^{M(t)} \Phi(m)\,{\rm d}m,
 \label{eq:SNIInum}
\end{equation}
\begin{equation}
 M_{\mathrm{SNII,tot}} = M_{\star}\int_{M(t+\Delta t)}^{M(t)} M f_{\mathrm{rec}}(m,Z) \Phi(m)\,{\rm d}m.
 \label{eq:SNIImass}
\end{equation}
In equations (\ref{eq:SNIInum}) and (\ref{eq:SNIImass}), $M_\star$ is the mass of
the star particle at birth, $\Phi(M)$ is the \citet{Chabrier2001} initial mass
function, $M(t)$ is the mass of a star that leaves the main sequence at an age $t$
and $f_{\mathrm{rec}}(M,Z)$ the amount of mass given back to the ISM by stellar
evolution \citep[which depends on stellar mass and metallicity, see
e.g.][]{Portinari1998}. We set the minimum main sequence mass for a type II
SN explosion to $8\,{\rm M_{\odot}}$ and adopt an IMF upper limit of $100\,{\rm M_\odot}$.

The calculation of SNIa events proceeds in a slightly different way.
We parametrize the temporal distribution of SNIa events using a delay time distribution (DTD) and derive the number of type Ia SN events as \citep[see][]{Vogelsberger2013}
\begin{equation}
 N_{\mathrm{SNIa}} = \int_{t}^{t+\Delta t} {\rm DTD}(t')\,{\rm d}t'.
 \label{eq:SNIanum}
\end{equation}
The form of the DTD is poorly constrained. In this work we define it as follows 
\begin{equation}
 {\rm DTD}(t) = \Theta(t - t_8) N_0 \left(\frac{t}{\tau_8}\right)^{-s} \frac{s-1}{\tau_8},
 \label{eq:SNIDTD}
\end{equation}
where $\tau_8 = 40\,{\rm Myr}$ approximates the main sequence life time of an $8\,{\rm M_\odot}$ star, $N_0 = 2.6\times 10^{-3}$ SN ${\rm M_\odot^{-1}}$,
$s = 1.12$ and $\Theta$ is the Heaviside function that parameterizes the delay between the birth of the 
stellar population and the first SNIa event \citep[see][]{Maoz2012}. 
This functional form agrees with theoretical models that link the SNIa rates to the orbital energy and angular momentum loss rate due to gravitational wave emission \citep{Greggio2005} and is also consistent with the previous implementation of type Ia mass and metal return to the ISM adopted in cosmological simulation of galaxy formation with \arepo\ \citep{Vogelsberger2013}. 
Finally, each SNIa releases the same amount of ejecta, 
$M_{\mathrm{SNIa}} \simeq 1.37\,{\rm M_\odot} {\rm SN^{-1}}$ \citep{Thielemann2003}, and therefore the total mass return follows
\begin{equation}
 M_{\mathrm{SNIa,tot}} = M_{\mathrm{SNIa}} N_{\mathrm{SNIa}}.
 \label{eq:SNIamass}
\end{equation}
Finally, we would like to stress that since for stellar ages less than $t_8$ there are no SNIa events, the two types of SNe are occurring as two distinct, temporally-separated channels.

Since SN explosions are discrete events, our model aims to mimic their discrete nature using the following procedure \citep[see 
also][]{Fire2}.
We first impose a time-step constraint for each stellar particle based on its age (i.e. evolutionary stage) 
as 
\begin{equation}
 \Delta t_\star = \min\left(\Delta t_{\rm grav}, \Delta t_{\rm evol}\right),
 \label{eq:deltastar}
\end{equation}
where 
\begin{equation}
 \Delta t_{\rm evol} = \min\left(\Delta t_{\rm SNII}, \frac{t_{\rm age}}{300}\right)\,{\rm yr},
\end{equation}
and $t_{\rm age}$ is the age of the star and 
\begin{equation}
\Delta t_{\rm SNII} = \frac{\tau_8}{N_{\rm SNII}}.
\label{eq:deltaSN}
\end{equation}
In the previous equation $N_{\rm SNII}$ is the expected number of SN events for the stellar particle over $\tau_8$. For our choice of the IMF \citep{Chabrier2001} and of the lower mass of type II SN progenitor ($8\,\msun$) about $10^{-2}\,{\rm SN\,\msun^{-1}}$ are expected. 

Both $\tau_8$ and $N_{\rm SNII}$ in equation~(\ref{eq:deltaSN}) are computed self-consistently by our stellar evolution model for each star particle, as the main sequence life time also depends on the metallicity of the star \citep[see][]{Vogelsberger2013}.
Imposing the limit derived in equation (\ref{eq:deltastar}) ensures that the expectation value for the number of SN events per timestep is of the order of  unity\footnote{There is of course a dependence on the mass of the stellar particle and, for the least resolved of our simulations, more than one SNII event may occur per time step. These are allowed (and dealt with self-consistently by our Poisson sampling) in order not to impose a too restrictive time step condition. For higher resolution simulations multiple SN explosions do not typically take place.}. To determine whether a SN event takes place, we integrate  equations (\ref{eq:SNIInum}) and (\ref{eq:SNIanum}) to calculate the expected number of SNe ($\lambda$) over the time step, $\Delta t_\star$. 
This value is taken as the expectation value of a Poisson distribution, which is sampled to obtain the actual number of discrete SN events per time step (usually either zero or one since $\lambda \ll 1$ at our fiducial resolution and because of the time step limit). 
More details about the Poisson sampling of SN events are given in Appendix \ref{sec:appA}.

\subsubsection{Accounting for PdV work in the Sedov-Taylor phase}
The initial energy conserving Sedov-Taylor phase lasts until the SN approaches the cooling radius where the post-shocked
gas reaches temperature of about $10^6\,{\rm K}$ at which point radiative losses
become important.
This cooling radius is given by \citep[see, e.g.][]{Cioffi1988, Hopkins2014, Hopkins2018} 
\begin{equation}
r_{\mathrm{cool}} = 28.4 \; E_{51} ^{2/7} \left< n\right>^{-3/7} f(Z)\,\,\mathrm{pc},
\end{equation}
where $n$ is the average gas density within $r_{\rm cool}$ in ${\rm cm^{-3}}$, and $f(Z)$ is a function of gas metallicity defined in equation (\ref{eq:fz}) below.
If we do not fully resolve the cooling radius (i.e., if SN
mass/energy is being returned on scales larger than $r_{\mathrm{cool}}$), as it
is usually the case in our simulations, then we need to explicitly account for
momentum that is generated by the $P{\rm d}V$ work by the hot post-shocked gas
during the adiabatic Sedov-Taylor expansion phase of the SN blast.

Several schemes have been devised to account for this process \citep[see e.g.][]{Kimm2014,Gatto2015,Kimm2015,Kim2017,Rosdahl2017,Hopkins2018,Smith2018}. In our model, we do so by boosting the momentum 
imparted to each gas cell $i$ influenced by SN feedback as
\begin{equation}
 \Delta p_i = \tilde{w}_i \min\left[p_{\mathrm{SN,tot}}\sqrt{1 + \frac{m_i}{\Delta m_i}}, p_t\right],
\end{equation}
where $\tilde{w}_i$ is a weight function partitioning the energy and momentum
injection among gas cells (see equations~\ref{eq:weight1} and \ref{eq:weight2}), $m_i$ is the mass of the gas cell,
$\Delta m_i = \tilde{w}_i (M_{\mathrm{SNII,tot}} + M_{\mathrm{SNIa,tot}})$ and
$p_t$ is the so-called terminal momentum, i.e. the final value of the momentum
of the SN blast at $\approx$ the cooling radius, when the evolution of the
blast wave transitions from the Sedov-Taylor phase to a momentum-conserving
phase. The terminal momentum per SN is defined as
\begin{equation}
 p_t = 4.8\times 10^5 E_{\mathrm{SN,tot}}^{13/14} \left(\frac{\left<n_{\rm H}\right>}{1\,{\rm cm^{-3}}}\right)^{-1/7} f(Z)^{3/2}\,\,{\rm M_\odot}{\rm km\,s^{-1}},
 \label{eq:pterm}
\end{equation}
and this expression is determined from high-resolution simulations of
individual SN blast waves \citep[see][equation~4.7]{Cioffi1988}.  In
equation (\ref{eq:pterm}) $\left<n_{\rm H}\right>$ is the local gas hydrogen number density around the star (determined in an SPH-like fashion, see equations~\ref{eq:neigh} and \ref{eq:rhoj}) and 
\begin{equation}
 f(Z) = \min\left[\left(\frac{\left<Z\right>}{Z_\odot}\right)^{-0.14}, 2\right],
 \label{eq:fz}
\end{equation}
with $\left<Z\right>$ being the average SPH-weighted gas metallicity around the star particle computed in the same aperture as the local gas density. $Z_\odot = 0.0127$ is the value we adopt for solar metallicity \citep{Asplund2009}.
We note that to get the value for the terminal momentum,
equation~(\ref{eq:pterm}) is multiplied by $N_{\rm SNII} + N_{\rm SNIa}$, the number
of SN events occurring at any given time step.

\subsubsection{Supernova energy and momentum coupling}
We spread the SN energy and momentum injection over a number of 
nearest gas particles using weight functions. To do so we 
identify for each star particle a predefined effective number of neighbours:
\begin{equation}
 N_{\rm ngb} = \frac{4\pi}{3}  h^3 \sum_{i}W(|\mathbf{r}_{i} - \mathbf{r}_{s}|, h).
 \label{eq:neigh}
\end{equation}
Here, $h$ is a search radius (coupling radius), $W$ is the standard
cubic spline SPH kernel \citep{Monaghan1985} and $\mathbf{r}_{i}$ and
$\mathbf{r}_{s}$ are the position vectors of the $i$-th gas neighbour and of
the star particle, respectively. Equation~(\ref{eq:neigh}) defines an implicit
relation for the search radius $h$, which is solved iteratively until the
predetermined number of neighbours $N_{\rm ngb}$ is found.  In our tests
`                       $N_{\rm ngb} = 64$ and variations of $\pm 1$ (effective) neighbour are allowed. 

Additionally, to avoid artificial numerical effects described in Sec.~\ref{sssec:limiter}
we also impose a feedback limiter, namely, a maximum radius for the coupling of energy and 
momentum from the stars, $R_{\rm SB}$. The introduction of this second spatial scale requires a special 
differentiated handling of the weights for, on one hand, the redistribution
of mass and metallicity from the stars (which must by definition be conserved) and,
on the other hand, for the distribution of energy and momentum (which might not be occurring
beyond $R_{\rm SB}$, see Sec.~\ref{sssec:limiter}). In practice, we define the effective scale for 
coupling as $h_{\rm coupling} = \min(h, R_{\rm SB})$, where $R_{\rm SB}$ is introduced
in \eref{eq:limiter} as the feedback limiter. If $h$ is larger than the feedback limiter radius,
two sets of weights are computed for the two different values of the mass ($h$) and of the
feedback ($h_{\rm coupling}$) coupling radii. Conversely, if $h$ is smaller than 
$R_{\rm SB}$, the two sets of weights have the same values. 
We now describe the procedure that we use to compute such weights.

Once $h_{\rm coupling}$ has been determined, weights are defined in such a way 
that each gas cell within $h_{\rm coupling}$ receives SN energy and momenta\footnote{Metals and mass returned from SNe are also coupled to the ISM, but with a set of weights computed within the unlimited coupling radius $h$.} proportionally to the fraction of the $4\pi$ solid angle that it covers as seen from the star position. Namely, we
define a weight $w_i$ such that \citep[see also][]{Fire2, Smith2018}
\begin{equation}
 w_i \equiv \frac{\Delta \Omega_i}{4\pi} = \frac{1}{2}\left\{1 - \frac{1}{[1 + A_i / (\pi |\mathbf{r}_{i} - \mathbf{r}_{s}|^2)]^{1/2}}\right\},
 \label{eq:weight1}
\end{equation}
where $|\mathbf{r}_{i} - \mathbf{r}_{s}|$ is the distance between the gas cell and the star particle and $A_i$ is 
the area of the gas cell defined as $A_i \equiv \pi \Delta x_i^2$ (recall that $\Delta x_i$ is the cell size). To ensure
that the correct amounts of energy and momentum are imparted each cell receives a fraction  
\begin{equation}
 \displaystyle\tilde{w_i} \equiv \frac{w_i}{\sum_i w_i}
 \label{eq:weight2}
\end{equation}
of the amount of $p_{\mathrm{SN,tot}}$ and $E_{\mathrm{SN,tot}}$ determined in equations (\ref{eq:SNmom}) and (\ref{eq:SNenergy}). Momenta are directed radially away from the star position. Feedback momentum and energy are injected in the reference frame in which the star particle is at rest,
and the final momentum and total energy of the gas cell are later transformed to the appropriate values 
for the frame of reference adopted in the simulation. \rev{We highlight that gas particles receiving feedback from supernovae are \textit{never} hydrodynamically decoupled after mass, energy and momentum are injected.}

\subsubsection{Maximum size for the SN coupling radius}
\label{sssec:limiter}
The pressure associated with an expanding SN blast wave decreases with time or, equivalently, 
the radius reached by the shock wave during its expansion. The impact of a SN blast wave will have on the surrounding ISM gas 
becomes negligible (and actually the SN remnant ceases to exist as an individual entity and blends 
into the ISM) when the pressure associated with the SN ejecta becomes comparable to the ambient 
ISM pressure or alternatively and the velocity of the shock wave propagating into the ISM becomes equal to the turbulent velocity of the ambient gas. 

At the resolution achieved in our simulations, however, we are unable to model single stars. Rather, each stellar particle formed represents a coeval single stellar population, or stellar cluster, in which multiple SN events occur as a result of the presence of associations of massive (namely OB) stars. This clustered SNe will inflate large bubbles in the galaxy's ISM  known as superbubbles \citep{Weaver1977} with a size of a few hundreds of parsec across, thus influencing the ambient ISM on scales larger than those of an individual SN remnant. 

The size of a superbubble can be estimated as \citep[see][]{MacLow1988,Weaver1977}
\begin{eqnarray}
R_{\rm SB} = 1.67\times 10^2 \nonumber \\
\displaystyle\left(\frac{L_{\rm SN}}{10^{37}\,\erg\,{\rm s^{-1}}}\right)^{1/5} &
\displaystyle\left(\frac{\Delta t}{10^{7}{\rm yr}}\right)^{3/5}
\displaystyle\left(\frac{n_{\rm H}}{1\,\cm^{-3}}\right)^{-1/5}\pc,
\label{eq:limiter}
\end{eqnarray}
where $n_{\rm H}$ is the hydrogen number density (we have assumed a hydrogen mass fraction of $0.76$ for the estimation), $\Delta t$ is the life time of the OB association ($\sim 40\,\Myr$ for a lower SN progenitor mass of $8\,\msun$) and $L_{\rm SN}$ is the mechanical SN energy, which can be defined as
\begin{equation}
L_{\rm SN} = 3.17\times 10^{36} N_{\rm SN} \left(\frac{E_{\rm SN}}{10^{51}\,\erg}\right)
\left(\frac{\Delta t}{10^{7}\,{\rm yr}}\right)^{-1}
\,\erg\,{\rm s^{-1}},
\end{equation}
where $N_{\rm SN}$ is the total number of SNe in the single stellar population and $E_{\rm SN}$ is the energy of an individual SN event.

Beyond the distance estimated in equation~(\ref{eq:limiter}) the energy/momentum injected by SNe will be unable to have a strong impact on the ISM properties, although the superbubble might be able to break out and vent material outside the star-forming disc when its size exceeds a few disc scale heights \citep[see][and the discussion of Figs~\ref{fig:gasevo} and \ref{fig:gasevo2} below]{MacLow1988}. In general,  typical values of $R_{\rm SB}$ 
are several hundreds of parsecs, depending on gas ambient density and numerical resolution. For instance, assuming $n_{\rm H} = 1\; \cm^{-3}$, the corresponding superbubble radius is in the range $\simeq 0.3-1\;{\rm kpc}$ for the  numerical resolution of the runs presented later in Sec.~\ref{sec:results} (high to low resolution, respectively). We note, however, that the numerical convergence of our model improves when $R_{\rm SB}$ is kept constant instead of adjusting this value on the fly. Therefore, a flag in our feedback implementation allows the model to either follow equation~(\ref{eq:limiter}) and adopt the limiter to density and resolution or, alternatively, to keep the value constant. Motivated by the better convergence behaviour, we choose the latter for the current work, adopting $R_{\rm SB}=0.86\;{\rm kpc}$  as our default value. We have tested that varying this in the range $0.3-3\;{\rm kpc}$ does not have a significant impact on our results. 

\subsection{Radiative feedback from young massive stars}\label{sec:radiation}

Radiation, especially from young and massive stars, has an important impact on both, the thermal and the dynamical state of the ISM gas. It can alter its ionization state and therefore the gas temperature by photoionizing it. Moreover, each absorbed photon imparts additional momentum to the gas via radiation pressure. The imparted momentum can be further enhanced by multiple photon scattering and can then become relevant at high densities, such those found in giant molecular clouds, the birth clouds of stars. 

This process is particularly important because of its timing. 
Radiative feedback can be responsible for the dispersal of such clouds \textit{before} any SN goes off \citep{Murray2010, Lopez2011, Walch2012}, thereby rendering SN feedback more effective due to the reduced gas densities in which SNe subsequently explode. This process sometimes goes under the name of early stellar feedback \citep[see e.g.][]{Stinson2013}. Some studies have also proposed radiative feedback as the key mechanism behind the launching of galactic-scale outflows \citep{Murray2005}, but this aspect is still strongly debated \citep{Krumholz2012, Krumholz2013}. We describe our implementation of the radiative feedback, and specifically of the photoionization of gas and treatment of radiation pressure, in the subsections below.

\subsubsection{Photoionization} \label{sec:photoion}
Young massive stars are a
copious source of ionising radiation. This radiation can in turn impact the ionisation
state of the surrounding gas, leading to the emergence of H{\sc ii} regions. In
our model we capture the formation of such regions as follows. 

\begin{table*}
\caption{Structural parameters of the Milky Way-type galaxy considered in this work. Columns indicate (from left to right):
mass of the dark matter halo ($M_{\rm halo}$), circular velocity of the halo at the radius at which its average density is equal to 200 times critical density for closure ($v_{200}$), halo concentration ($c$), bulge mass ($M_{\rm b}$), 
bulge (\citealt{Hernquist1990} sphere) scale length ($a$), stellar disc mass ($M_{\rm d}$), stellar disc scale length ($r_{\rm d}$), 
stellar disc scale height ($h$), gas disc mass ($M_{\rm g}$), gas disc scale length ($r_{\rm g}$), and gas fraction within $R_\odot = 8.5\,\kpc$ (computed as the ratio between gaseous and stellar disc masses; $f_{\rm gas}$).}
\centering
\begin{tabular}{ccccccccccc}
\hline
\multicolumn{1}{c}{$M_{\rm halo}$} & 
\multicolumn{1}{c}{$v_{\rm 200}$} & 
\multicolumn{1}{c}{$c$} & 
\multicolumn{1}{c}{$M_{\rm b}$} & 
\multicolumn{1}{c}{$a$} & 
\multicolumn{1}{c}{$M_{\rm d}$} & 
\multicolumn{1}{c}{$r_{d}$} & 
\multicolumn{1}{c}{$h$} & 
\multicolumn{1}{c}{$M_{\rm g}$} & 
\multicolumn{1}{c}{$r_{g}$} &
\multicolumn{1}{c}{$f_{\rm gas}$} \\
\multicolumn{1}{c}{($\msun$)} & 
\multicolumn{1}{c}{(${\rm km\,s^{-1}}$)} & 
\multicolumn{1}{c}{\,} & 
\multicolumn{1}{c}{($\msun$)} & 
\multicolumn{1}{c}{(kpc)} & 
\multicolumn{1}{c}{($\msun$)} & 
\multicolumn{1}{c}{(kpc)} & 
\multicolumn{1}{c}{(pc)} & 
\multicolumn{1}{c}{($\msun$)} & 
\multicolumn{1}{c}{(kpc)} &
\multicolumn{1}{c}{($R < R_\odot$)}\\ \hline
\\
$1.53\times10^{12}$ & 169 & 12 & $1.5\times10^{10}$ & 1.0 & $4.73\times10^{10}$ & 3.0 & 300 & $9\times10^9$ & 6.0 & 0.10\\
\hline
\end{tabular}
\label{tab:ics}
\end{table*}

\begin{table}
\caption{Parameters defining the resolution level at which the initial conditions are sampled. Columns indicate (from left to right): resolution level name, gravitational softening length  for star particles ($\epsilon_{\star}$), minimum gravitational softening length for gas cells ($\epsilon_{\rm g}$; softening is adaptive and scales logarithmically with the cell size), target 
mass of gas cell ($m_{\rm g}$), mass of a stellar bulge particle ($m_{\rm b}$), and mass of a stellar disc particle ($m_{\rm d}$).}
\centering
\begin{tabular}{lccccc}
\hline
\multicolumn{1}{l}{Resolution} &
\multicolumn{1}{c}{$\epsilon_{\star}$} &
\multicolumn{1}{c}{$\epsilon_{\rm g}$} &
\multicolumn{1}{c}{$m_{\rm g}$} &
\multicolumn{1}{c}{$m_{\rm b}$} &
\multicolumn{1}{c}{$m_{\rm d}$} \\
\multicolumn{1}{l}{level} &
\multicolumn{1}{c}{(pc)} &
\multicolumn{1}{c}{(pc)} &
\multicolumn{1}{c}{($\msun$)} & 
\multicolumn{1}{c}{($\msun$)} & 
\multicolumn{1}{c}{($\msun$)} \\
\hline
\\
low & 50.0 & 21.4 & $9.0\times 10^4$ & $1.5\times 10^5$ & $1.2\times 10^5$\\
intermediate & 21.4 & 10.0 & $1.1\times 10^4$ & $2.0\times 10^4$ & $1.5\times 10^4$\\
high & 7.2 & 3.6 & $1.4\times 10^{3}$ & $2.3\times10^3$ & $1.9\times 10^{3}$\\
\hline
\end{tabular}
\label{tab:res}
\end{table}

We define the ionising photon rate emitted by a stellar particle as
\begin{equation}
N_\star = \frac{L_\star}{\langle h\nu \rangle} = \frac{\gamma_{\star} M_\star}{\langle h\nu \rangle},
\label{eq:nion}
\end{equation}
that is the luminosity of the star divided by the average photon energy emitted above $13.6\,{\rm eV}$ and we estimate
the star particle luminosity given its mass $M_\star$ by assuming a mass-to-light ratio $\gamma_\star$. We choose
$\langle h\nu \rangle = 17\,{\rm eV}$ and $\gamma_\star = 10^3\,{\rm L_\odot}/{\rm M_\odot}$ as our fiducial values, corresponding to the
peak emission of a black-body Planck spectrum with temperature $T \sim 40,000\; \rm K$ \citep{Rybicki_Lightman}, consistent with massive OB stars. 

The mass of gas that can be photoionised by a young stellar particle (i.e. the mass within the so-called Str\"omgren radius) is usually (much) smaller than the mass of gas contained in within $h_{\rm coupling}$, the scale at which the model couples the feedback energy. We therefore carry out the photoionization probabilistically. Cells are assigned a probability of being photoionized $p = n_{\star} / (\alpha_{\rm rec} n_{\rm H}^2 V)$, where $\alpha_{\rm rec} \simeq 2.6\times10^{-13}\,{\rm cm^{3}s^{-1}}$ is the hydrogen recombination rate, $n_{\rm H} = X\rho/m_{\rm p}$ is the average hydrogen number density of the cell -- with $X$ being the hydrogen mass fraction and $m_{\rm p}$ the proton mass -- $V$ is the cell volume and $n_{\star} = \tilde{w}_i N_{\star}$ is the rate of ionizing photons injected into the cell. In other words, the probability is given by comparing the number of recombinations expected in the cell to the total photon number emitted by the source scaled by the same solid angle weighting scheme adopted to model the supernova feedback.

A random number $p'$ is then selected from a uniform distribution in the 
range [0,1] and cells where $p' < p$ are tagged for photoionization. 
For those, we impose a temperature floor $T_{\rm phot} = 1.7\times10^4\,\K$ and 
disable their radiative cooling for a duration $t_{\rm off}$ equal to the 
star particle time step. The temperature floor is consistent with a gaseous
medium made of hydrogen and photoionized by  $\langle h\nu \rangle = 17\,{\rm eV}$ photons. 
In addition to the probability criterion 
outlined in the previous Section, a gas cell has to fulfill two further
requirements to be declared eligible for photoionization: (i) its $u_{\rm 
therm}$ must be less than $1.2\times u_{\rm phot}$ (i.e. the thermal energy per unit 
mass corresponding to $T_{\rm phot}$), (ii) must be able to cool (i.e $t_{\rm off} = 0$). If any of these two conditions apply, the gas particle is considered to be already (photo)ionised \rev{-- either because they are at a high enough temperature or are kept ionised by another stellar particle --} and no action is taken. Finally, we point out that,while $t_{\rm off}$ is initialized to the \textit{star} time step ($\Delta t_\star$), its value is updated according to the duration of the cell (individual) time step, so that cooling is disabled precisely for a duration equal to $\Delta t_\star$. \rev{The heating of eligible gas cells due to photoionization will generate some over-pressurization of the gas compared to their colder (and mostly neutral) surroundings. This will lead to the expansion of the photoionized regions, which in the spirit of the model should be capturing the expansion of the ${\rm H_{\sc II}}$ regions around young stellar systems, thus imparting momentum to the gas. However, please note that this momentum generation originates from a different physical mechanism than a direct transfer of momentum from the radiation field to the gas (i.e. radiation pressure), which is accounted for in \namemodel\ as described below.}

\subsubsection{Radiation pressure} 
Radiation coming from young stars has an
additional impact on the dynamical state of the gas because of the pressure it
can exert on it. This represents a source of momentum, which is important
in dense, optically-thick regions, allowing a preprocessing of the
gas environment in which SNe will subsequently explode
\citep[e.g.][]{Agertz2013, Fire, Stinson2013}.

In our model, we account for this momentum source from young stars by considering an injection 
of momentum around each star particle of the form
\begin{equation}
\Delta p = \frac{L_\star}{c} (1 + \tau_{\rm IR}) \Delta t,
\label{eq:radpress}
\end{equation}
where $\Delta p$ is the total momentum injected by the star over the time step $\Delta t$,
$L_\star$ is the star luminosity (see equation~\ref{eq:nion}), $c$ the speed of light and 
$\tau_{\rm IR} = \kappa_{\rm IR} \Sigma_{\rm gas}$ is the optical depth of the gas to 
infrared radiation, which takes into account the multiple scattering of infrared photons in
dense gas. $\kappa_{\rm IR}$ and 
$\Sigma_{\rm gas}$ are the opacity in the infrared band and the gas column density, respectively. We set $\kappa_{\rm IR} = 10 (Z/Z_\odot)\,{\rm cm^{2}g^{-1}}$ \citep{Fire2}, although its precise value is uncertain and smaller $\kappa_{\rm IR}$ (by a factor of $2$) are also reasonable \citep{Agertz2013}. This would impact the total amount of momentum imparted by radiation. However, in our simulations radiation does not appear to be the dominant feedback channel (see Section~\ref{sec:feedchan}).
The total radiation momentum determined in equation~(\ref{eq:radpress}) is coupled to the 
gas contained within $h_{\rm coupling}$ in the same way as it is done for SN feedback 
(see equations \ref{eq:weight1}-\ref{eq:weight2}) and directed radially away from the star
position. The same maximum value for the coupling radius adopted for SN feedback 
is adopted in both photoionization and radiation pressure calculations.

\begin{table*}
\caption{Parameters adopted in this works for physical processes implemented in our ISM and stellar feedback model with their fiducial value and description.}
\centering
\begin{tabular}{llll}
\hline
Parameter & Fiducial value & Units & Description\\
\hline
\multicolumn{4}{c}{Star formation} \\
\hline
$\epsilon$               & $0.01$    & --               & Star formation efficiency    \\
$\rho_{\rm th}$          & $100$     & ${\rm cm^{-3}}$  & Star formation density threshold   \\
\hline
\multicolumn{4}{c}{Supernova feedback} \\
\hline
$f_{\rm SN}$             & $1$       & -- & SN energy relative to fiducial value                 \\
$E_{\rm SN}$             & $10^{51}$ & ${\rm erg}$ & Fiducial energy per SN          \\
$N_{\rm ngb}$            & $64$      & --  & Effective neighbour number                 \\
$\Delta N_{\rm ngb}$     & $1$      & --  & Neighbour number tolerance           \\
\hline
\multicolumn{4}{c}{Radiative feedback} \\
\hline
$\kappa_{\rm IR}$         & $10 (Z/Z_\odot)$     & ${\rm cm^2\,g^{-1}}$ & Gas infrared opacity \\
$t_{\star,{\rm max}}$    & $5$      & ${\rm Myr}$ & Maximum stellar age for ionizing radiation       \\
$\gamma_{\star}$         & $10^3$      & ${\rm L_\odot\, M_\odot^{-1}}$ & Stellar mass-to-light ratio\\
$T_{\rm phot}$           & $1.7\times 10^4$  & $\K$ & Temperature of photoionized gas\\
$\langle h\nu\rangle$           & $17$ & ${\rm eV}$ & Average ionizing photon energy \\
\hline
\end{tabular}
\label{tab:parameters}
\end{table*}

The last quantity to be determined to completely specify the model is an
estimate of the column density of the gas $\Sigma_{\rm gas}$ within $h_{\rm
coupling}$. We adopt a standard Sobolev approximation and
compute for each star particle $j$ a Sobolev length as 
\begin{equation}
\ell_j = h_{{\rm coupling},j} + \frac{\rho_j}{||\nabla \rho_j||},
\end{equation}
in which $\rho_j$ and $\nabla \rho_j$ are the gas density and the associated
gradient determined by adopting a standard SPH approach (see 
equation~\ref{eq:rhoj} below and also \citealt{Fire2}). The column density is computed as 
\begin{equation}
\Sigma_{{\rm gas},j} =  \langle \rho_{s}\rangle_{j}\ell_j,
\end{equation}
and $\langle \rho_{s}\rangle_{j}$ is given by
\begin{equation}
 \langle\rho_{s}\rangle_j = \sum_{i}W(|\mathbf{r}_{i} - \mathbf{r}_{j}|, h) m_i,
 \label{eq:rhoj}
\end{equation}
where $W$ is the cubic spline SPH kernel (see also equation~\ref{eq:neigh}), $\mathbf{r}_{i}$ and $\mathbf{r}_{j}$ are the position vectors of the $i$-th gas neighbour and of the star particle, respectively, $m_i$ the mass of the gas cell and $h$ appears in equation~(\ref{eq:neigh}). 

\subsection{Feedback from OB and AGB stellar winds}\label{sec:winds}
Stars also contribute to feedback through stellar winds. For our purposes 
two classes of stars are important: massive, short-lived OB stars and asymptotic giant branch (AGB) stars.
The former class helps in pre-processing the gas around young star particles before the onset of SNe \citep{Matzner2002,Krumholz2009} -- and in particular in dispersing the dense gas clouds from which stars are born --  a role similar to radiative feedback. In this sense they represent an additional early stellar feedback channel. 

Moreover, AGB winds are an attractive source of feedback on their own, as they continue to 
act at later times than radiation and are associated to older stellar populations.  
Their mass return can further contribute to fuel later episodes of star formation in galaxies
with their own subsequent injection of associated energy and momentum.
It is worth noting that a detailed accounting of the mechanical power carried by winds of massive stars shows that their momentum injection rate per unit stellar mass formed is comparable to that of the SNe \citep{Agertz2013}, with the difference that most of this power is released before the first SN explosion ($t \lsim 5\,\Myr$) and rapidly declines thereafter. Therefore, stellar winds represent a feedback channel that should be explicitly modelled in galaxy formation simulations, given its potential in enhancing the effect of subsequent SN explosions because of the reduction in the ambient density. 

We implement
these two contributions in our model as follows. In a first step we compute the mass loss due to these two processes.
For OB stars we parametrize the cumulative mass loss per unit stellar mass as 
\begin{equation}
 m_{\rm closs} = 
 \begin{cases}
f(t) & \text{if} \qquad  t < 1\\
g(t) & \text{if} \qquad 1 < t < 3.5\\
h(t) & \text{if} \qquad 3.5 < t < 100\\
 \end{cases}
,
\end{equation}
where \citep[see][]{Fire2}
\begin{equation}
 \begin{cases}
f(t) = 4.763 \times 10^{-3} (0.01 + \tilde{Z})  t\\\\
g(t) = 4.763 \times 10^{-3} (0.01 + \tilde{Z})  \displaystyle\frac{t^{2.45 + 0.8 \log(\tilde{Z})} - 1}{2.45 + 0.8  \log(\tilde{Z})} \\
\qquad\qquad\qquad+ f(1) \\\\
h(t) = g(3.5) -4.57 \times 10^{-2} \left[\left(\displaystyle\frac{t}{3.5}\right)^{-2.25} - 1\right] \\ \qquad\qquad\qquad +  4.2\times10^{-6} (t - 3.5)\\
 \end{cases}
.
\end{equation}
In the previous expressions $\tilde{Z}$ is the metallicity of the stellar particle in solar units (and capped to 1.5), time is 
expressed in Myr, and for $t > 100\,{\rm Myr}$ $M_{\rm closs} = h(100)$. Over a time step $\Delta t$
the mass lost by a star particle with an initial mass equal to $M_\star$ is $M_{\rm loss} = M_\star[m_{\rm closs}(t + \Delta t) - m_{\rm closs}(t)]$. Moreover,
OB stellar winds leave each stellar particle with the same chemical composition as the particle itself.
We consider OB stars wind injection only for stellar life times smaller than the one associated with the minimum mass of type II SNe progenitors ($8\,{\rm M_{\odot}}$).
Finally, AGB stars are treated by the stellar evolution model already present in 
\arepo\ \citep{Vogelsberger2013}, so we consider those values for their contribution to the metal and total mass return.

\begin{figure*}
\includegraphics[width=0.49\textwidth]{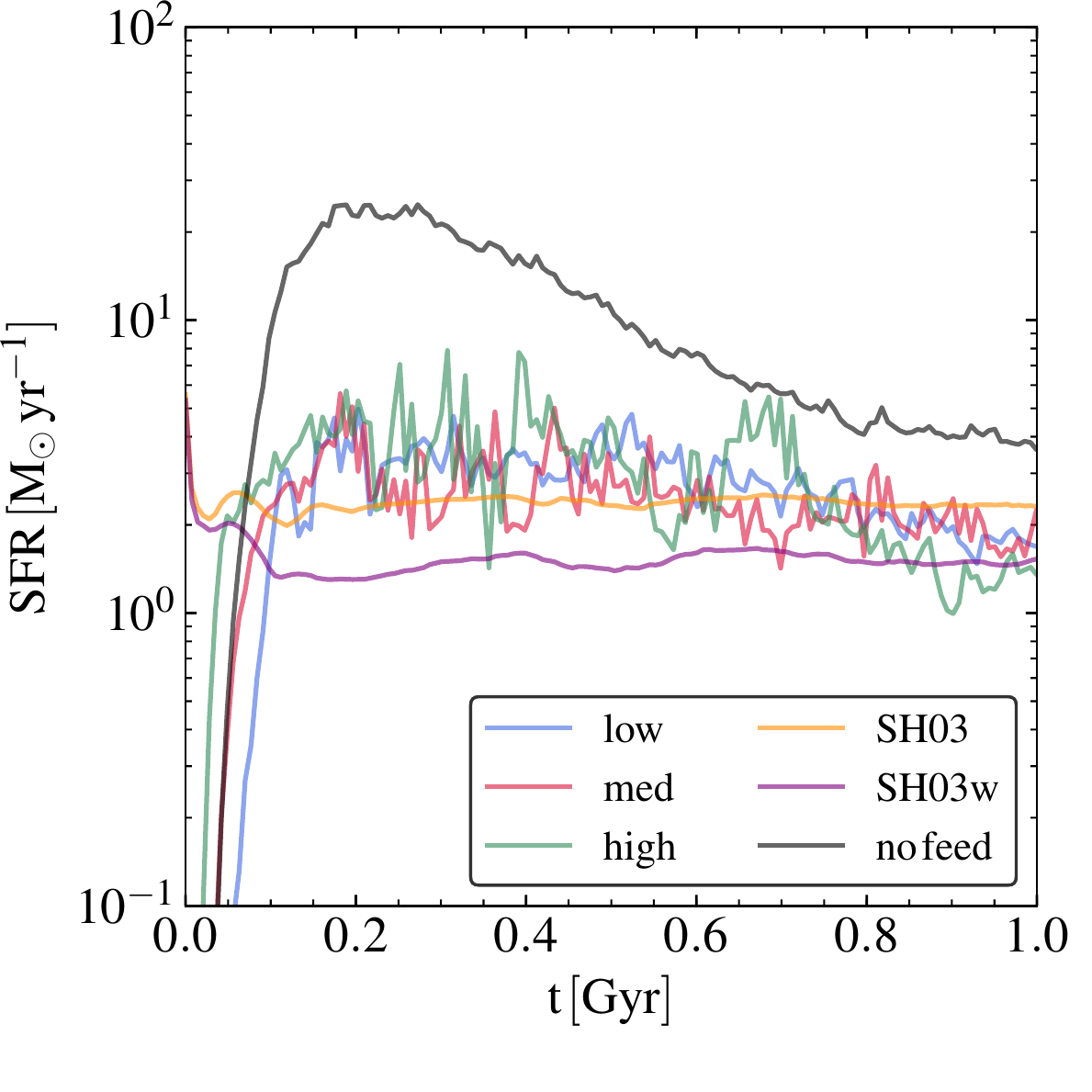}
\includegraphics[width=0.49\textwidth]{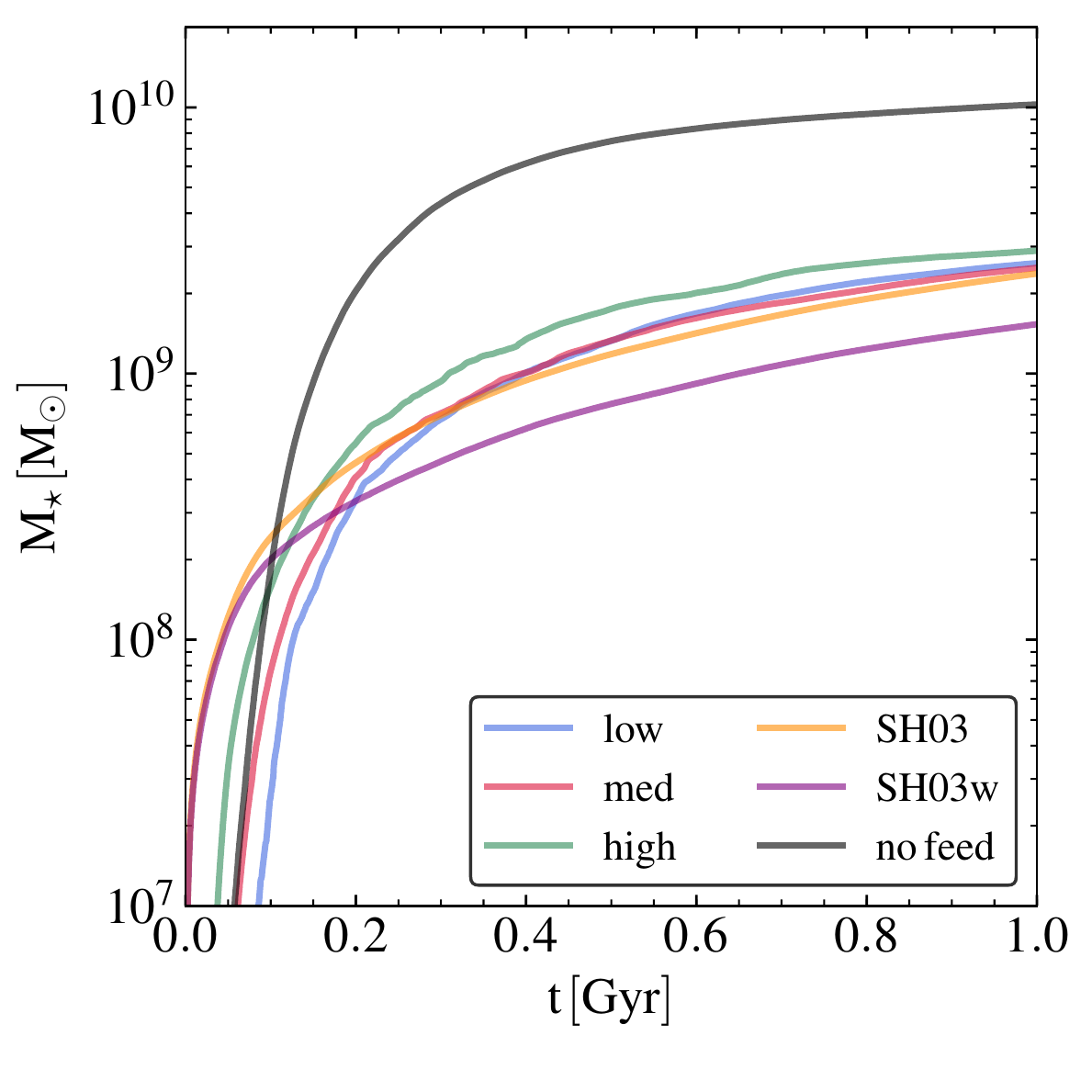}
\caption{\textit{Left}: Star formation history of the Milky Way galaxy at low (blue), intermediate (red) and high (green) resolution. The lines show the instantaneous star formation rate, computed as the sum of the star formation rates of each individual gas cell at a given time $t$. The self-regulation of star formation by feedback (the star formation rate is approximately constant over the simulated time span) is evident also in comparison with a simulation in which all the feedback processes are turned off (solid black line), which peaks at about $30\,{\rm M_\odot\,yr^{-1}}$ after $0.2\,\Gyr$ of evolution, about an order of magnitude larger SFR. \rev{A comparison with the \citet{Springel2003} model without (dark orange) and with (purple) galactic winds reveals that the predicted star formation rates of our model ($\sim 3\,{\rm M_\odot\,yr^{-1}}$) are rather similar. The inclusion if hydrodynamically decoupled winds reduces the overall SFR of the \citet{Springel2003} model by about 35 per cent.}  \textit{Right:} Cumulative stellar mass formed as a function of time for the Milky Way galaxy at low (blue), intermediate (red), and high (green) resolution and in the no feedback case (black) and with the \citet{Springel2003} model \rev{without (dark orange) and with galactic winds (purple)}. Although there is a residual dependence on resolution, the plot demonstrates the robust convergence properties in the regulation of star formation by feedback with about a factor of four to five more stellar mass formed in the no feedback case and an amount of stellar mass consistent with the one predicted by the \citet{Springel2003} model. \rev{As for the left-hand panel, the inclusion of galactic winds reduces the total stellar mass formed in the \citet{Springel2003} model by about 35 per cent.}
} 
\label{fig:sfr}
\end{figure*}

To compute the energy and momentum injection from stellar winds we set \citep{Fire2}
\begin{equation}
 E_{\rm winds} = \Delta t L_{\rm kin} = M_{\rm loss} \times \psi \times 10^{12}\,{\rm erg\,g^{-1}},
\end{equation}
where $M_{\rm loss}$ is the mass loss determined above and 
\begin{equation}
 \psi = \displaystyle\frac{5.94 \times 10^4}{1 + 
 \left(\displaystyle\frac{t_{\rm Myr}}{2.5}\right)^{1.4} + \left(\displaystyle\frac{t_{\rm Myr}}{10}\right)^5} + 4.83.
\end{equation}
The total momentum injection rate is thus determined as
\begin{equation}
 p_{\rm winds} = \sqrt{2 M_{\rm loss} E_{\rm winds}}.
\end{equation}
The mass, metallicity, momentum and energy are injected as in the SN case in the rest frame of the star and the final values of momentum and total energy of the gas cell are then transformed back into the reference frame of the simulations. The same maximum values to the coupling radii also apply. Finally, we would like to note that differently from SNe, the feedback injection from stellar winds is a continuous process and is implemented accordingly. However, OB wind injection and the associated feedback is only performed if the returned mass over a given time step is larger than $10^{-4}$ times the mass $M_\star$ of the stellar particle at birth. If this is not the case, the mass loss is accumulated until the first time step at which the mass return threshold for this channel is reached.
\rev{Our implementations of cooling, star formation, and stellar feedback are inspired by those in the FIRE-2 simulations and largely follow, as highlighted by the references throughout this Section, the methods laid out in \citep{Hopkins2018}, but with a number of modifications as summarized in detail in Section~\ref{sec:discussion}.}

\section{Isolated Galaxies}\label{sec:results}

\subsection{Initial conditions and model setup}
We test our new feedback implementation for isolated late-type galaxies. Our
main goal is to explore the ability of the new model to
regulate star formation, and to investigate its impact on the
properties of the ISM of the galaxy.

We construct the initial conditions following the technique described in \citet[][see also \citealt{Hernquist1993}]{Springel2005b}. We set up an equilibrium 
compound galaxy model representative of the Milky Way consisting of a dark matter halo, 
a bulge and a stellar and gaseous discs. Both the halo and the bulge are modelled with a
\citet{Hernquist1990} profile. The stellar and gaseous discs are exponential in the radial direction. While the stellar disc follows a $\sech^2$ distribution in the vertical direction, the vertical profile of the gaseous disc is computed self-consistently to ensure hydrostatic equilibrium at the beginning of the simulation. The initial gas temperature is set to $10^4\,\K$. 

For the structural parameters of these components we selected the same values that were used to construct the Milky Way galaxy model presented in \citet{Hopkins2012}. 
This model has a total mass of 
$1.6\times10^{12}\,{\rm M_{\odot}}$ of which $\simeq 1.5\times 10^{10} {\rm M_{\odot}}$ are contained in the bulge, $\simeq 4.73\times 10^{10}\,{\rm M_{\odot}}$ and $\simeq 9\times 10^{9}\,{\rm M_{\odot}}$ in the stellar
 and gaseous discs, respectively, resulting in a disc gas fraction of $f_{\rm gas} \simeq  10$ per cent within $R = 8.5\,\kpc$. 
The gas in the disc has a metallicity equal to the solar value of $0.0127$ \citep[see][]{Asplund2009}.
A summary of the structural parameters is listed in Table~\ref{tab:ics}.

Three different resolution levels (low, intermediate and high) are created, differing in the number of element used to sample each galaxy component. At the lowest resolution level we put $N_{\rm bulge} = 10^5$ particles in the bulge, $N_{\rm disc} = 4\times 10^5$ particles in the stellar disc and $N_{\rm gas} = 10^5$ gas cells in the gaseous disc. Each time we increase the resolution level, we do so by increasing the particle number by a factor of 8. The parameters defining each resolution level are listed in Table \ref{tab:res}. For efficiency reasons, the simulations model the dark matter halo as a static gravitational field. Although modelling the dark matter in a static way is an approximation, in the sense that the dark matter halo will not react to changes occurring in the system, the results are only marginally affected by this choice. For instance, the final stellar mass formed in our simulations with a static versus live halo agree within $\approx 10$ per cent.

\begin{figure*}
\includegraphics[width=0.49\textwidth]{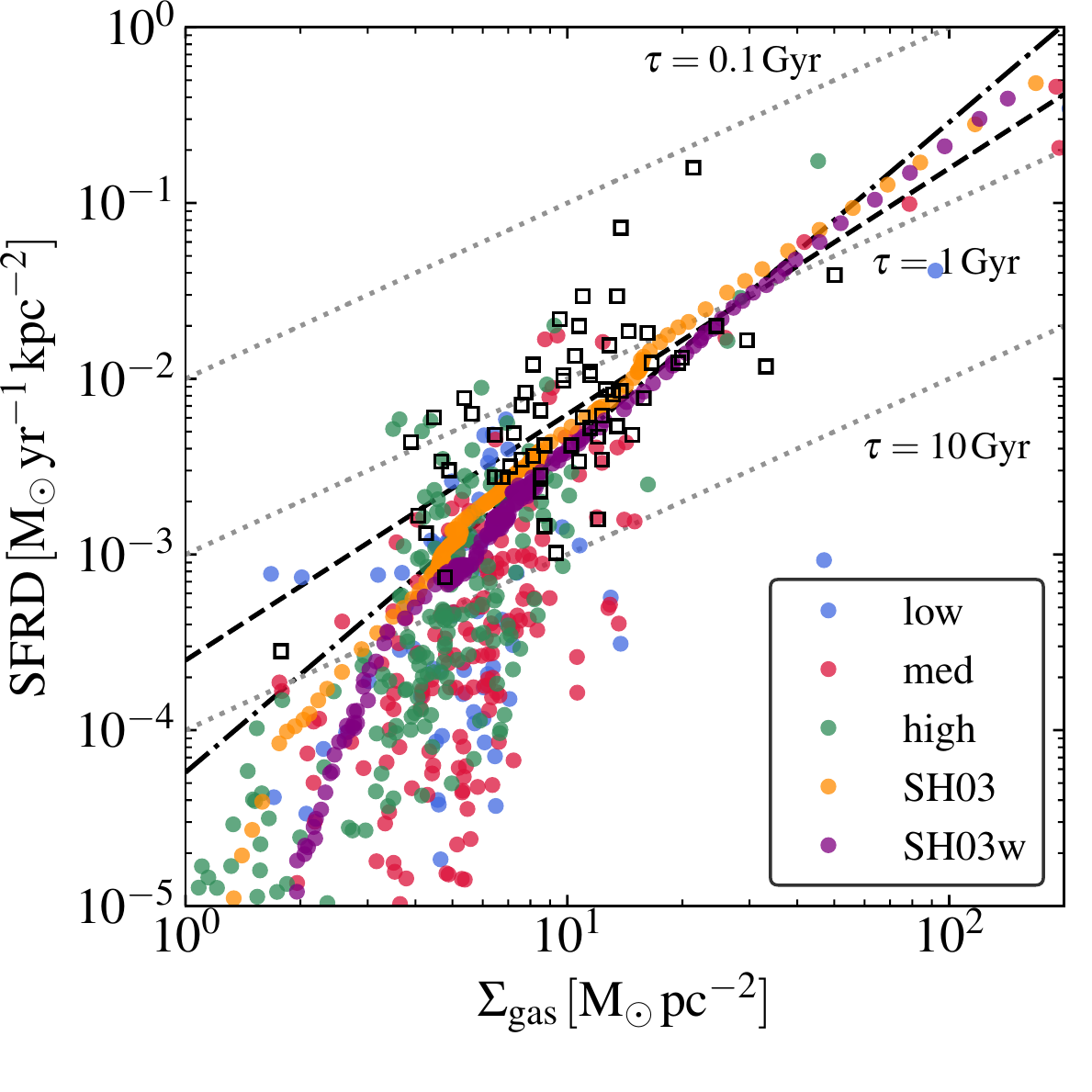}
\includegraphics[width=0.49\textwidth]{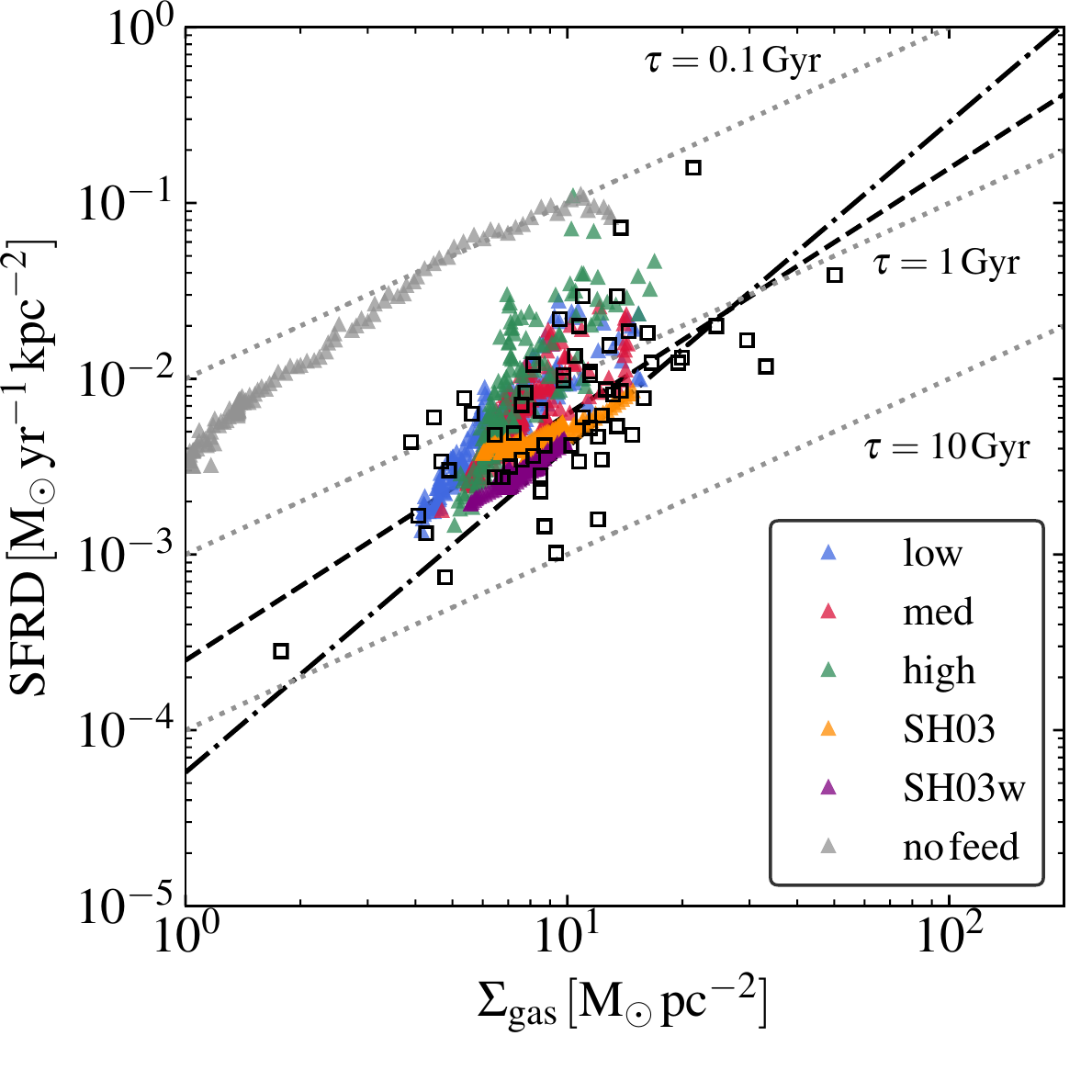}
\caption{\textit{Left:} Kennicutt-Schmidt relation for the simulated Milky Way-like galaxy at low (blue circles), intermediate (red circles) and high (green circles) resolution \rev{and for the \citet{Springel2003} model without (dark orange circles) galactic winds and with (purple) galactic winds} at $t = 0.7$ Gyr. Only gas particles in the disc region 
-- $R < 50\,\kpc$, $|z| < 5\,\kpc$ -- are considered in this plot. SFRDs and gas 
surface densities are computed in circular 
annuli centred on the disc that are $0.1 \,\kpc$ wide. Empty squares are the original 
\citet{Kennicutt1998} observations of ``normal'' (i.e. non-starbursting) 
galaxies and the dashed line gives the original Kennicutt relation. The dash-dotted 
line is a fit obtained by \citet{Bigiel2008}. Dotted lines are lines of 
constant gas depletion times put at $10$, $1$ and $0.1\,\Gyr$ from bottom to 
top. \textit{Right:} Time evolution of the Kennicutt-Schmidt relation for the Milky Way galaxy at low (blue triangles), intermediate (red triangles), and high (green triangles) resolution. The same spatial cuts as for the left panel are applied. 
Each triangle represent a given time in the simulation spaced by $\sim 
7\,\Myr$. Gas and star formation surface densities have been computed within the 
galactocentric distance enclosing $90$ per cent of the total star formation rate at any 
given time. For comparison, the same relation is shown with all the feedback processes turned off (grey triangles) and \rev{for two runs in which the \citet{Springel2003} model without and with galactic winds is used (dark orange and purple, respectively)}. Empty symbols and lines have the same meaning as in the left panel.
}
\label{fig:ksplot}
\end{figure*}

Finally, we enclose the system in a cubic volume with a side of $857~\kpc$ in which we inserted, with the procedure described in Section 9.4 of \citet{Springel2010}, a background grid consisting of low-density gas cells to avoid vacuum boundary conditions. The coarsest level of resolution of the background grid corresponds to that
of a grid composed by $16^3$ cells. Cells are allowed to be refined (de-refined) whenever they are above (below) a factor of two (half) of a predetermined target mass equal to the average gas mass \textit{before} the introduction of the grid. The de-refinement procedure is not applied to the background grid cells, which are identified as having a volume larger than 10 per cent of the average gas cell volume in the gaseous disc. The resulting initial conditions are then evolved for $\sim 1~\Gyr$ with the schemes described in Section \ref{sec:models}. In Table~\ref{tab:parameters} we list the fiducial values adopted for 
the numerical parameters associated with the different physical processes modeled in this work. 

\subsection{Star formation rates and the Kennicutt-Schimdt relation}

An essential aspect of any stellar feedback model is its ability to regulate star formation, which otherwise will be a runaway process occurring on much shorter time-scales (or at much larger rates) than observed in ``normal'' star-forming galaxies. In this Section, we will investigate the ability of our new model to achieve star formation regulation.

Figure~\ref{fig:sfr} (left-hand panel) presents the star formation history (SFH) for 
our Milky Way run. The solid lines in the figure give the evolution 
of star formation in the three different resolution runs: low (blue), intermediate (red) and high (green). \rev{For comparison, we also include
two runs with the \citet{Springel2003} model one with the basic model implementation (i.e. without galactic winds, dark orange) and one including decoupled galactic winds (purple, at intermediate resolution) as
well as a run without any feedback source (black, at intermediate resolution). The \citet{Springel2003} run including winds has a weak outflow strength as compared to the parametrizations used in the Illustris \citep{Vogelsberger2014a} and Auriga \citep{Grand2017} projects. This is to avoid that very strong stellar feedback prematurely terminates star formation given that in our isolated setup the galaxy lacks a fresh supply of gas due to cosmological gas accretion}. 
All the lines show the instantaneous star 
formation rate, computed as the sum of the star formation rates of each 
individual gas cell at a given time $t$.

Comparisons of the colored lines with the no-feedback run in Fig.~\ref{fig:sfr} indicate
that the star formation rate in the simulated Milky Way reaches a self-regulated state due to the presence of feedback, avoiding a large peak in the beginning of the simulation and the 
subsequent decline shown in the no-feedback case (black curve). This decline is due 
to the rapid gas consumption during the early stages of the no-feedback run. 
Instead, our new feedback model shows a
remarkably constant and stable SFR around 
$\sim 3\,{\rm M_\odot}\,{\rm yr^{-1}}$, in agreement with observations of this galaxy type. 
There is a small residual dependence on resolution at early times (see for example $t \leq 0.2$ Gyr),
but passed the initial stages and settling of the disc the three resolution levels agree quite well,
which is also clear from the total integrated stellar mass formed, shown on the right panel.

Most importantly, the predicted level of star formation broadly agrees with the 
results obtained with the \citet{Springel2003} model, which is the most commonly adopted model in {\sc arepo} thus far and which relies on an effective equation of state. 
Achieving a similar averaged SFR
together with self-regulation but without having to resort to the imposition of an
equation of state to treat the gas is one of the main accomplishments of our new model. 
\rev{The inclusion of decoupled galactic winds in the \citet{Springel2003} run contributes to further lowering the SFR of that model by about 35 per cent.}
Finally, we note that the inclusion of feedback suppresses star formation by factors as large as $\sim 10$, especially at early times. 

\begin{figure}
\includegraphics[width=0.46\textwidth]{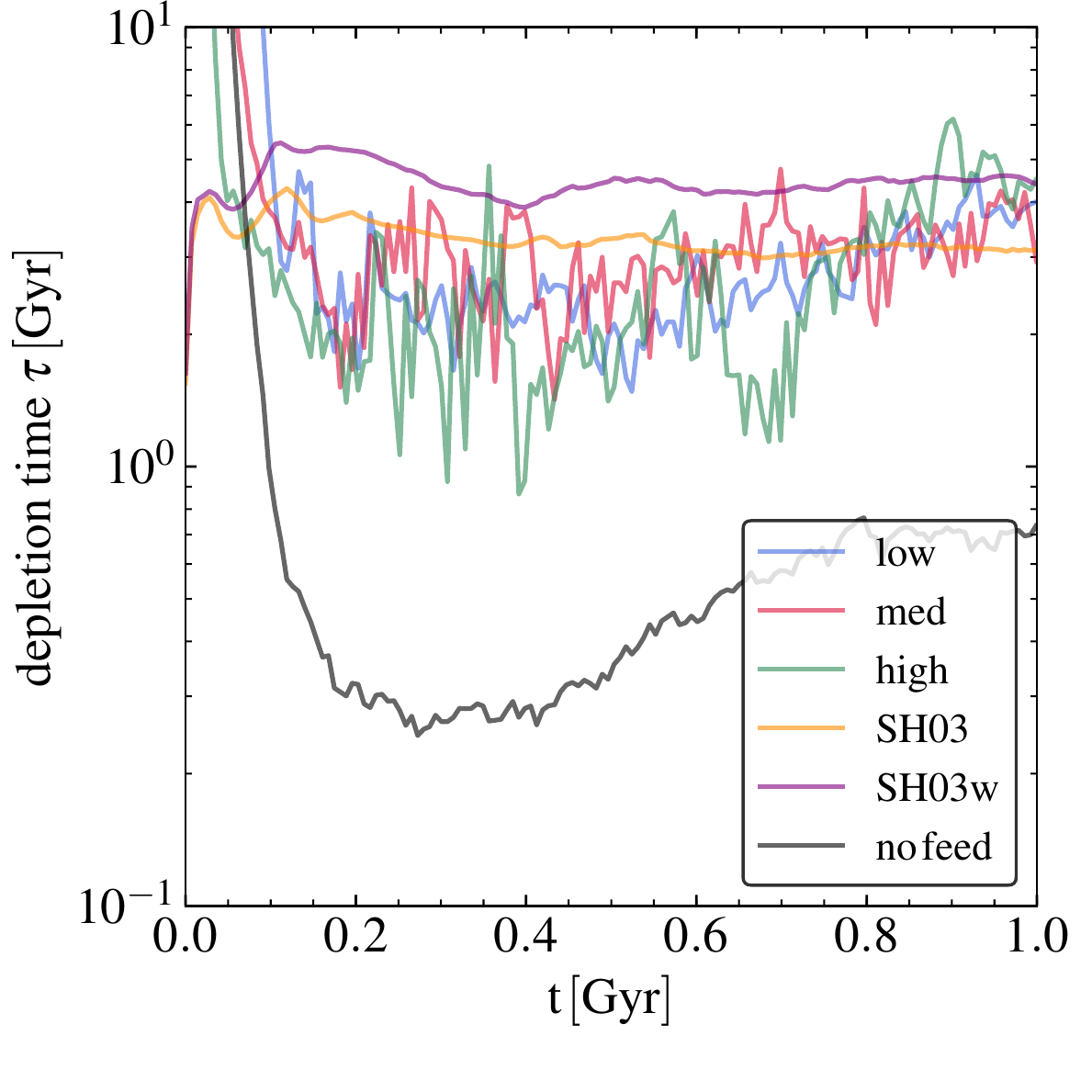}
\caption{Gas depletion time-scale evolution for the low (blue), intermediate (red), and high resolution (green) simulations of the Milky Way galaxy. The depletion time-scale is computed as the gas mass in the region $R \leq 40\,\kpc$ and $|z| \leq 1.5\,  \kpc$ divided by the galaxy star formation rate in that region. It can be seen how the stellar feedback keeps this quantity above a typical value of a few Gyr, independently of resolution. For comparison, the dark orange \rev{purple lines present the results obtained for the \citet{Springel2003} model without and with galactic winds, respectively. The \citet{Springel2003} model without winds predicts a constant depletion time of about $3\,\Gyr$ in line with, but slightly above, our findings, whereas in the model featuring winds the depletion time scale reaches $\sim 4\,\Gyr$}. The black solid line shows the trend for a simulation in which all feedback channels are switched off. In this case, the regulation of star formation by feedback is absent and the depletion time scale reduces accordingly by about one order of magnitude.}
\label{fig:tdepl}
\end{figure}

Figure~\ref{fig:sfr} (right-hand panel) is the cumulative version of the 
previous plot and shows the total amount of stellar mass formed by the 
different simulations as a function of time. In line with the results discussed 
for the instantaneous star formation rate, by the end of the simulations
the stellar mass formed in our new feedback model agrees within \rev{$\sim 13$ per cent for
the different resolution levels, within $\simeq 18$ per cent for the total mass produced by the \citet{Springel2003} model without winds} and shows an overall $\sim 3-4$ suppression factor 
compared to the no-feedback run. \rev{The addition of kinematically decoupled winds to the \citet{Springel2003} model pushes the difference in cumulative mass with \namemodel~ to within a factor of two.}

A related assessment of the performance of our model to regulate star formation
can be made by analyzing the Kennicutt-Schmidt relation \citep{Kennicutt1998} connecting the
projected density of gas, $\Sigma_{\rm gas}$ to the star formation rate density 
(SFRD). This is shown
in Fig.~\ref{fig:ksplot}, with the left panel highlighting different spatial locations for 
our simulated galaxy at $t=0.7$ Gyr and the right panel showing the disc-averaged
values (up to the radius enclosing 90 per cent of the star formation rate) at different $7$ Myr time intervals. 
Blue, red and green colors indicate the different resolution levels
as before, together with \rev{orange/purple and black used for the \citet{Springel2003} without/with galactic winds and no-feedback
runs, respectively.} The simulated relations 
have been obtained by considering the gas in a region centred on 
the disc and defined by the conditions $R < 50\,\kpc$, $|z| < 5\,\kpc$. For the left panel, selected gas cells have been binned into concentric annuli with width $0.1 \,\kpc$ over which the gas and star formation rate surface densities have been estimated.

Our simulations including feedback compare well with the original observational
sample by \citet{Kennicutt1998} for normal star-forming galaxies (empty squares), 
whereas the non-feedback run strongly over-predicts star formation. To guide the eye, 
we include with dashed and dash-dotted lines the best fit of the relation obtained by 
\citet[][dashed]{Kennicutt1998} and \citet[][dash-dotted]{Bigiel2008}. The dotted 
lines show constant depletion times of $10$, $1$ and $0.1\,\Gyr$ from bottom to top. 
As a result of feedback, star formation is self regulated and we find good agreement between
our new model predictions and the observations, \textit{regardless} of the numerical resolution. 

In other words, the resulting star formation in our model is rather  
inefficient with gas being converted into stars with a typical time 
scale of $1-2\,{\rm Gyr}$, rather than on a free fall time. Interestingly, in the left panel, there is also evidence of a break in the simulated relations at $\lsim 10\,{\rm M_\odot}{\rm pc^{-2}}$ \citep{Bigiel2008}. 
We note that the \citet{Springel2003} model successfully reproduces these trends, albeit completely free of \rev{or with minimal} scatter and show a less pronounced downturn of the relation below $\sim 10\,{\rm M_\odot}{\rm pc^{-2}}$. \rev{These trends hold independently of the inclusion of galactic winds in the model, which has only a minimal impact on the recovered slope and normalization of the simulated Kennicutt-Schmidt relation.}
Instead, our new feedback and star formation prescriptions show a level of scatter that compares well with observational estimates. This increased scatter is not surprising 
given the fact that the \citet{Springel2003} model was specifically formulated to explicitly 
reproduce the {\it median} observed relation.

\begin{figure}
\includegraphics[width=0.48\textwidth]{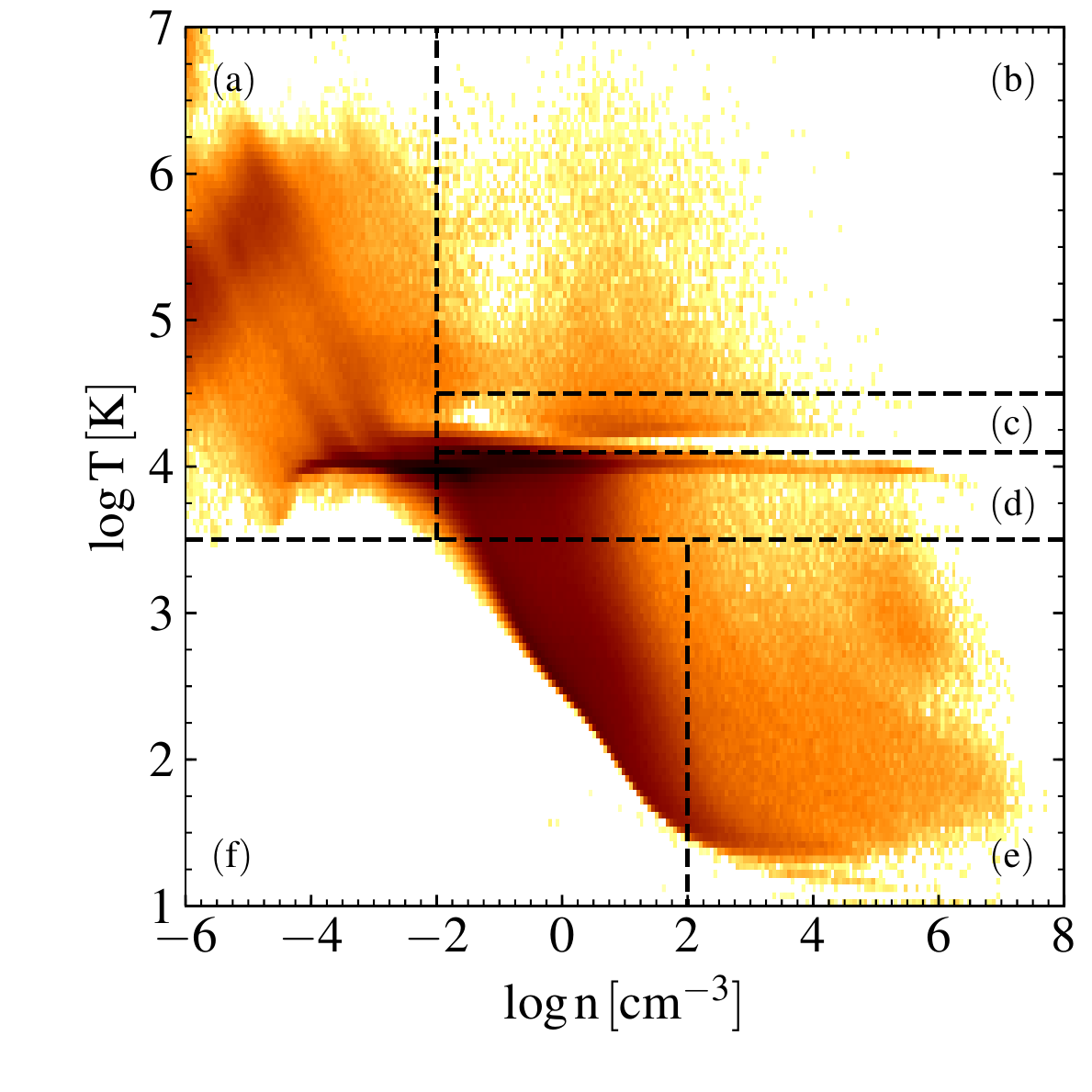}
\caption{Gas phase diagram for the Milky Way galaxy at high resolution at time $t=0.7\,\Gyr$. The diagram has been computed as a two-dimensional histogram where the mass of each gas cell in the simulation has been assigned to the equally spaced logarithmic bins in gas temperature and density. The colouring on the histogram is therefore proportional to gas mass contained in each bin (darker colour meaning larger masses). The plot shows the presence of a multiphase medium composed of warm gas at an approximately constant temperature of $\sim 10^4\,\K$ (d) and a colder phase that extends to lower temperatures and higher densities (f) and eventually reaches the threshold for star formation (e). The material between these two phases is thermally unstable and is the result of the competition between stellar feedback and cooling. In particular, only gas at a high enough density ($\sim 1\,\cm^{-3}$) can efficiently cool to low ($\lsim 100\,\K$) temperatures because it can effectively self-shield from the UV background. Also apparent is the presence of a more diffuse and hot gas phase (a), which is created by feedback pushing (hot) gas outside the galactic disc (b). The continuation of the warm ($T\sim10^{4}\,\K$) gas phase to higher densities ($\gsim 10-100\,\cm^{-3}$) is the result of gas photoionization from young stars (c). The gap between phases (c) and (d) results from our choice of $T_{\rm phot} = 1.7\times 10^4\,\K$.}
\label{fig:phasediag}
\end{figure}

\begin{figure*}
\includegraphics[width=0.49\textwidth]{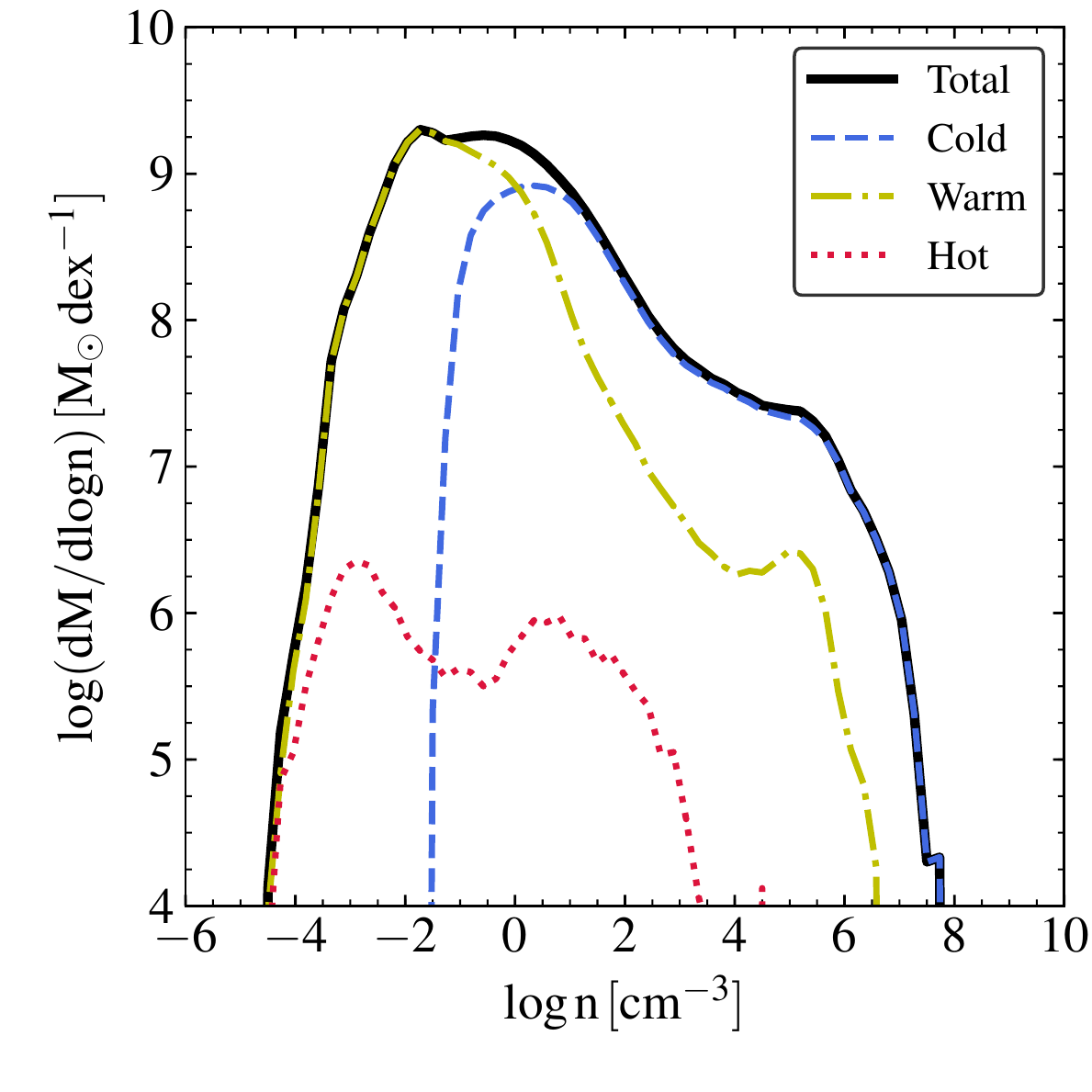}
\includegraphics[width=0.49\textwidth]{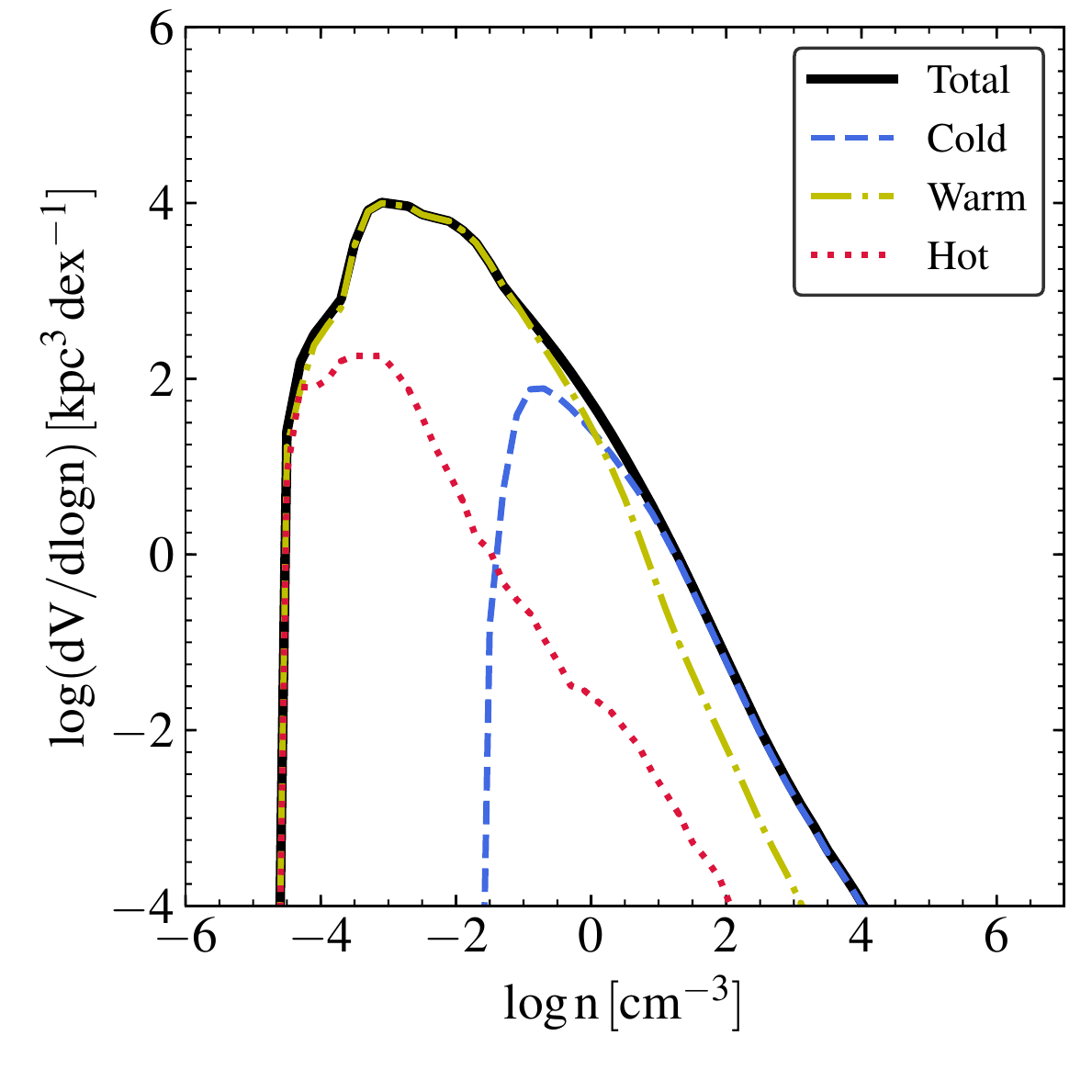}
\caption{Differential gas mass (left) and volume (right) distribution for the Milky Way galaxy at high resolution as a function of the gas density at time 
$t=0.7\,\Gyr$. Only gas in the disc has been selected according to the 
spatial cuts $R < 40\,\kpc$ and $|z| < 1.5 \,\kpc$. The solid black line shows the 
total distribution, whereas the dashed blue, dash-dot yellow and dotted red lines display the distribution of cold ($T < 2\times10^3\,\K$), warm ($2\times 10^3\,K < T <
4\times 10^{5}\,\K$) and hot ($T > 4\times10^5\,\K$) gas, respectively. Note how 
a multiphase structure of the ISM of the galaxy is formed as a consequence of 
feedback, as it is apparent by a hint of a bimodal character of the mass distribution. In particular at large densities the cold phase dominates the gas budget, 
whereas the warm phase is predominant at lower ($\sim 1 \,\cm^{-3}$) densities. 
Around $\lsim 10^3 \,\cm^{-3}$ and extending towards lower density values also a hot phase starts to appear. However, its contribution to the total mass budget is not dominant at any density. These trends are also present for the volume distribution, in which the hot phase shows a non-negligible contribution to the total volume at densities $\sim 10^{-3}\,\cm^{-3}$ and below.
}
\label{fig:densPDF}
\end{figure*}

\begin{figure*}
\includegraphics[width=0.49\textwidth]{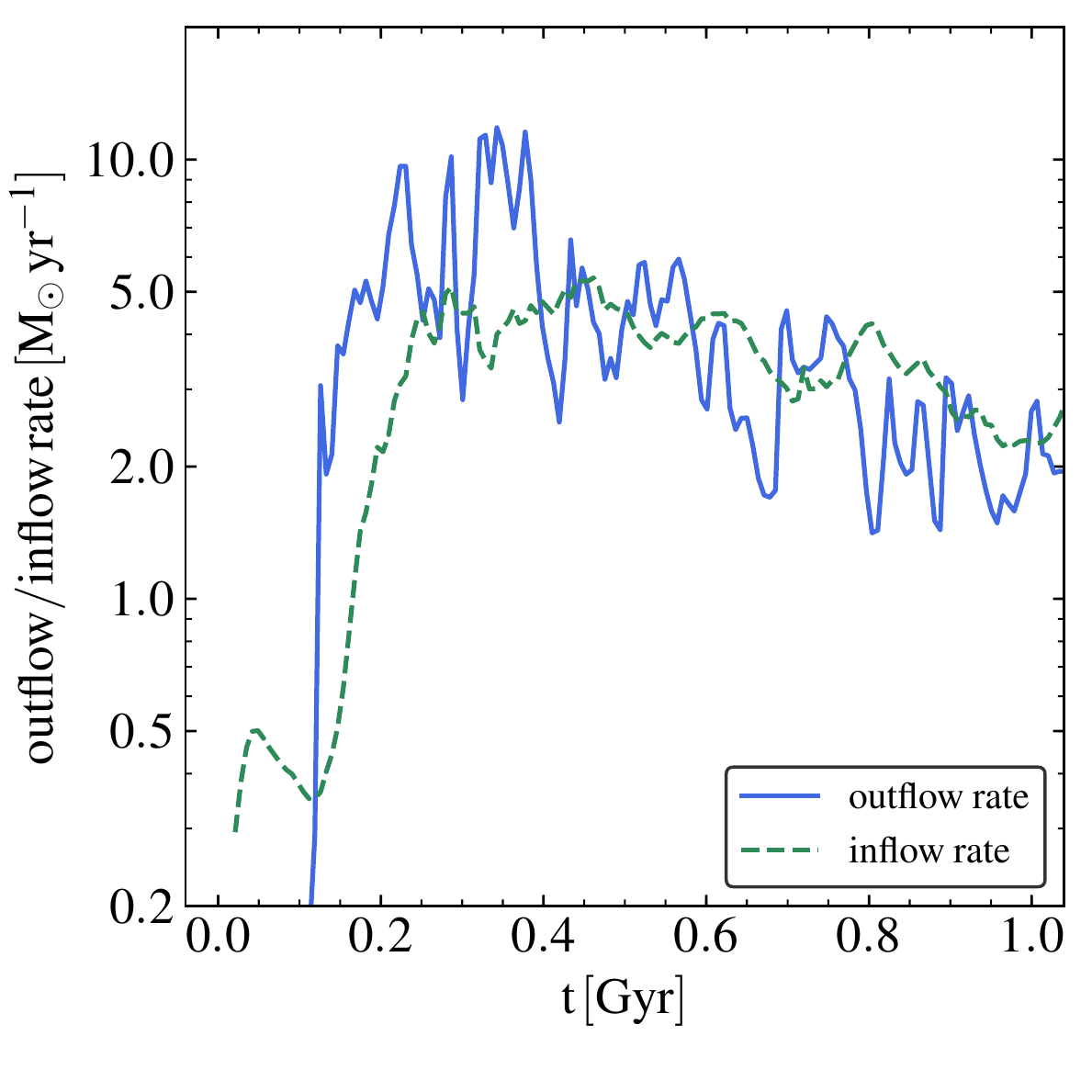}
\includegraphics[width=0.49\textwidth]{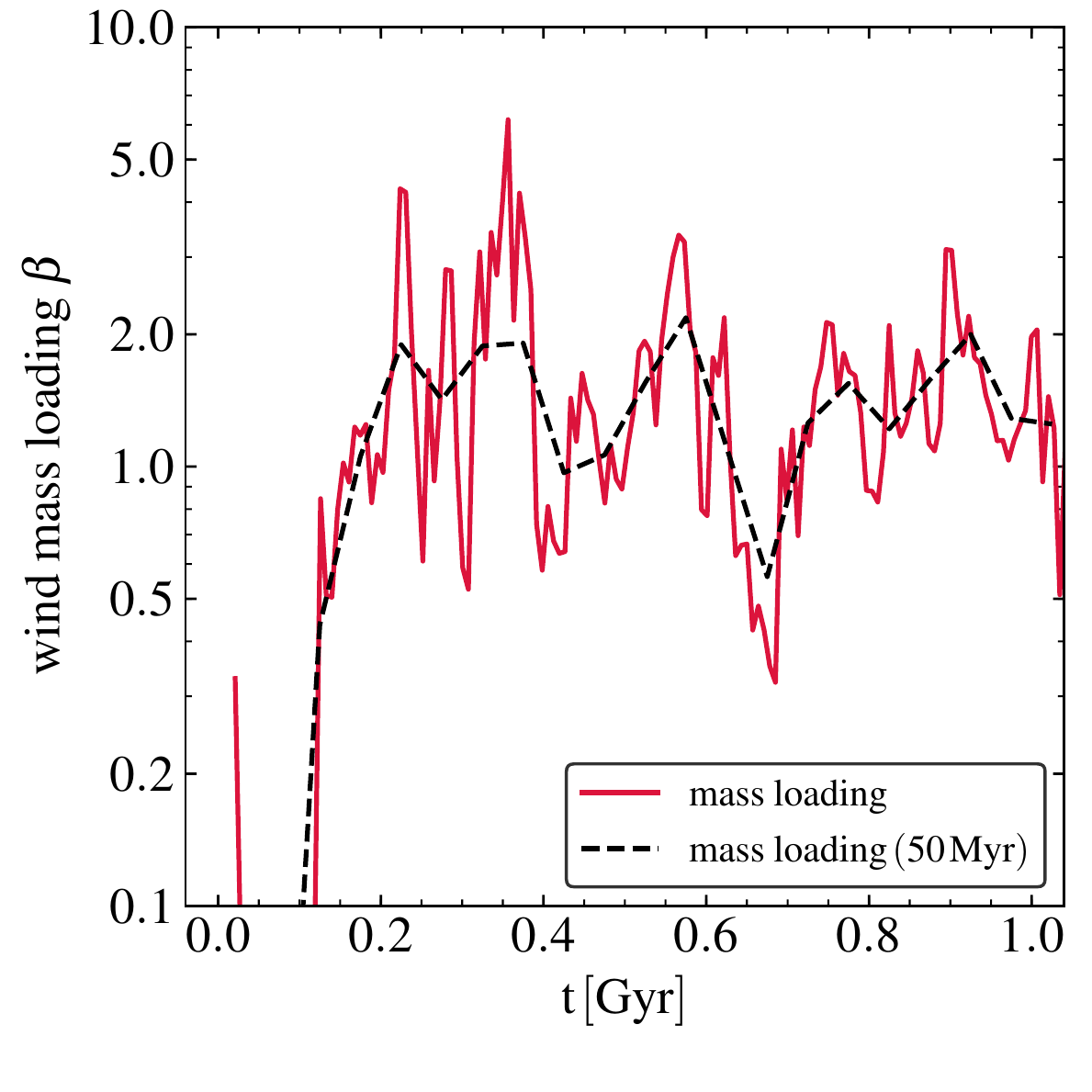}
\caption{\textit{Left:} Gas outflow (solid blue) and inflow (dashed green) rates as a function of time for the high resolution Milky Way simulation. 
Mass outflow (inflow) rates have been determined by considering gas within a galactocentric distance $R \leq 40\,\kpc$ contained in two slabs located $\pm 2\,\kpc$ above and below the midplane of the galaxy and $\Delta z = 300\,\pc$ 
thick with outflowing (inflowing) velocities $v_z$ with respect to the slab. For each selected gas particle a rate has been estimated as $mv_z/\Delta z$ and the single contributions were summed up. 
\textit{Right:} Wind mass loading (solid red) as a function of time for the high resolution Milky Way simulation. The mass loading for outflowing gas is the outflow rate presented divided by the galaxy star formation rate at any given time in the same spatial region. The dashed black line shows the same quantity calculated over a time-scale of $50\,\Myr$, which makes the average trend more visible. Our feedback model is able to launch gaseous outflows with a 
significant ($\beta \sim 1$, although there are higher spikes) mass loading. These outflows generate a fountain flow, as it is apparent by the similarity in value of the inflow and outflow rates 
that regulate star formation and keep it going for about a Gyr, the time span simulated in this run.}
\label{fig:outflow}
\end{figure*}

The large impact of feedback on the depletion time-scale $\tau$ can also be seen in more detail in Fig.~\ref{fig:tdepl},
following the same color coding as before. $\tau$ is defined as 
the ratio between the gas mass and the instantaneous SFR in the galaxy, and acts
as a proxy for the typical time over which a galaxy will consume its gas reservoir due to star 
formation processes. To compute both quantities a spatial cut defined by the 
conditions $R \leq 40\,\kpc$ and $|z| \leq 1.5\, \kpc$ has been adopted. 
Depletion time scales in the new feedback runs are up to $10$ times larger compared
to the no-feedback case and reasonably converged with resolution, especially 
once the runs have settled into self-regulation ($t \gsim 0.3-0.4$ Gyr). The resulting timescales after feedback is included agree well with observational estimates 
\citep{Sancisi2008,Bigiel2011,Kennicutt2012,Leroy2013,Putman2017}.

Interestingly, results from the \namemodel\ model are in line with those from 
the \citet{Springel2003} model \rev{without winds, where $\tau \simeq 3\,\Gyr$ and is relatively constant
over time (dark orange line).} \rev{The inclusion of galactic winds in the \citet{Springel2003} model (purple) further pushes up $\tau$ above $\simeq 4\,\Gyr$. This is caused by a combination of the ISM pressurization due to the adoption of an effective equation of state and of the gas ejection due to the galactic wind presence. We hasten to add that the agreement between the depletion time scale in \namemodel\ and in the no wind implementation of the \citet{Springel2003} model} is by no
means straightforward, as in our model star formation suppression and 
self-regulation is achieved self-consistently 
by following the different gas phases, whereas in the \citet{Springel2003} approach
this is entirely due to the pressurization of gas resulting from the 
implemented effective equation of state for all gas elements with 
densities above the star formation density threshold \rev{(which represent an equilibrium two-phase ISM model where the hot gas phase is heated by supernova energy input). For the typical parameters of our simulations the star formation density threshold of the \citet{Springel2003} model is $\sim 0.1\,\cm^{-3}$.} As we will see in Sec.~\ref{sec:winds}, the generation of winds in our model also
follows self-consistently from the gas treatment whereas they need to be explicitly
added in the \citet{Springel2003} model. 

\subsection{ISM structure}

As discussed above, an important feature of our ISM model is the natural emergence of a multiphase ISM during the evolution of the simulation. This subsection is devoted to the presentation of the properties and the structure of this multiphase ISM. Fig.~\ref{fig:phasediag} shows the phase diagram of all the gas cells in the high resolution Milky Way simulation at $t = 0.7\,\Gyr$. The figure presents the 
phase diagrams in terms of a temperature versus density plot and is constructed as a two-dimensional histogram. Darker shades correspond to increasingly larger gas mass values.

To guide the eye, we highlight the multiphase structure of the gas in 
Fig.~\ref{fig:phasediag} through dashed lines and labels to indicate
the different regions. A large fraction
of the ISM is composed of warm gas at a fairly constant temperature of $\sim 10^4\,\K$, 
covering a density range from about $10^{-2}\,\cm^{-3}$ to $\sim 10^{2}\,\cm^{-3}$ (d), and a colder phase, starting off at $10^{-2}\,\cm^{-3}$ and reaching low ($\lsim 100\,\K$) temperatures and high densities (e and f). Indeed, only gas which is 
sufficiently dense ($n > 10^{-2}\,\cm^{-3}$) can effectively self-shield from 
the UV background and effectively cool and become dense enough  to form stars (e). 
All the material present in the region of the phase diagram in between these two 
loci is thermally unstable, and populates these region of the diagram because of 
stellar feedback. 

Two other features of the diagram are interesting as well. The first 
is the continuation of the warm gas phase locus to higher density ($\gsim 
1\,\cm^{-3}$) and with approximately the same $\simeq 10^{4}\,\K$ temperature (c). 
This results from the photoionization of the gas described in 
Section~\ref{sec:photoion}. The second one is the presence of gas at low density ($\lsim 10^{-2} \,\cm^{-3}$) and relatively high ($\sim 10^5-10^6\,\K$) temperatures, coincident with the hot gas phase (a). This is largely the result of the ejection of hot gas from within the star forming disc due to the momentum and energy delivered by SN feedback (b).

A more quantitative determination of the contribution of each gas phase to the
ISM is presented in Fig.~\ref{fig:densPDF}, which shows the gas density 
probability density function (PDF) for our high resolution Milky Way galaxy 
at $t = 0.7\,\Gyr$. The left and right panels correspond to the 
mass- and volume-weighted versions of this figure, respectively. Gas 
belonging to the disc is selected in the region $R < 40\,\kpc$ and $|z| < 1.5 
\,\kpc$. The PDF is plotted for all gas cells in the disc region (solid black 
lines) and for the cold ($T < 2\times10^3\,\K$, dashed blue lines), warm ($2\times 
10^3\,K < T < 4\times 10^{5}\,\K$, dash-dot yellow lines) and hot ($T > 4\times10^5\,\K$, dotted red lines) ISM phases, to better highlight the different contribution to the PDFs and their relative importance.

The multiphase nature of the ISM can be identified in both representations
in Fig.~\ref{fig:densPDF} through the broad distribution of possible gas densities.
A typical $n \sim 1\,{\rm cm^{-3}}$ marks the transition between 
the warm and cold dominated gas phases. At high densities 
($n \geq 10^2 \; \rm cm^{-3}$), cold gas dominates in both distributions. Across all densities, a significant fraction ($\simeq 32$ per cent) of the total mass is in this phase, although it occupies very little volume ($\simeq 0.5$ per cent of the total).
For densities $\lsim 10^4\,\cm^{-3}$ the hot phase, heated by SN feedback starts to appear, being
more dominant in the volume-weighted distribution than in mass-weighted one, especially at very low densities ($\lsim 10^{-3}\,\cm^{-3}$). However, the global contribution of the hot phase is always very small both in a mass- ($\simeq 0.1$ per cent) and in a volume-weighted ($\simeq 1.6$ per cent) sense.

\section{Gas dynamics and galactic-scale fountain flows}\label{sec:outflows}

An essential aspect of the \namemodel\ model is the natural generation of 
gaseous outflows, which are locally correlated with star formation and
are also fully coupled to the hydrodynamic equations. 
Figure~\ref{fig:outflow} (left-hand panel) displays the measured gas outflow 
(blue solid line) and inflow (green dashed line) rates onto the disc as a 
function of time for the high resolution Milky Way simulation. To compute the 
rates only gas at a distance $R \leq 40\,\kpc$ and contained in two slabs of 
thickness $300\,\pc$ and located at a distance $\pm 2\,\kpc$ from the midplane 
of the galaxy has been considered. For each snapshot, after removing from the 
gas the average vertical gas velocity $\left< v_z\right>$, each cell contained in the 
slab contributes to the total rate as $mv_z/\Delta z$, where $v_z$ is the 
outflow/inflow velocity with respect to the disc plane. 

Our implementation of stellar feedback is 
able to launch outflows at a rate of a few ${\rm 
M_\odot}\,\yr^{-1}$ throughout the entire simulation. The outflow rate is larger 
at early times, reaching a maximum of $\sim 10\,{\rm M_\odot}\,\yr^{-1}$ at $t \simeq 
0.35-0.4\,\Gyr$, and subsequently declining. Particularly noticeable is 
the bursty appearance of the mass outflow rate that cycles through its maxima and minima 
on a time scale of $\sim 100\,\Myr$. Comparing these outflows rates to the instantaneous
SFR of the galaxy\footnote{The SFR per bin is determined within the same galactocentric distance cut and at heights $|z| < 2\,\kpc$.} we compute the mass loading factor parameter $\beta$ (right-hand panel),
which results in $\beta \sim 1$ approximately constant with time. 
 The solid red and dashed black lines show the raw mass loading factor obtained from the snapshots 
 and a smoothed version estimated over a time-scale of $50\,\Myr$, respectively.

The left panel of Fig.~\ref{fig:outflow} also includes the evolution of gas inflows with time
(green dashed curve). Peaks in the values of the inflow rates are 
usually correlated, albeit offset, with respect to peaks in the outflow rate or $\beta$ 
hinting at a galactic fountain origin of the accreting gas. Therefore, ejected gas in
our runs is not entirely lost for the galaxy but instead 
cycles back in a galactic fountain flow, a cycle that has been shown to be important for
the metallicity and angular momentum build up of the disc and hot corona 
\citep[e.g. ][]{Marinacci2010, Ubler2014, Christensen2016, AngelesAlcazar2017}. 

\begin{figure*}
\includegraphics[width=0.305\textwidth]{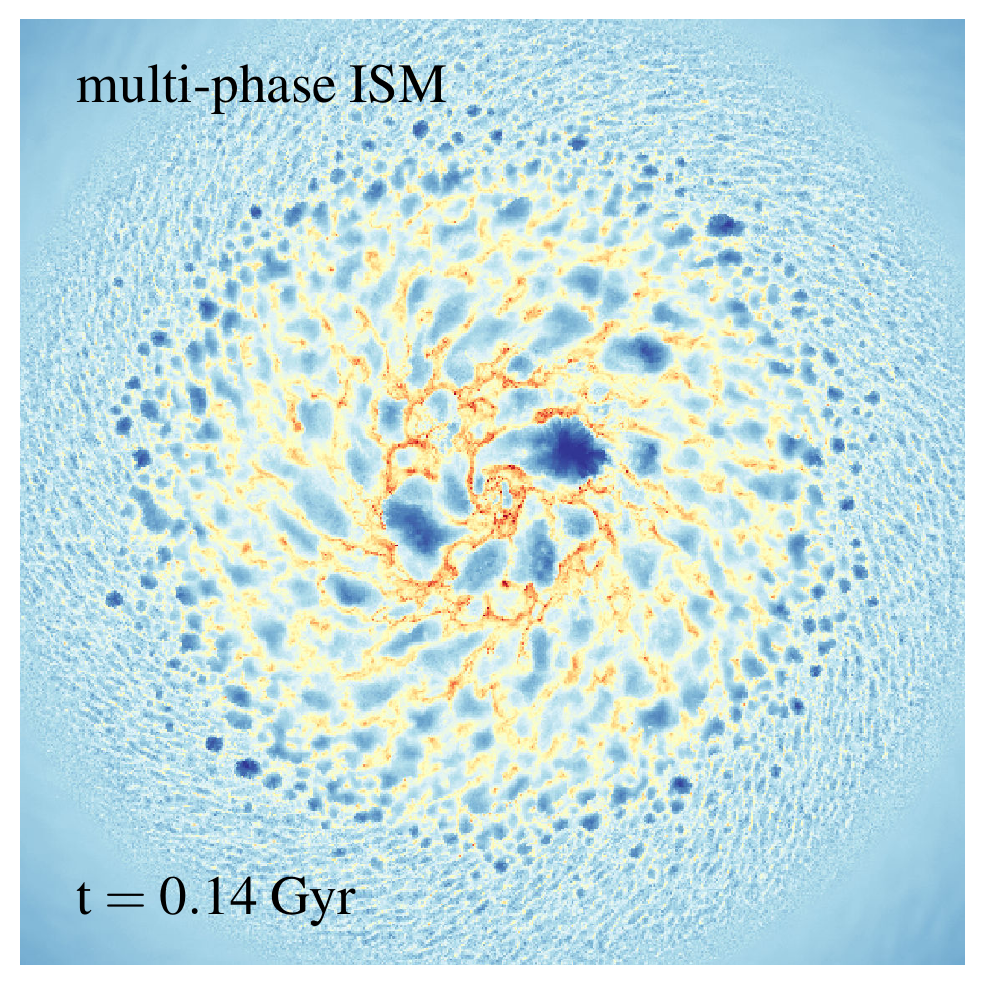}
\includegraphics[width=0.305\textwidth]{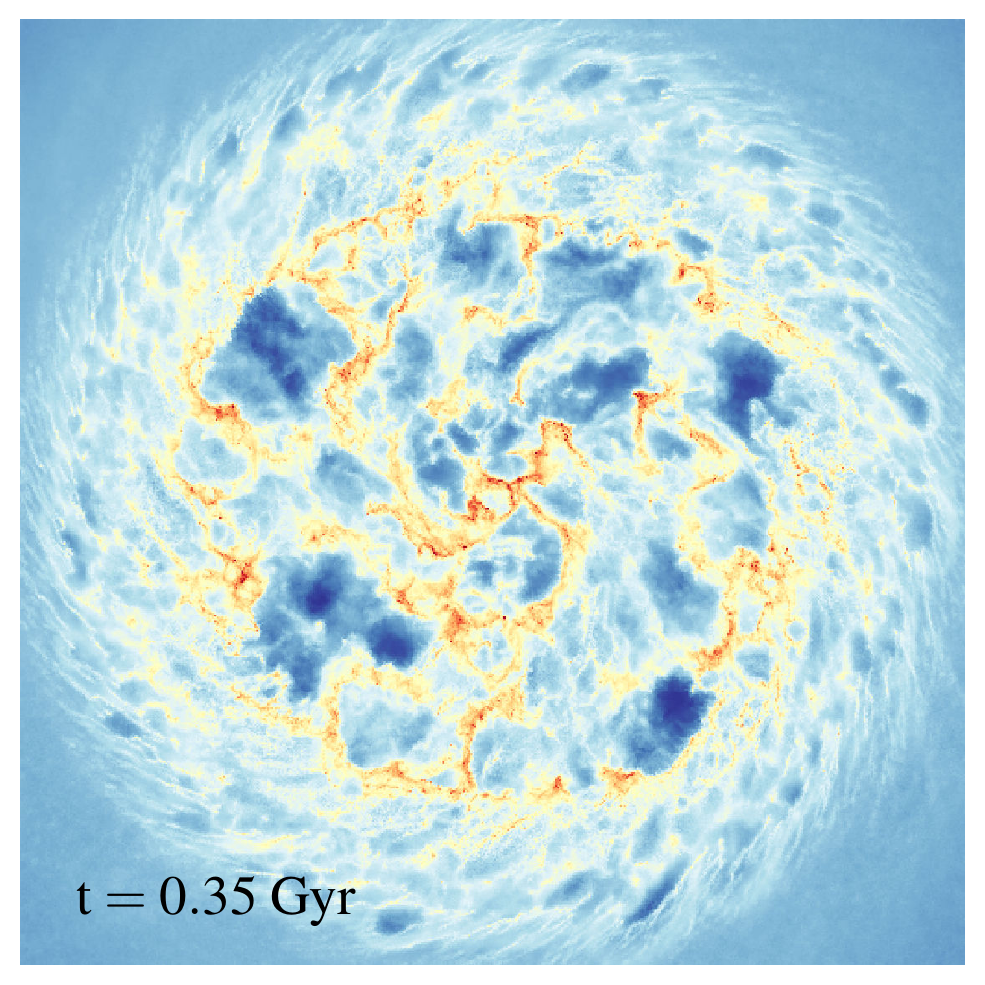}
\includegraphics[width=0.305\textwidth]{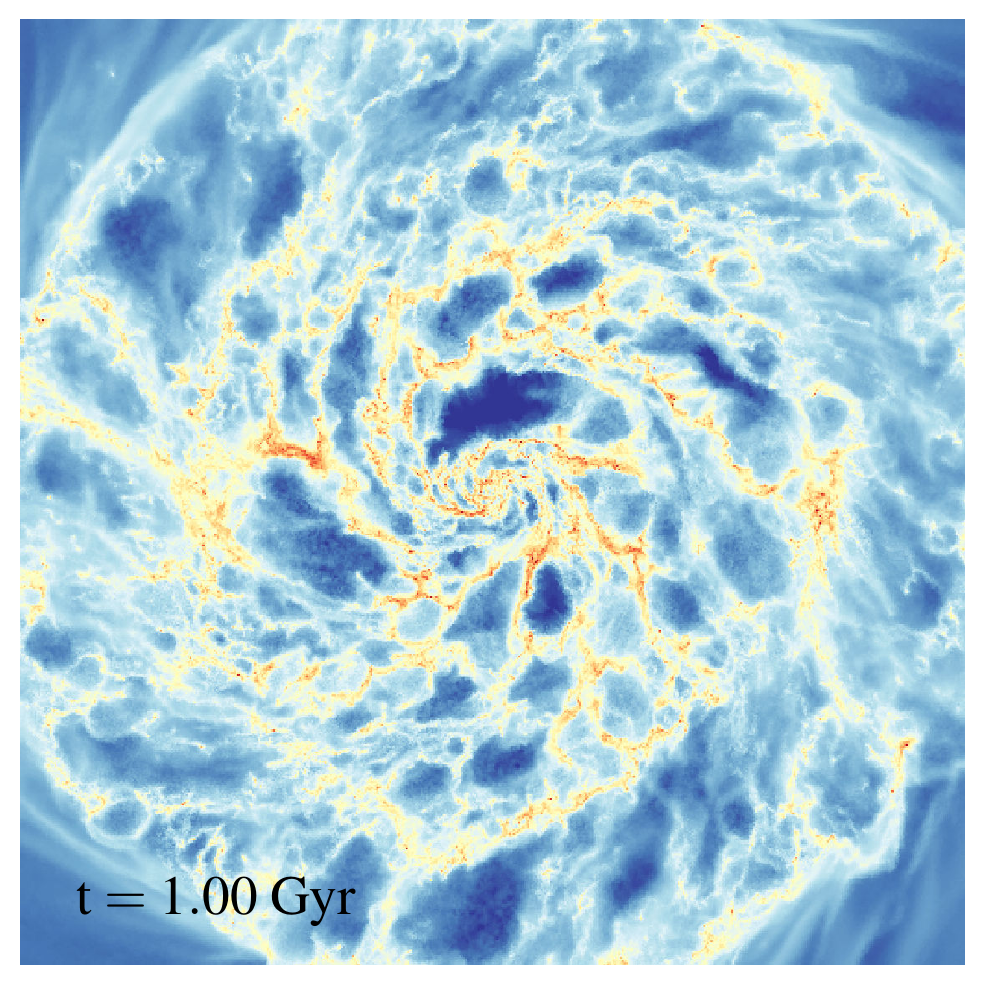}
\includegraphics[width=0.305\textwidth]{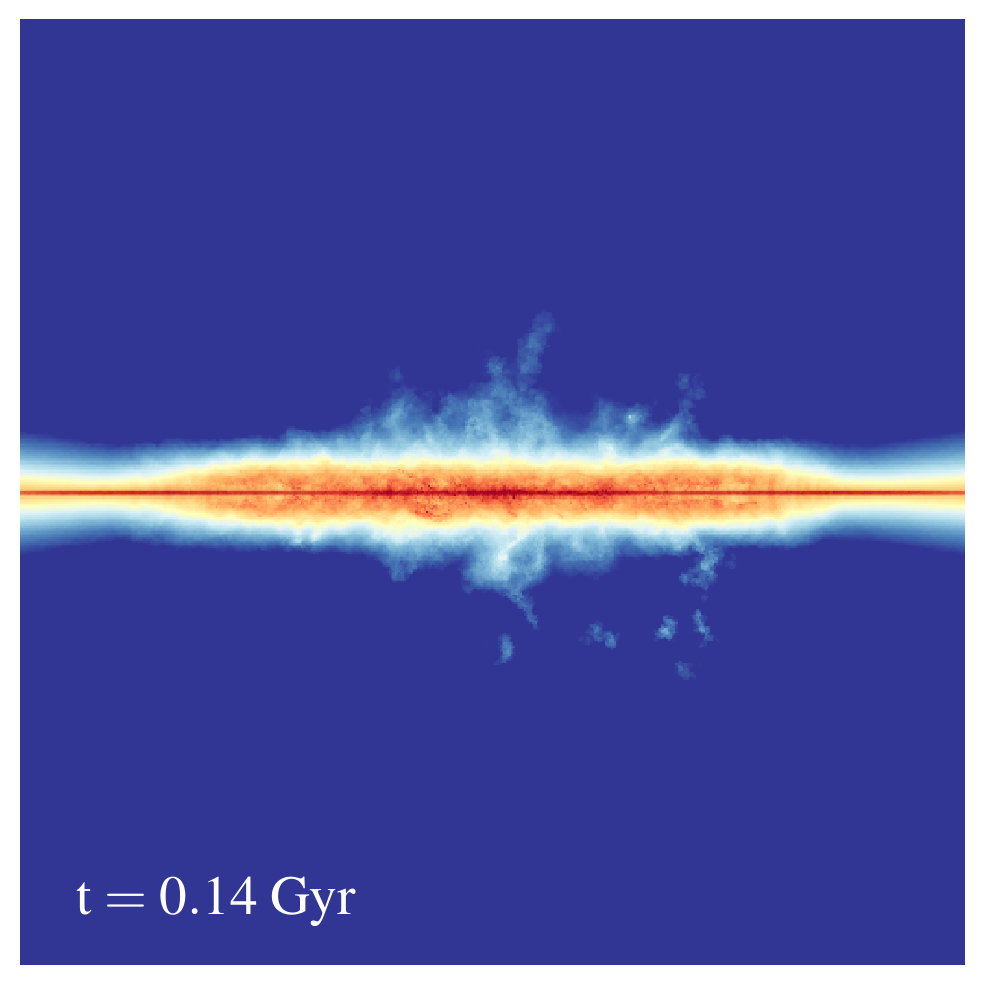}
\includegraphics[width=0.305\textwidth]{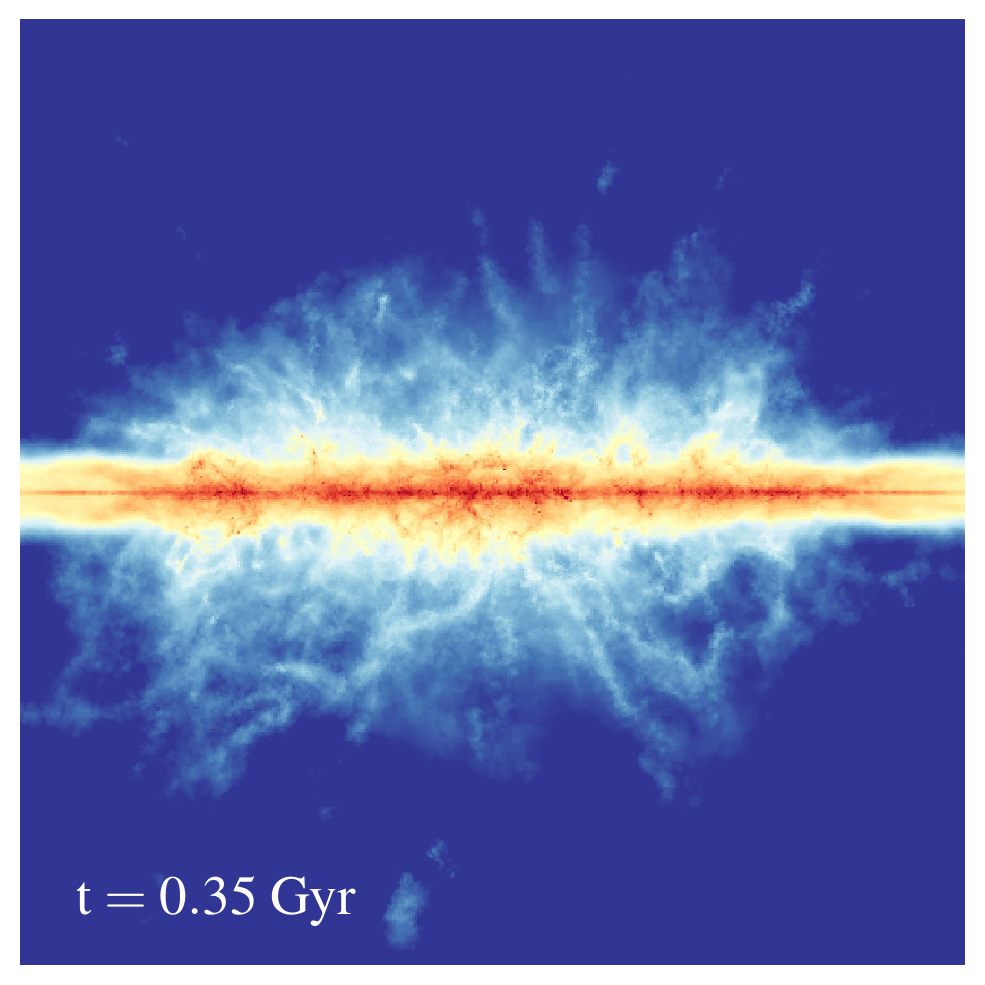}
\includegraphics[width=0.305\textwidth]{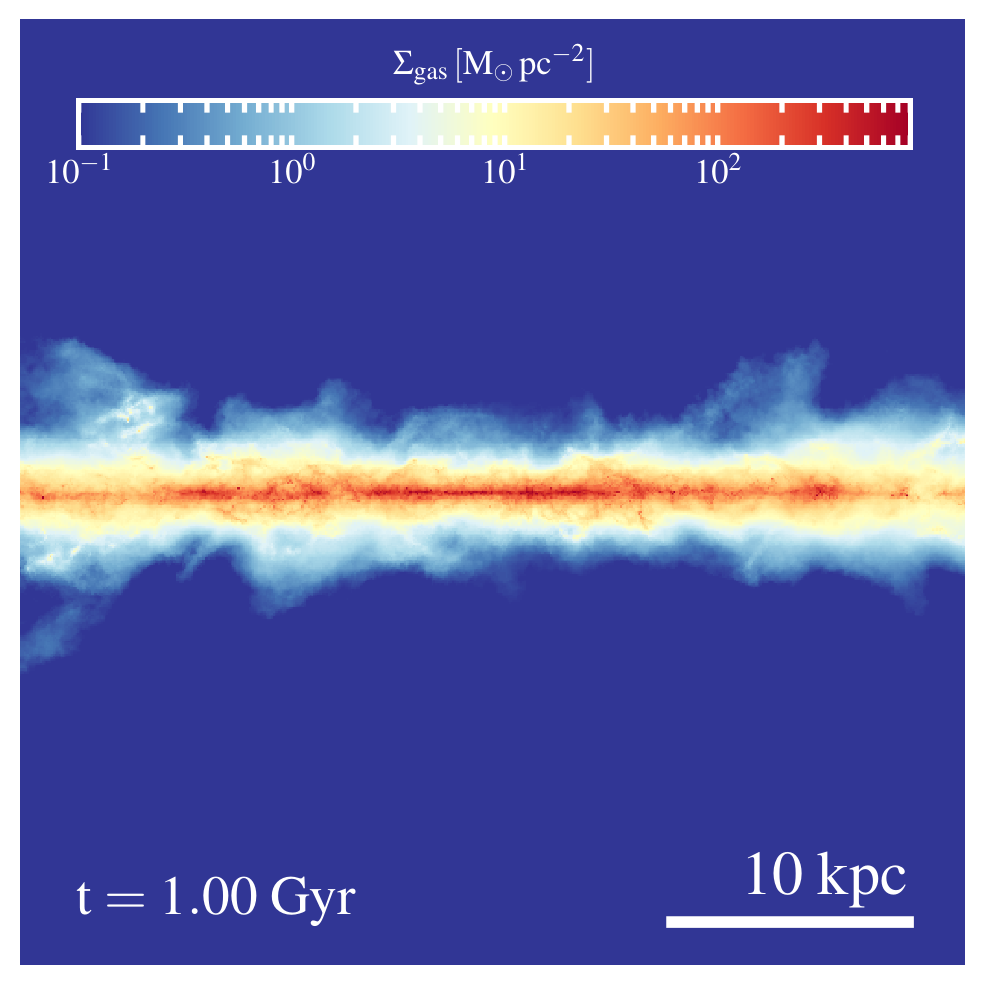}
\includegraphics[width=0.305\textwidth]{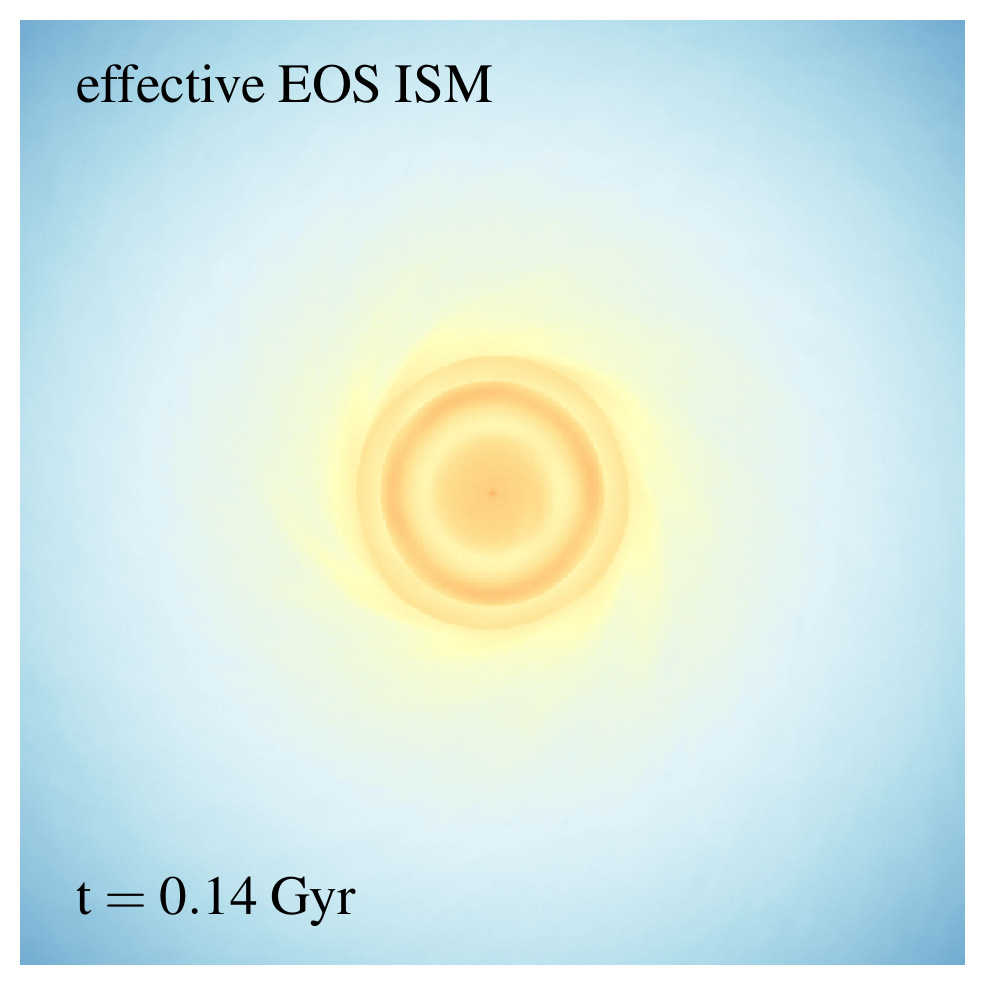}
\includegraphics[width=0.305\textwidth]{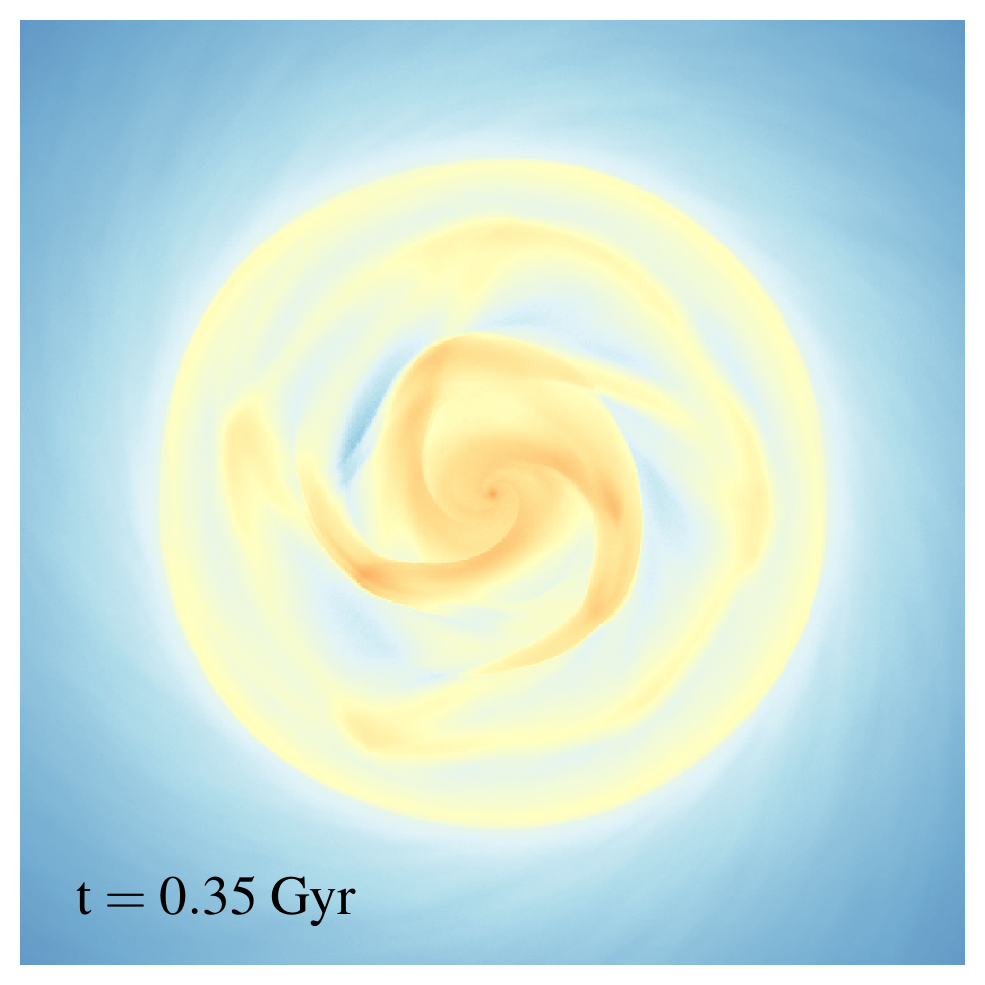}
\includegraphics[width=0.305\textwidth]{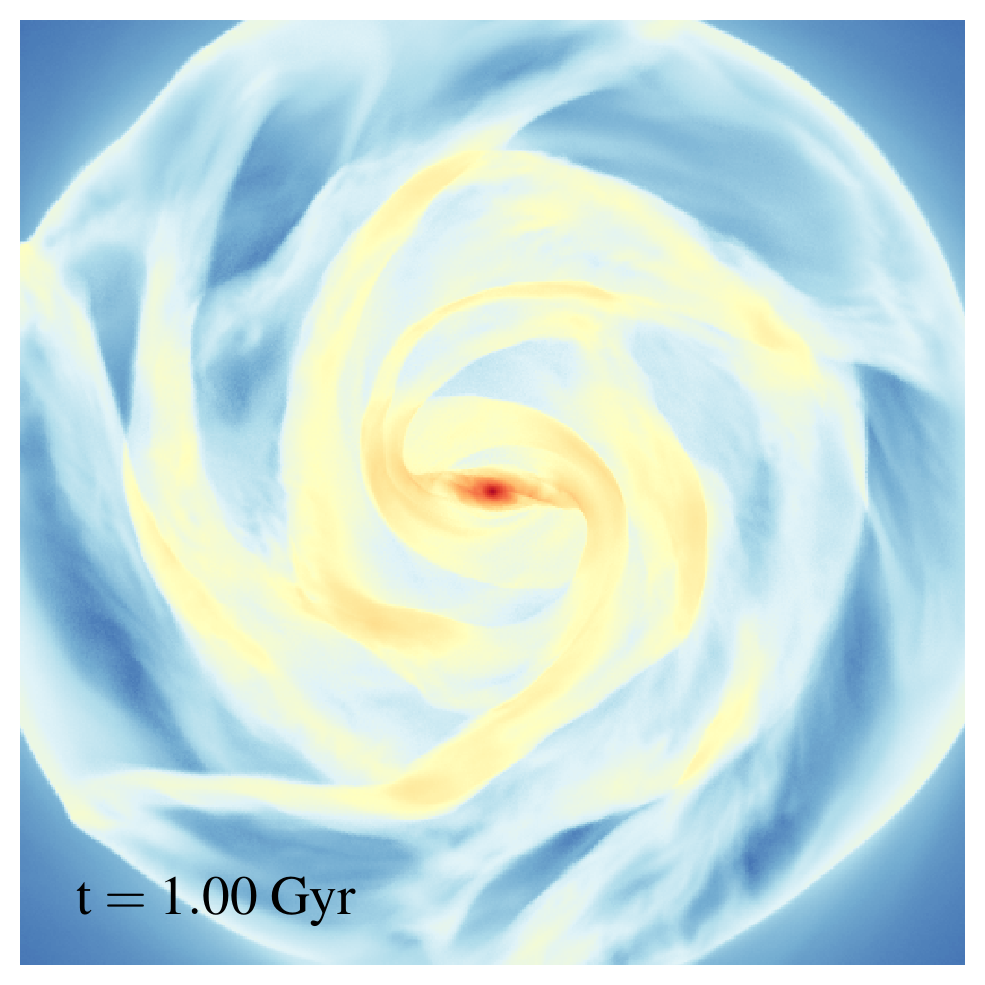}
\includegraphics[width=0.305\textwidth]{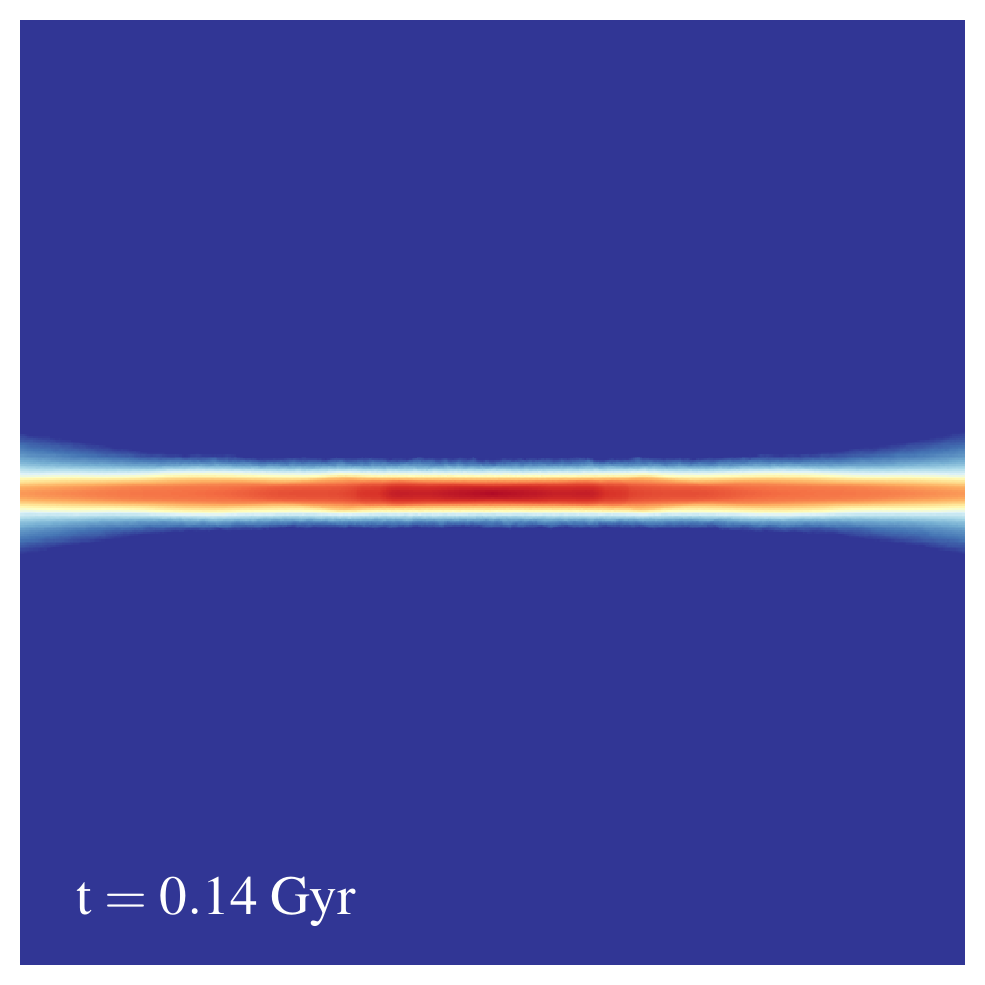}
\includegraphics[width=0.305\textwidth]{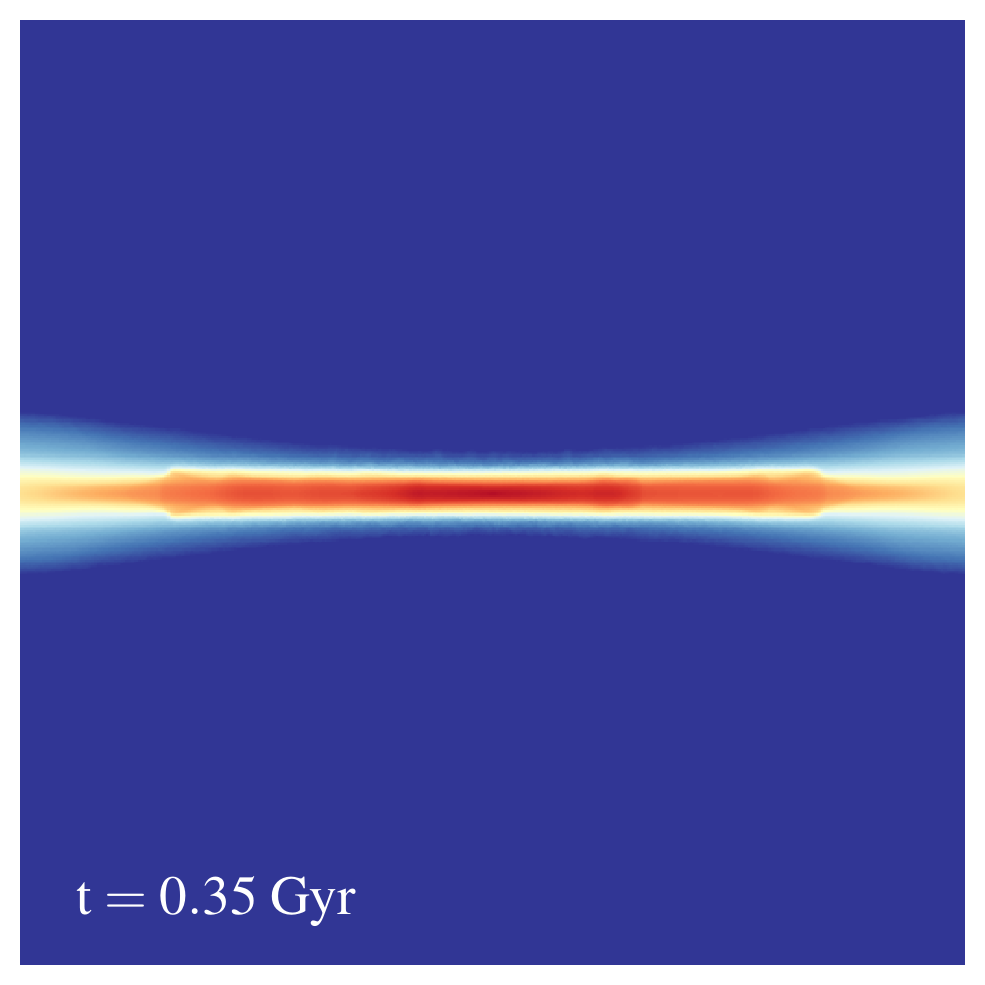}
\includegraphics[width=0.305\textwidth]{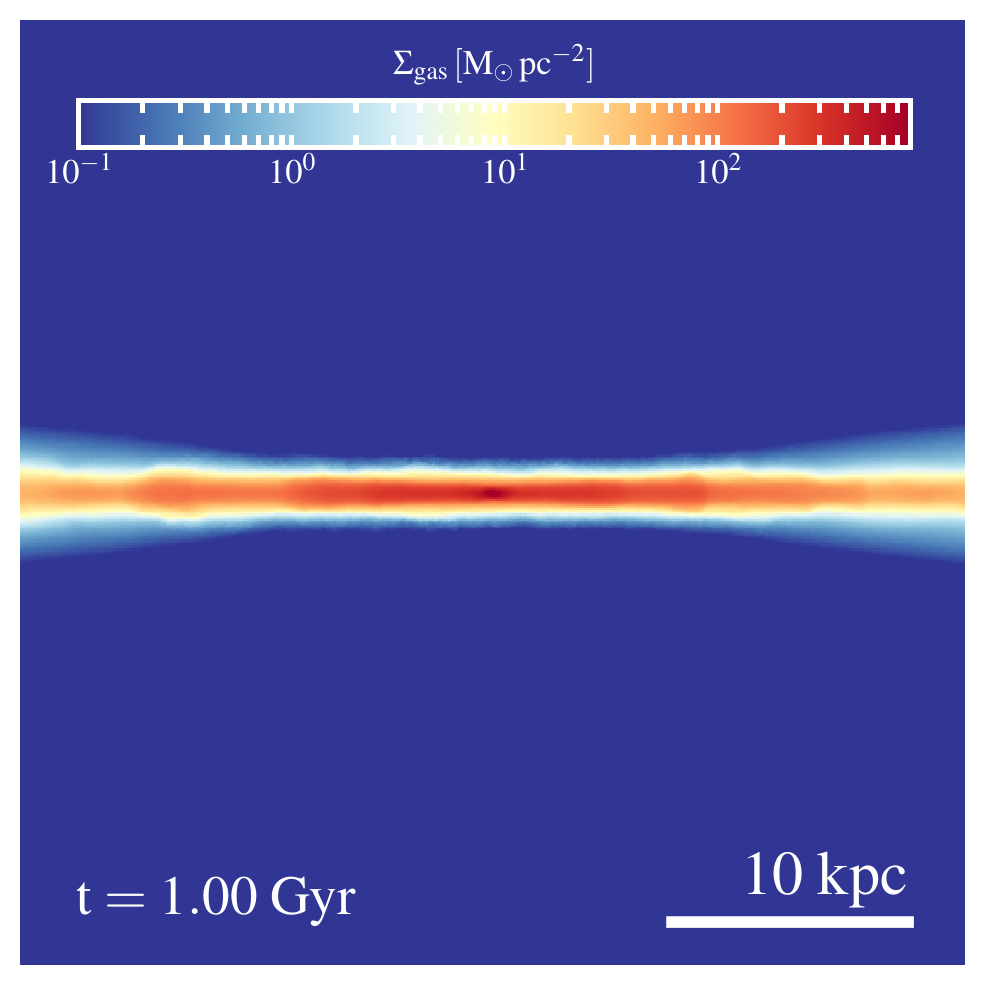}
\caption{\textit{Top:} Gas column density in a face-on (first row) and edge-on
(second row) projection for the Milky Way galaxy at high resolution. Projections 
have been computed at times $t=0.14, 0.35, 1\,\Gyr$ from left to right column. 
Feedback and in particular supernova explosions significantly affect the gas distribution within the disc leading to complex structures in the ISM. In the edge-on projection it can be clearly appreciated that galactic-scale fountain flows are generated.
\textit{Bottom:} The same column density projections (third and fourth row) for the high resolution Milky Way run simulated with the \citet{Springel2003} model. From these panels it is evident that the resulting ISM within the galaxy is much less structured because of the imposed effective equation of state. Furthermore, since supernovae are not explicitly modelled no SN cavities are present and no gaseous outflows are generated, as is evident from the edge-on projections.}
\label{fig:gasevo}
\end{figure*}

\begin{figure*}
\includegraphics[width=0.305\textwidth]{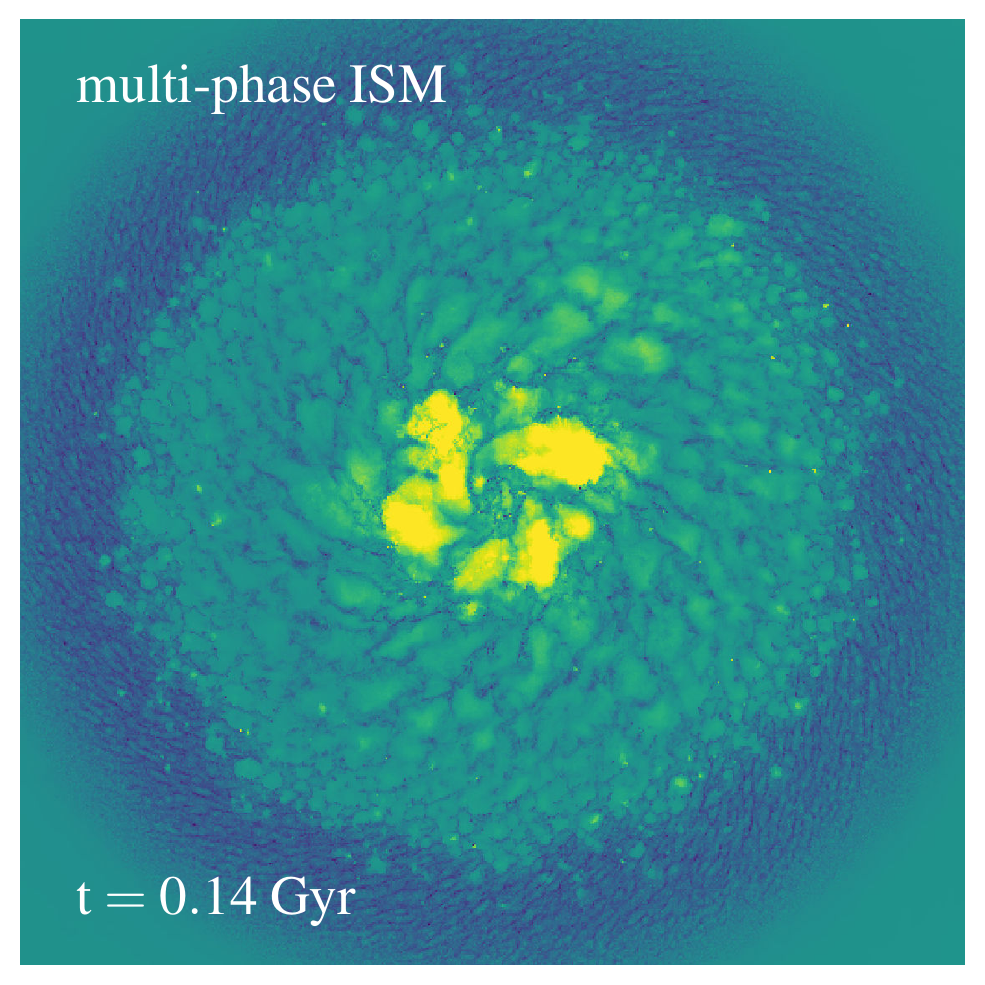}
\includegraphics[width=0.305\textwidth]{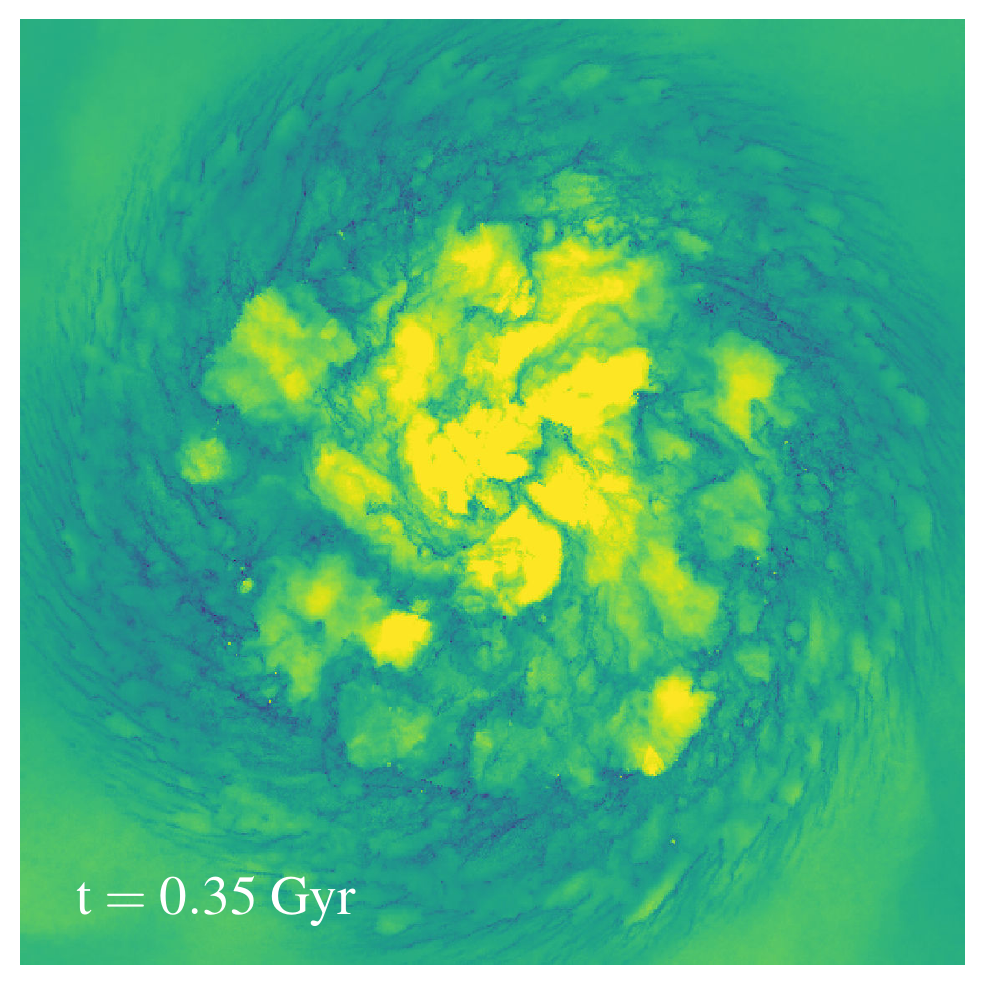}
\includegraphics[width=0.305\textwidth]{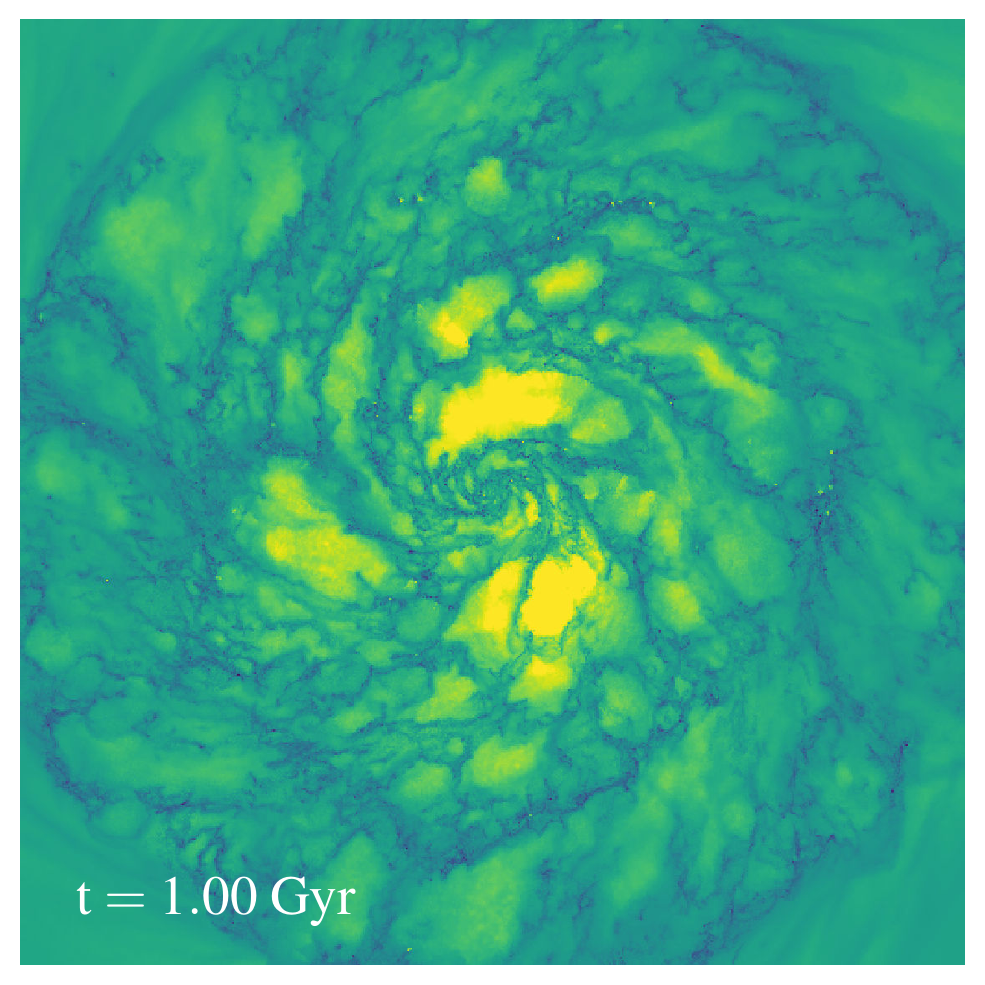}
\includegraphics[width=0.305\textwidth]{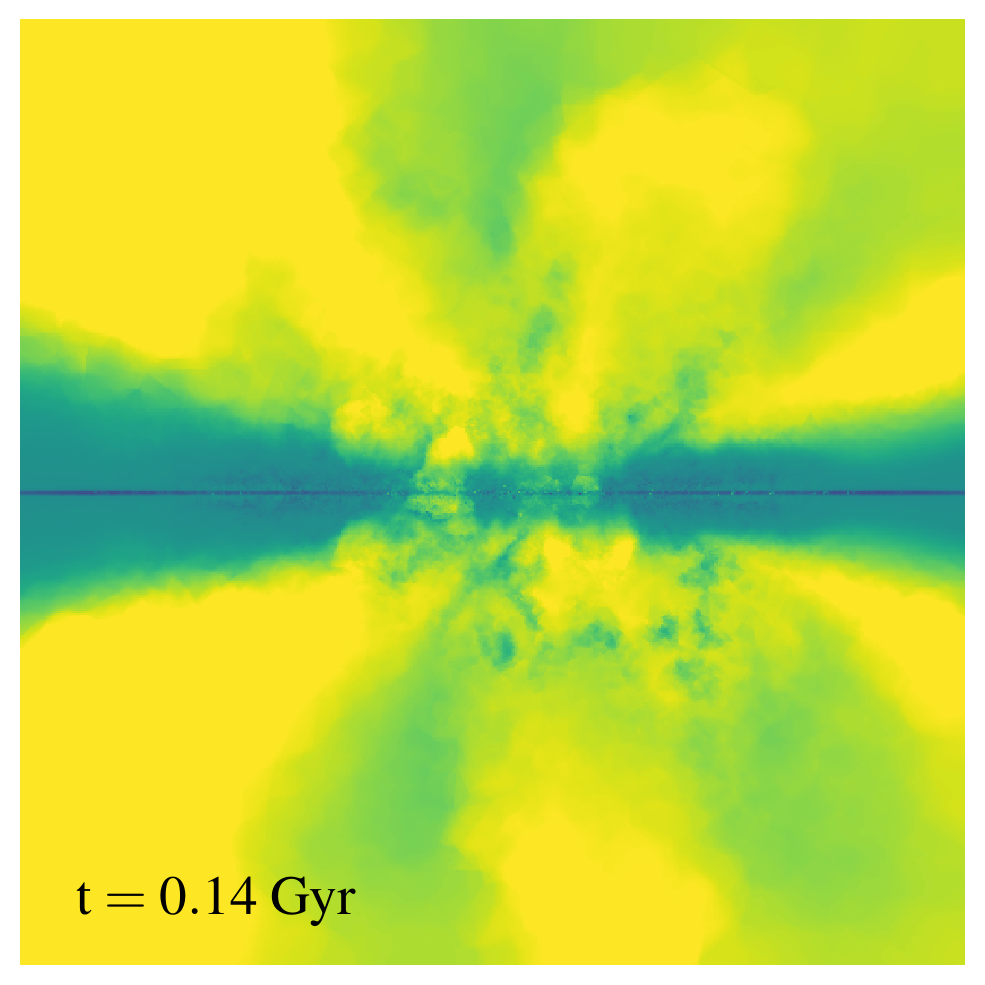}
\includegraphics[width=0.305\textwidth]{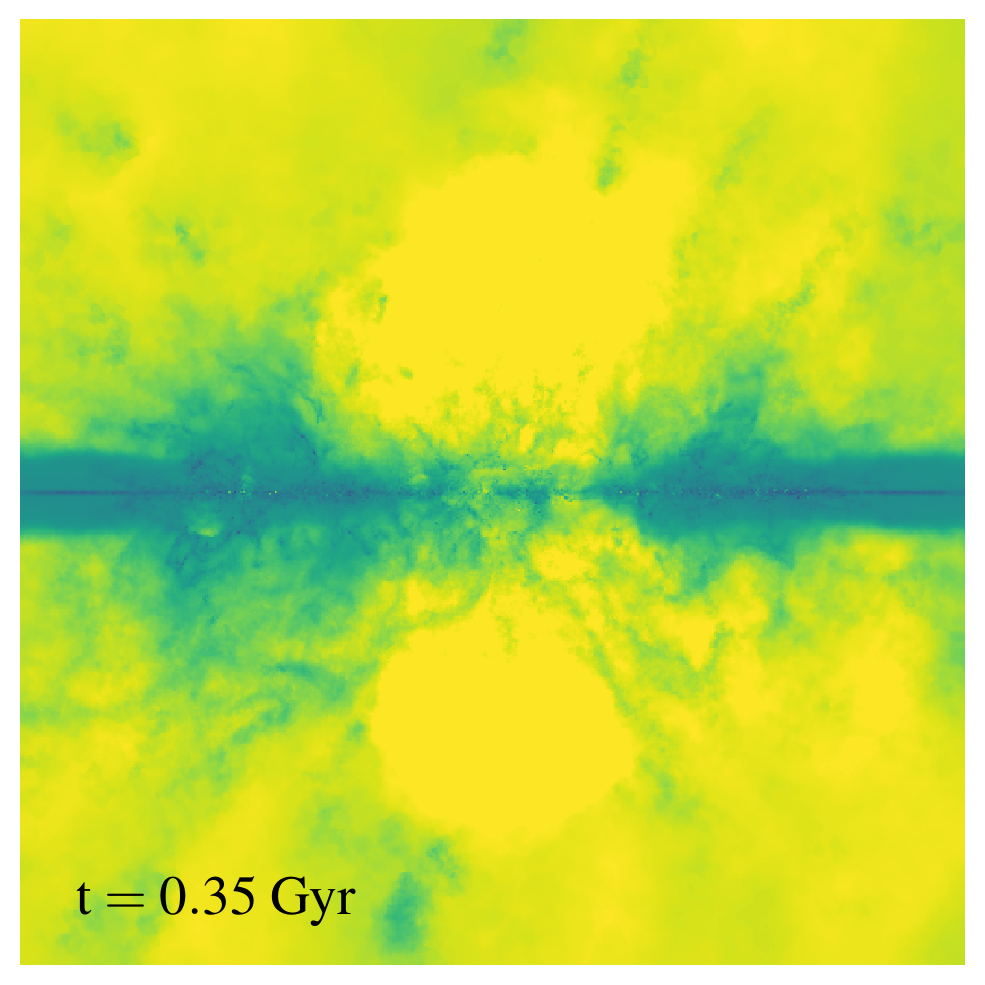}
\includegraphics[width=0.305\textwidth]{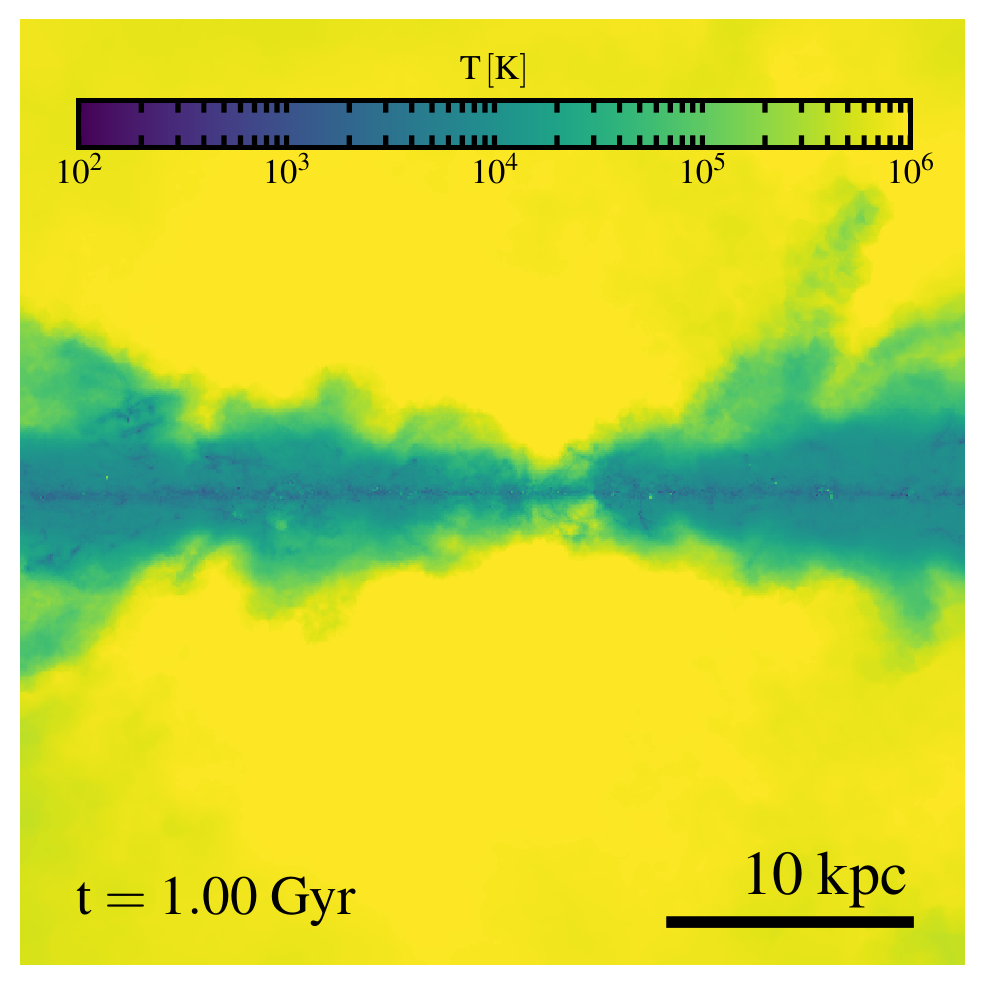}
\includegraphics[width=0.305\textwidth]{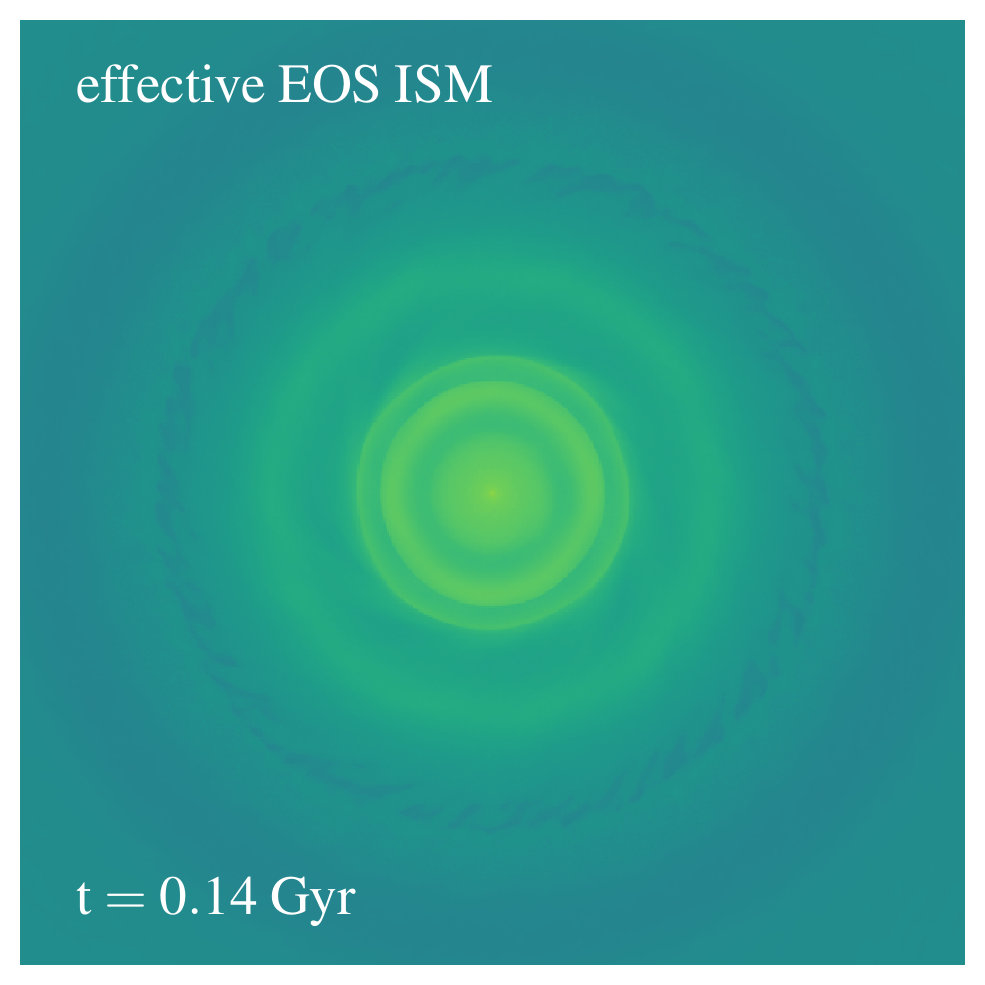}
\includegraphics[width=0.305\textwidth]{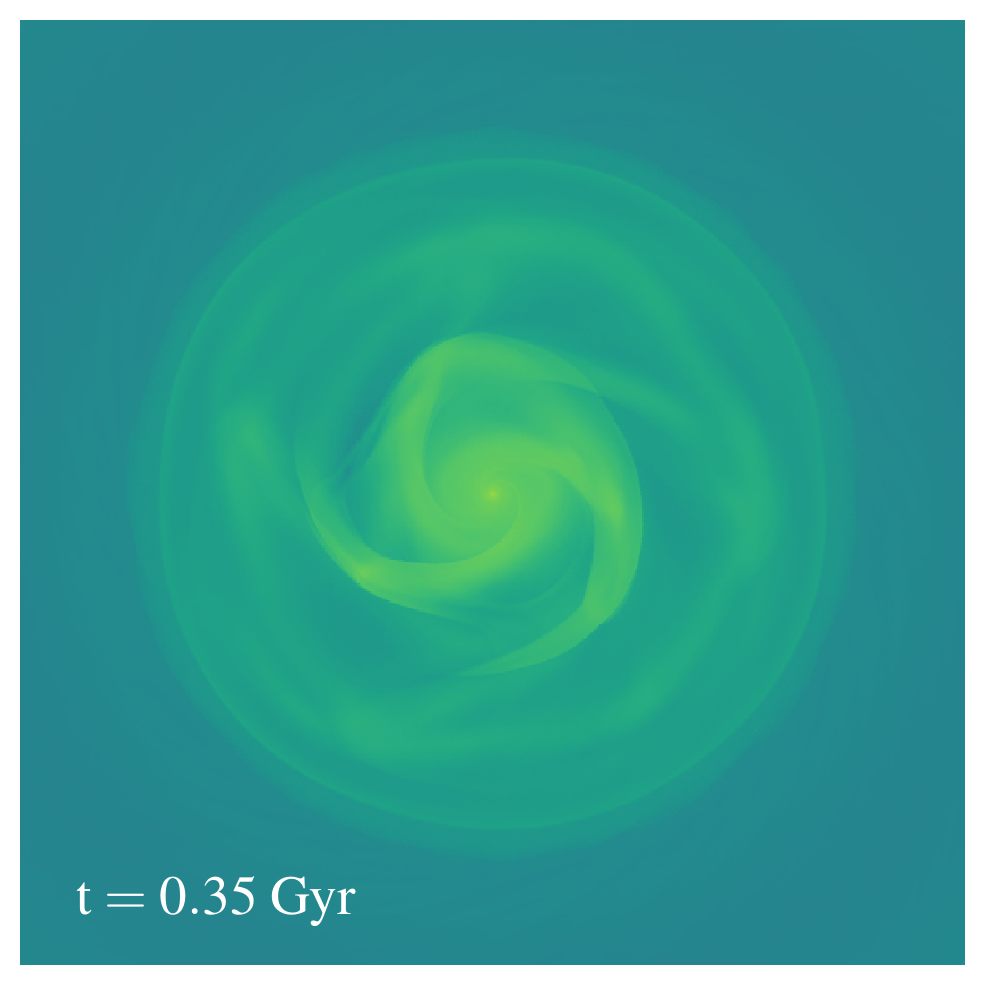}
\includegraphics[width=0.305\textwidth]{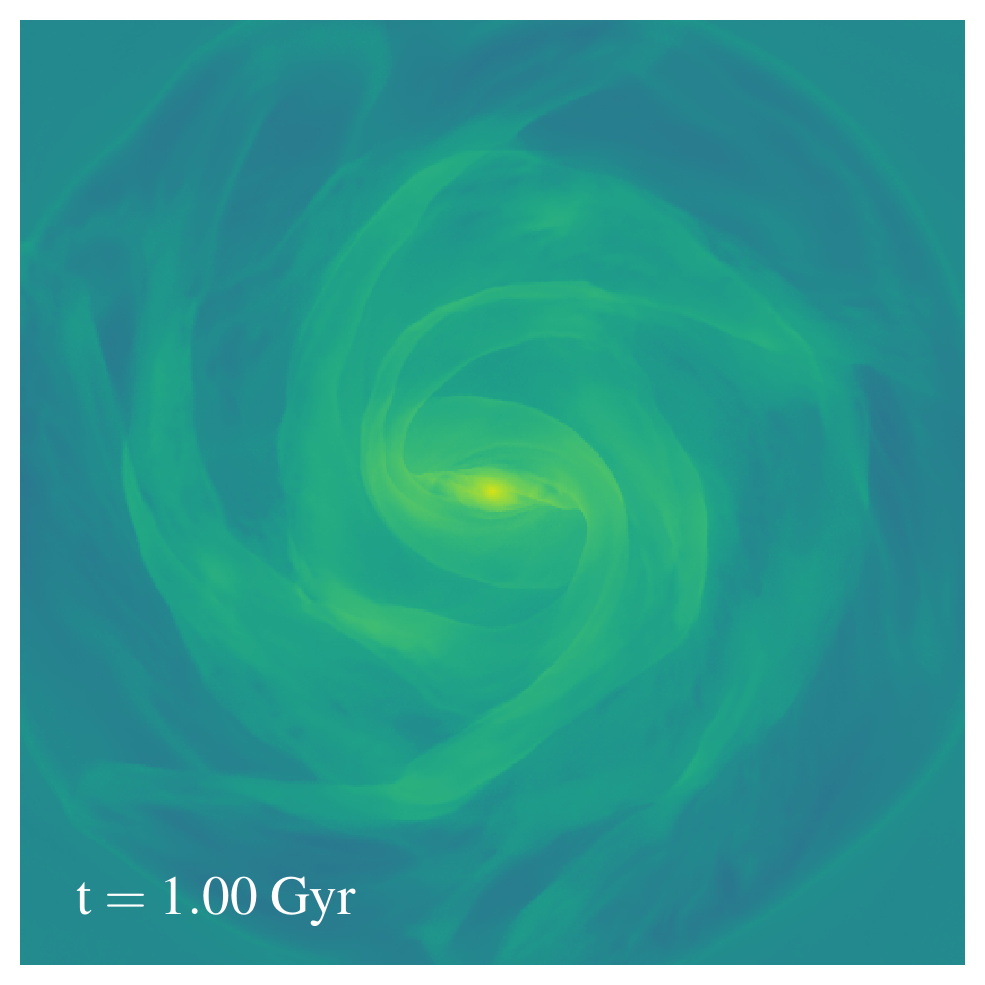}
\includegraphics[width=0.305\textwidth]{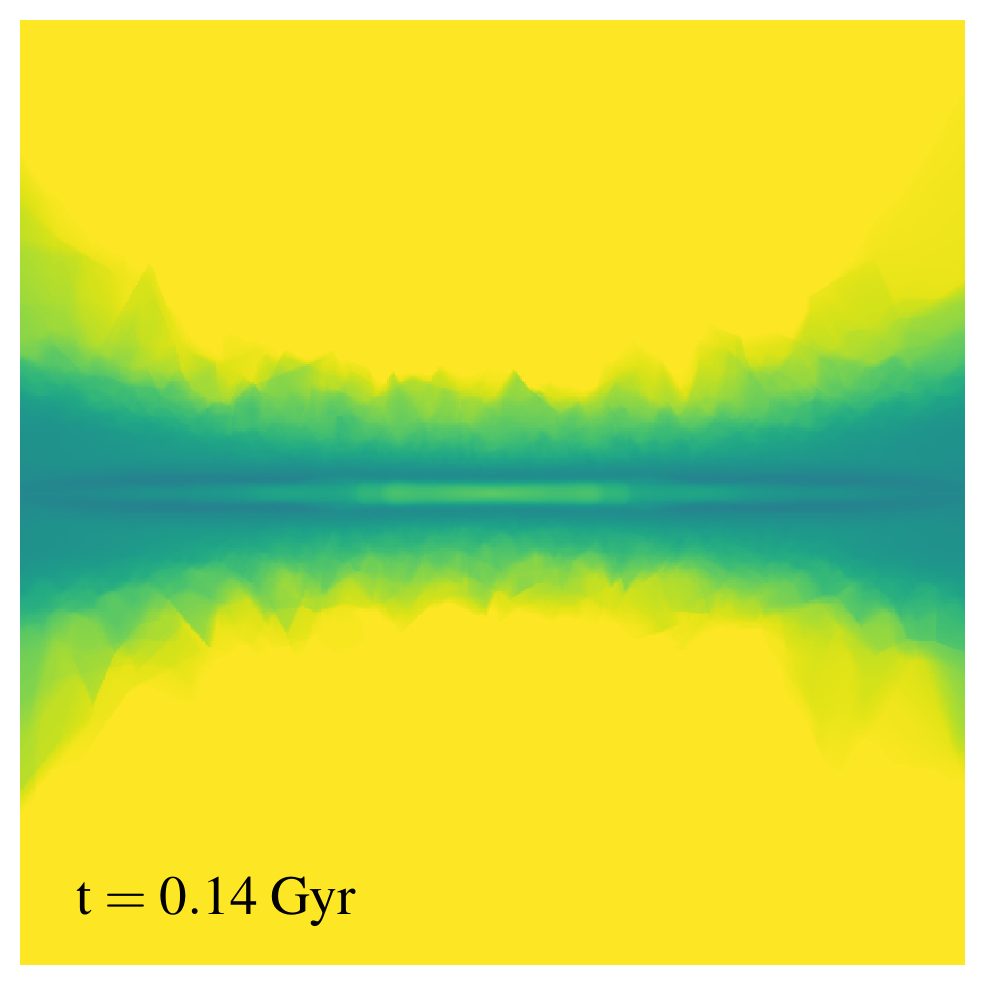}
\includegraphics[width=0.305\textwidth]{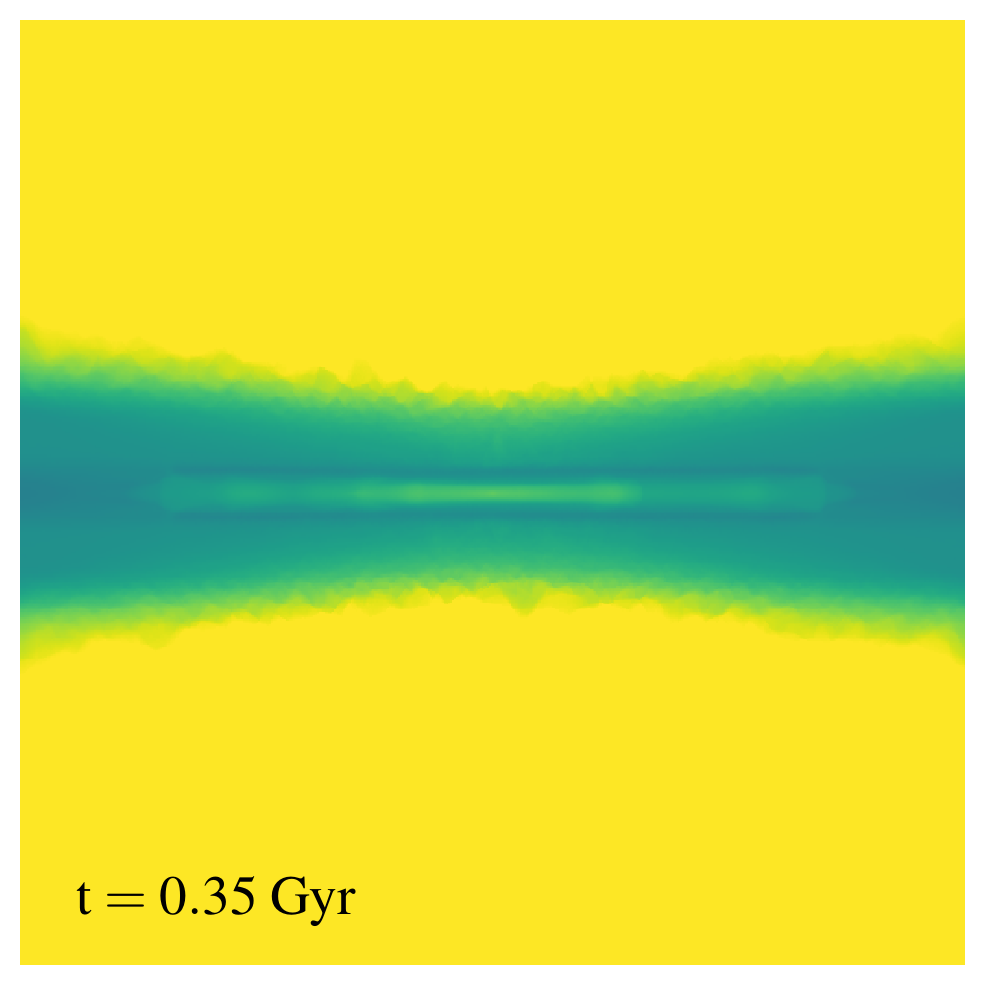}
\includegraphics[width=0.305\textwidth]{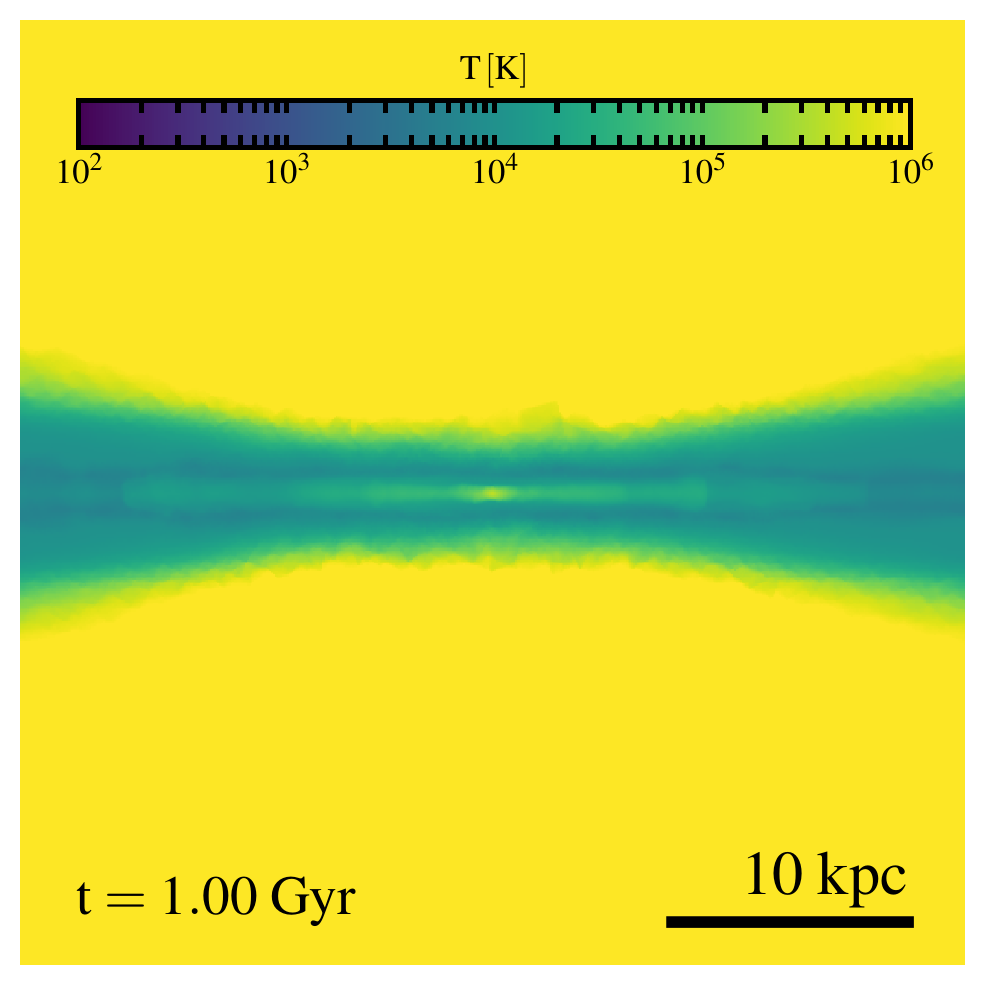}
\caption{\textit{Top:} Density-weighted gas temperature in a face-on (first row) and edge-on (second row) projection for the Milky Way galaxy at high resolution. Projections have been computed at times $t=0.14, 0.35, 1\,\Gyr$ from left to right column. Clearly visible is the multiphase ISM structure created by feedback and in particular the hot gas cavities within the disc created by supernovae that eventually break out.
\textit{Bottom:} The same projections for the high resolution Milky Way simulation run with the \citet{Springel2003} model. Note the difference in gas morphology (its more uniform appearance) and in temperature with respect the previous plot. In particular, the temperatures of the gas within the disc are higher because of the imposed effective equation of state.}
\label{fig:gasevo2}
\end{figure*}

\begin{figure*}
\includegraphics[width=0.305\textwidth]{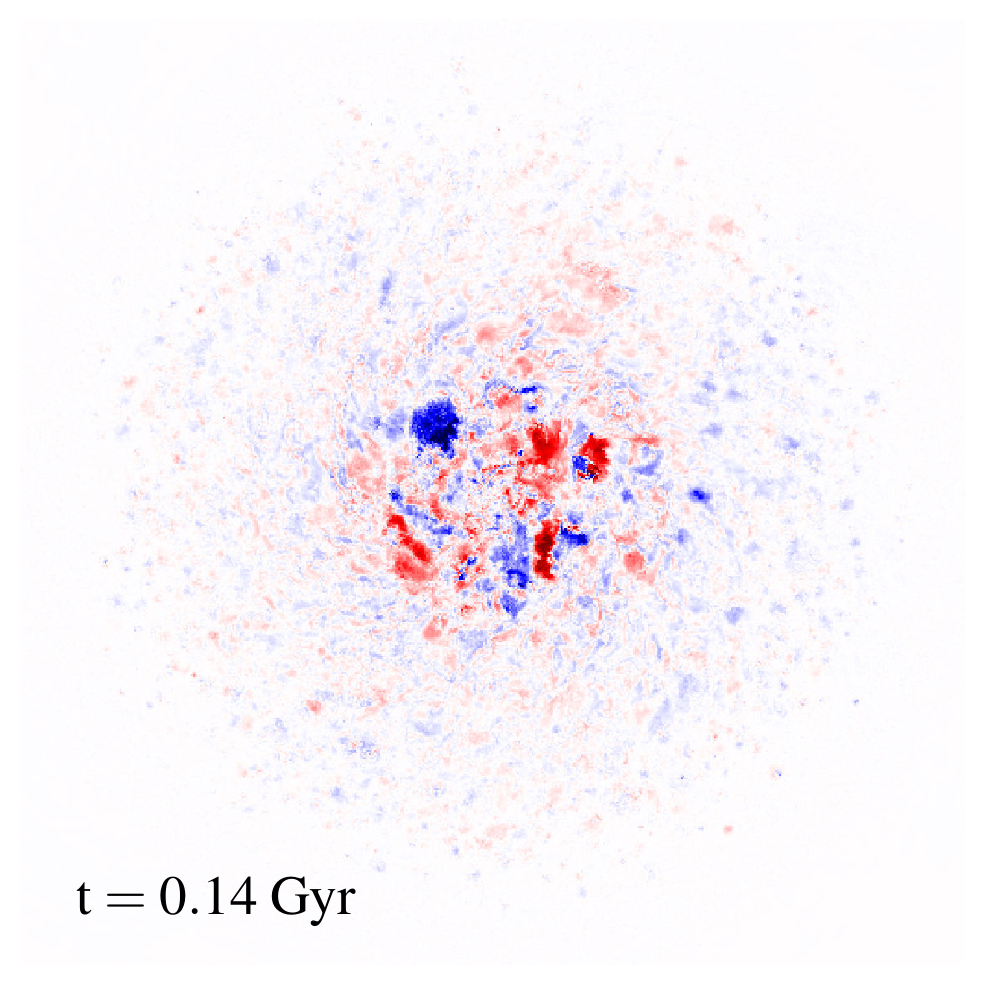}
\includegraphics[width=0.305\textwidth]{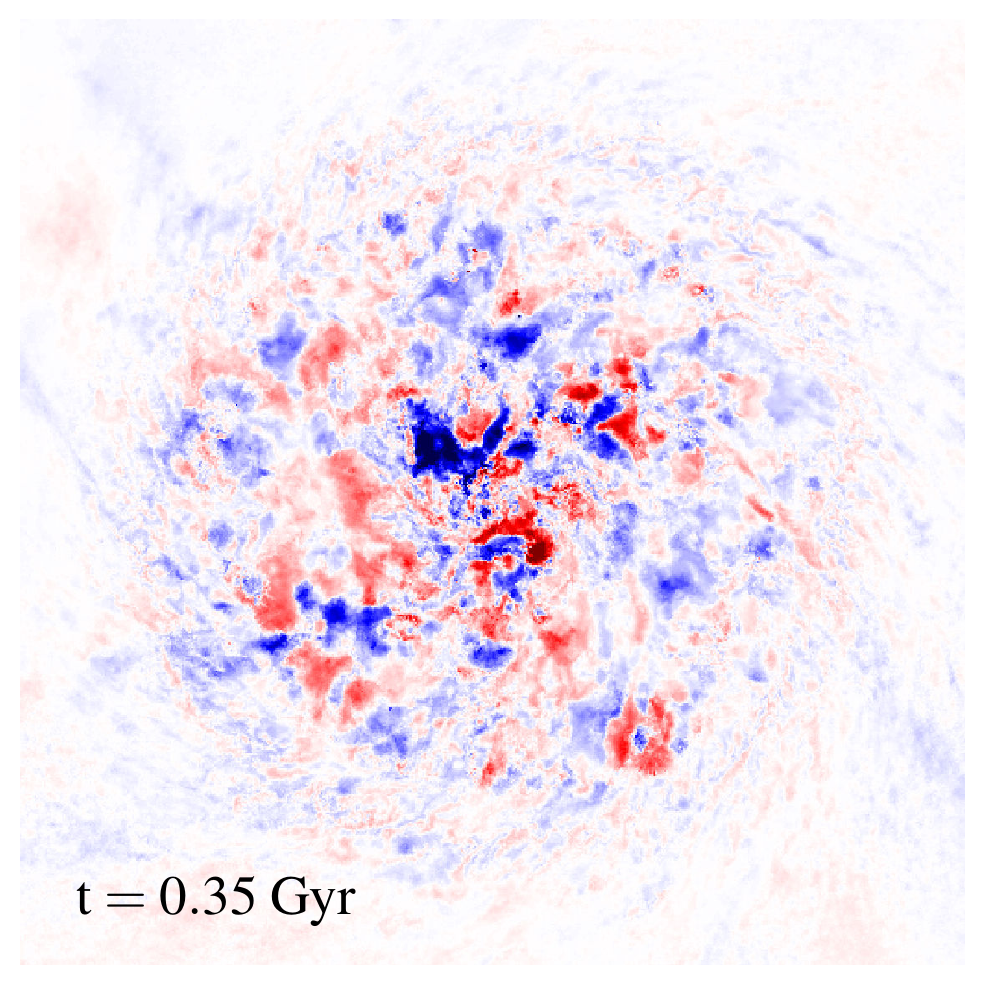}
\includegraphics[width=0.305\textwidth]{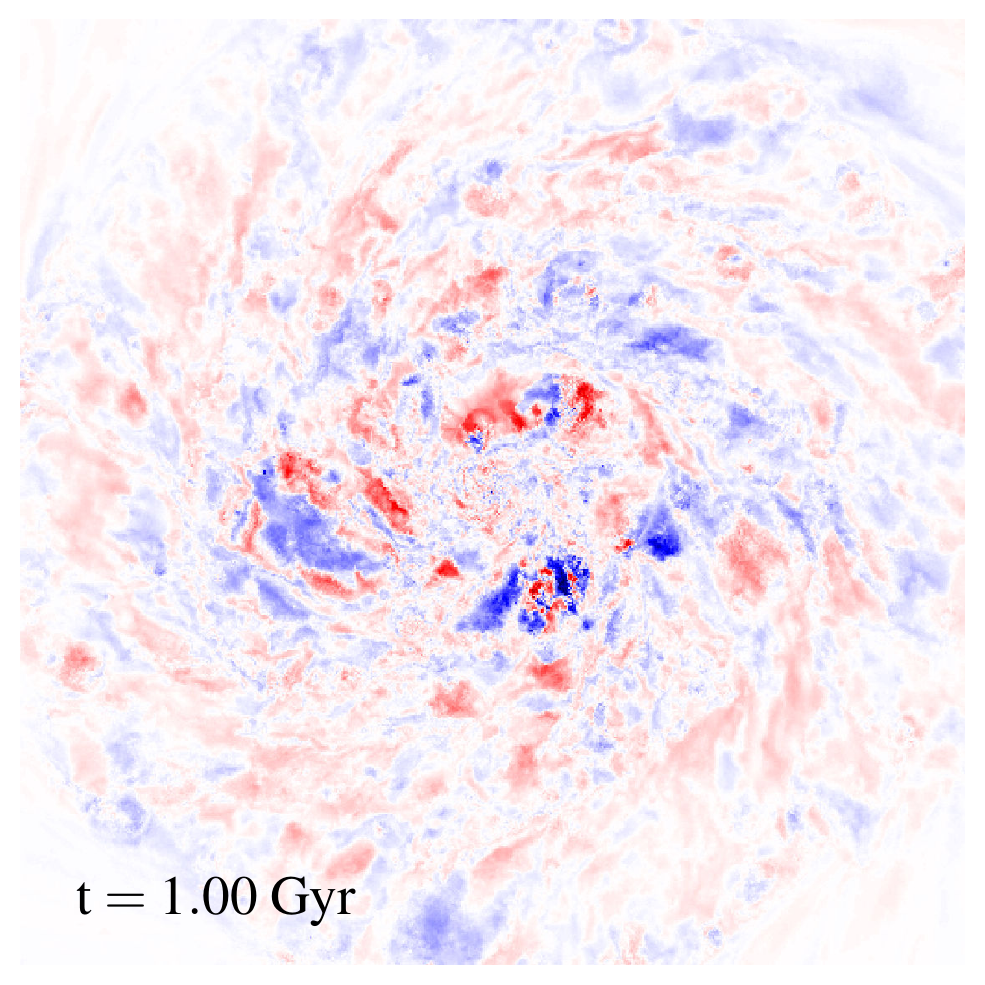}
\includegraphics[width=0.305\textwidth]{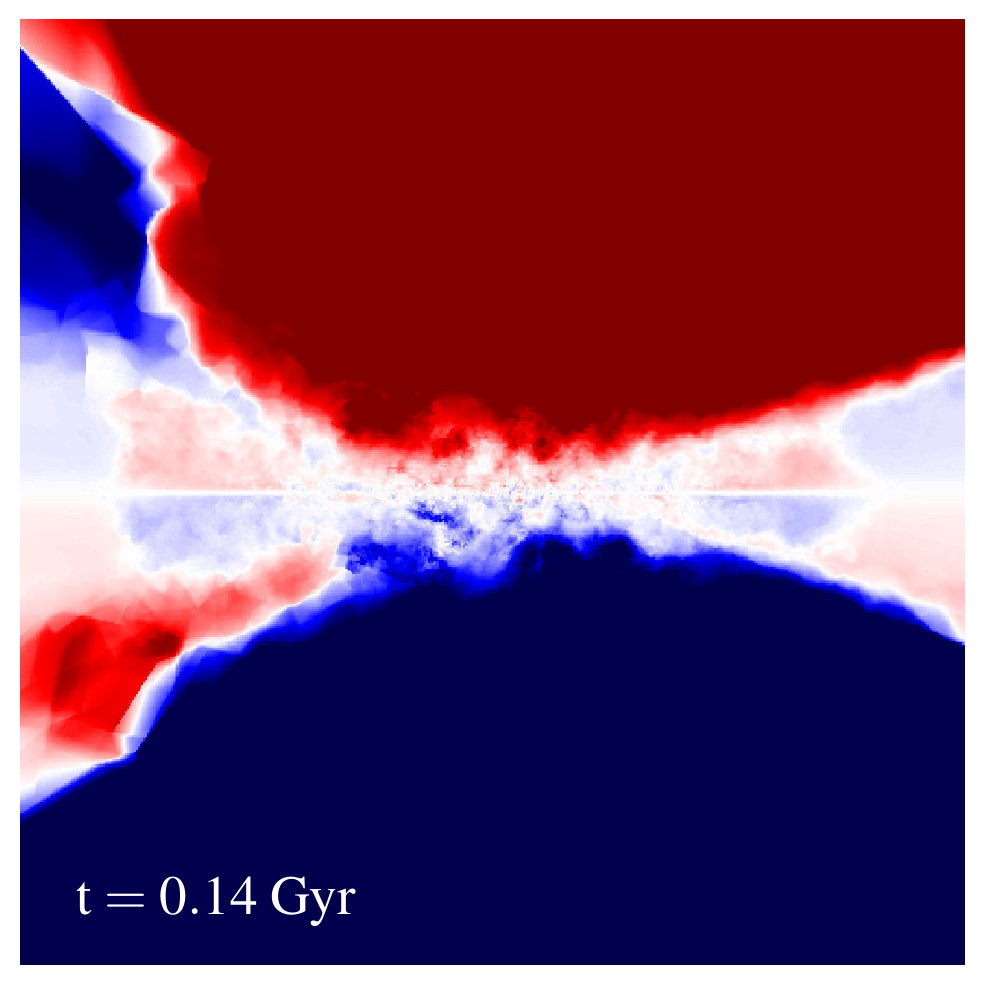}
\includegraphics[width=0.305\textwidth]{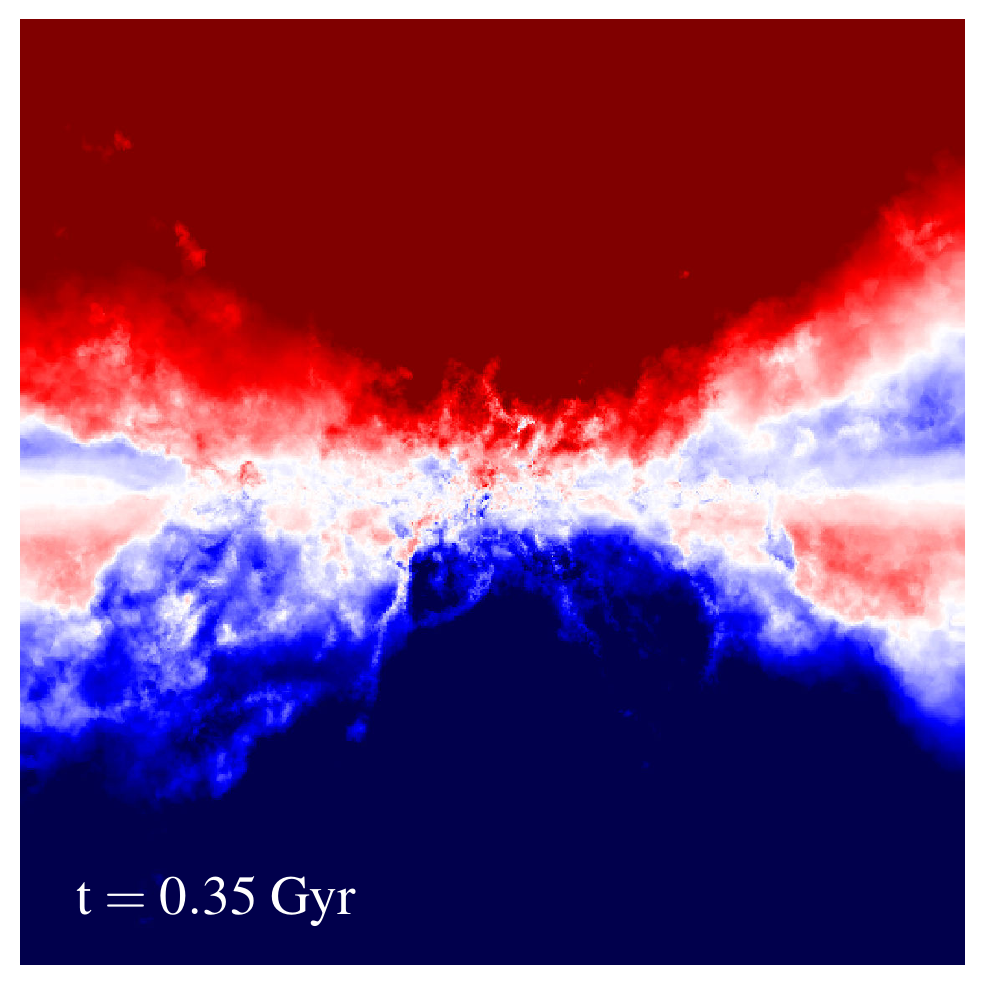}
\includegraphics[width=0.305\textwidth]{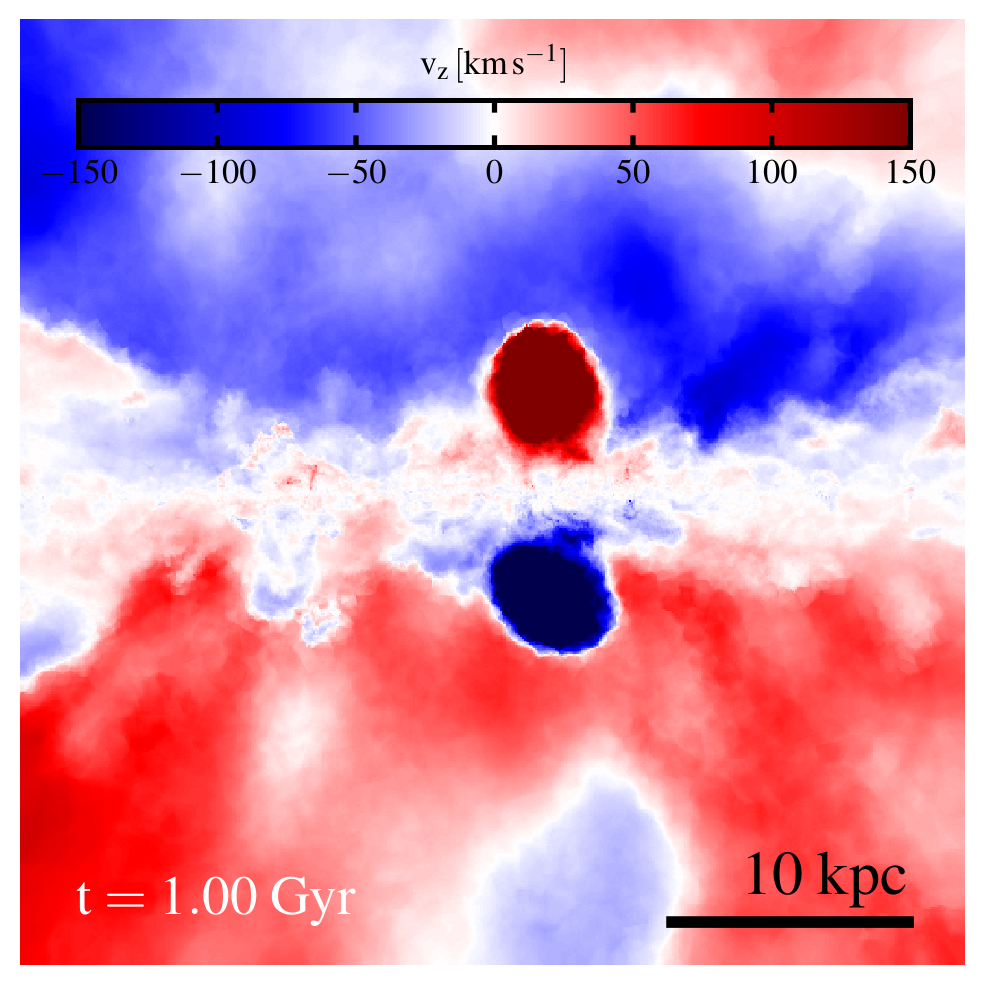}
\caption{Density-weighted gas velocity perpendicular to the disc plane in a face-on (first row) and edge-on (second row) projection for the Milky Way galaxy at high resolution. Projections have been computed at times $t=0.14, 0.35, 1\,\Gyr$ from left to right column. The face-on view highlights how gas that was contained in SN-generated cavities is expelled either above or below of the disc with velocity in excess of $\sim 150\,\kms$. The edge-on view clearly shows how the outflow of gas occurs over the whole galactic disc and that it is strongest at the center, where star formation is most active. However, at late times, substantial inflow of gas towards the disc can be observed, which is consistent with a galactic fountain circulation pattern.} 
\label{fig:gasvel}
\end{figure*}

\begin{figure*}
\includegraphics[width=0.975\textwidth]{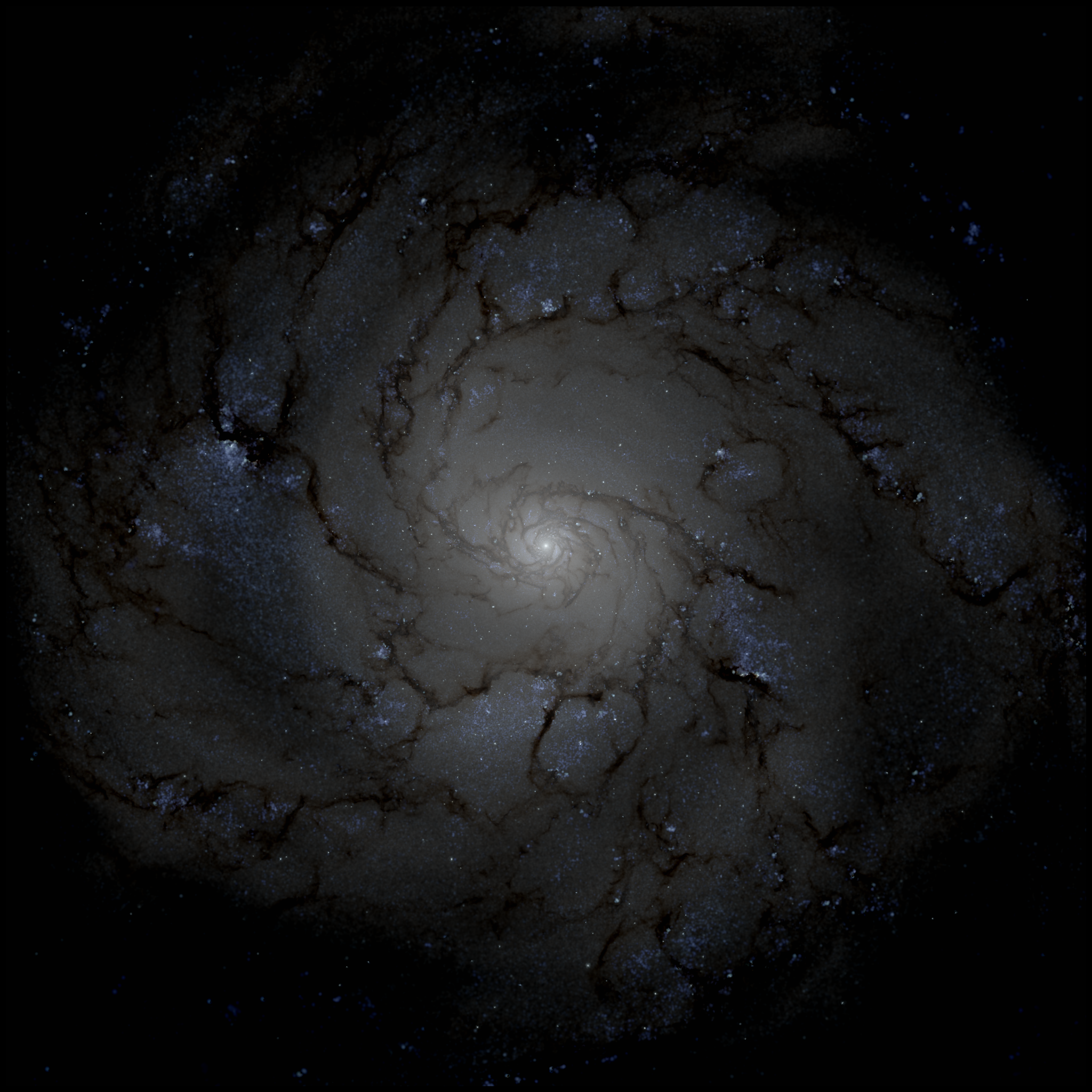}
\ifpdf
\\ 
\hspace{0.015cm}
\fi
\includegraphics[width=0.975\textwidth]{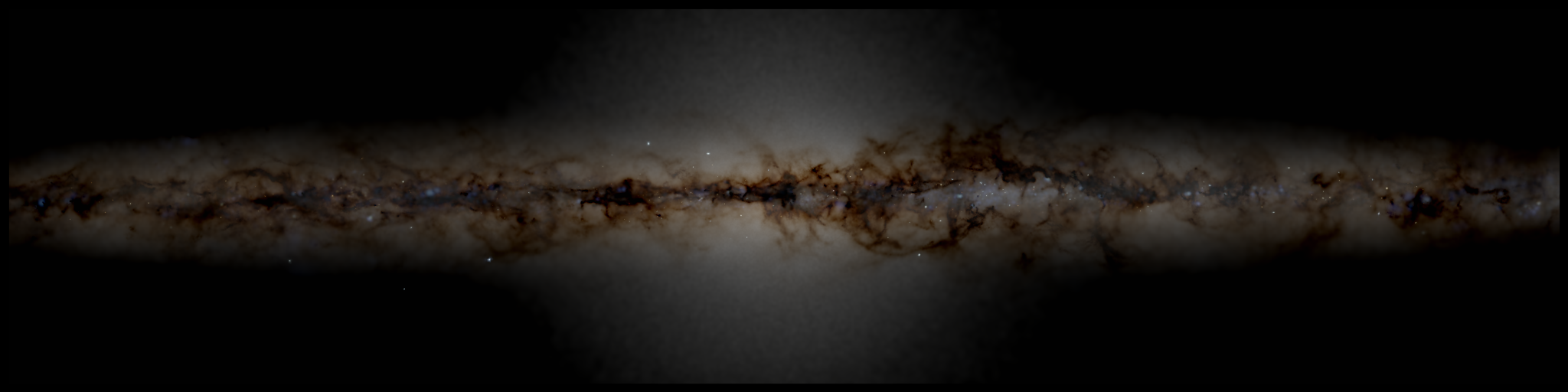}
\caption{Stellar light (ugr-band) images of the simulated galaxies face on (top) and edge on (bottom).
Light is assigned to stellar populations assuming a Chabrier IMF using the Starburst99 stellar population synthesis models.
Simple line-of-sight attenuation is included assuming a constant dust-to-metal ratio with a Milky Way attenuation curve.  } 
\label{fig:starsevo}
\end{figure*}

As a result of capturing this complexity in the gas phases, the structure 
of the gas in the galaxies simulated with \namemodel\ is also visually 
very different from previous models in {\sc arepo}. We show images of 
the evolution of the gas column density (Fig.~\ref{fig:gasevo}) and density-weighted 
gas temperature (Fig.~\ref{fig:gasevo2}) of the fiducial Milky Way galaxy run with
our new feedback model at high resolution (first two rows) 
and of the same galaxy run with the \citet{Springel2003} model (third and fourth row). 
Each figure shows face-on and edge-on projections 
taken at different times ($t=0.14, 0.35, 1\,\Gyr$, from left to right). Each 
panel is $40\,\kpc$ across and in projection depth, with a total number of 
$512^2$ pixels for a spatial resolution of $\approx 80\,\pc$. 

Stars tend first to 
form along the spiral pattern in the gaseous disc, where the density is 
highest. Particularly noticeable is the effect of stellar feedback in these 
regions, where cavities of low-density gas are carved by the combined action of 
radiation feedback and SN explosions. As time progresses, the gas 
distribution becomes increasingly structured, with dense gas regions in which 
star formation is active that are surrounded by cavities of lower density gas. 
Low-density bubbles are propelled from the star-forming disc into the halo region by the previous 
generation of stars. The whole process reaches self-regulation at later times. 
The feedback-driven gaseous outflows are more easily seen in the edge on 
projections, in which the amount of gas above and below the disc -- and also the disc 
thickness -- tends to increase as a function of time. 

In contrast, none of these features are present in the simulations run 
with the \citet{Springel2003} model. In this case, the use of an effective 
equation of state is able to regulate star formation at the price of losing 
information on the ISM structure. In particular, the appearance of the ISM is 
smoother and no SN cavities are visible. Furthermore, since the momentum 
injection from stellar feedback is not directly modelled \rev{(we recall that in the most basic implementation of the \citealt{Springel2003} model supernova feedback acts only as a heating source for the hot phase of the ISM)}, gaseous outflows are not generated, as can be clearly appreciated in the edge-on projections. 
This is the reason why in cosmological applications the \citet{Springel2003} 
model has to be complemented with a prescription for hydrodynamically decoupled 
galactic winds \rev{(see also their Section 4)}. 

Similar trends are also visible in the temperature plots (Fig.~\ref{fig:gasevo2}), 
for which it is worth 
noticing that the densest gas regions feature the lowest temperatures, whereas 
the gas cavities carved by feedback are filled with shock-heated hot ($\sim 10^6\,\K$) gas, as naturally expected in this multiphase gas media. This hot gas is eventually 
vented outside the disc in a superbubble blowout. 
These features are smoothed over in the case of the \citet{Springel2003} simulation, 
which also develops high temperature gas region in the disc midplane
as a result of the imposed effective equation of state.

The development of a gas circulation cycle over galactic scales can be further inspected
in the kinematical face-on and edge-on maps shown in Fig.~\ref{fig:gasvel}, constructed
following the same specifications from the previous two figures. 
The highest vertical velocities $\gsim 150\,\kms$ are reached in the low-density gas 
cavities generated by stellar (and in particular SN) feedback, which is best
appreciated in the face-on projection (top row) of Fig.~\ref{fig:gasvel}. 
This is consistent with the picture that the gas originally contained in such cavities has been displaced by the momentum injected by star particles and accelerated to 
velocities compatible with its ejection above or below the disc mid-plane. The 
fact that a large-scale outflow is generated in the \namemodel\ simulation is clearly visible in the edge-on projections, in which most of the gas that is a few $\kpc$ 
above or below the disc plane has outflow velocities in excess of $150\,\kms$. 

However, the kinematics of the gas is more complex than a simple ejection occurring 
across the whole star-forming disc. The outflow is strongest near the galactic centre, where star formation is most effective, but at late times 
it can be seen that gas is also accreting onto the disc. As argued before in the discussion of
Fig.~\ref{fig:outflow} it implies that outflows do not completely escape the galaxy, but rather 
create a galaxy-wide gas circulation compatible with a galactic fountain flow \citep{Bregman1980}. 
The accreted (recycled) gas thus becomes the fuel that sustains star formation at later times. 

These results, combined with the discussion of Fig.~\ref{fig:outflow}, suggest that the produced galactic-scale winds are still rather weak overall ($\beta \sim 1$) and do not seem to be able to produce baryonic loss from the halo, at least not at the halo mass scale ($M_{\rm tot} \sim 10^{12}\,M_\odot$) and redshift ($z\sim 0$) probed by these calculations. However, cosmological simulations indicate that strong outflows that lead to baryonic loss are needed to produce sensible galaxies in the $\Lambda$CDM cosmology \citep[e.g.][]{Schaye2015,Muratov2015,Pillepich2018b,Tollet2019}. Testing the ability of this model of doing so below $M_{\rm tot} \sim 10^{12}\,M_{\odot}$ (i.e. the dwarf galaxy mass scale) and in cosmological applications is therefore needed to assess its potential in producing realistic galaxies in the full cosmological context. We plan to address this important issue in future work.

\section{Contributions of the different feedback channels}\label{sec:feedchan}

The spatial distribution of newly formed stars in our model sets the
location and conditions under which feedback is injected. Thus, we start our
analysis by showing in Fig.~\ref{fig:starsevo} face-on (top) and edge-on 
(bottom) stellar light maps for the fiducial Milky Way. 
Light is assigned to stellar populations assuming a \citet{Chabrier2001} IMF using the {\sc starburst99} stellar population synthesis models. 
Simple line-of-sight attenuation is included assuming a constant dust-to-metal ratio with a Milky Way attenuation curve.
Younger stellar populations (identified by their blue colors) are not uniformly distributed, but rather clustered in regions of recent star formation activity. The gas distribution (identified by the darkened features) is also distinctly non-uniform. The edge-on panels show clear dust lanes, with a large fraction of the dense, star-forming gas being constrained to a disc plane. Newly born stars are mostly born within a very thin disc plane owing to the low ($\lsim 10^3\,\K$) temperatures reached by the dense ISM gas from which they are born. 

\rev{In \namemodel, consistent with other models in the literature, the clustering of newly formed stars is the result of tying the active star formation regions to cold and relatively high-density ($n\gsim 100\, \cm^{-3}$) gas. Indeed, in the simulations performed with the \citet{Springel2003} model, that features a lower ($n \sim 0.1\, \cm^{-3}$) star formation density threshold, the distribution of younger stellar particles is more uniform, albeit dependent on gas density. The presence of galactic outflows in the \citet{Springel2003} model does not significantly alter this result.}

\rev{The clustering of star formation might have an impact on the injection of supernova feedback and the response of the gas to it. Some studies \citep[e.g.][]{Buck2019} also suggests that the strength of the clustering signal of young stellar particles can be used to derive constraints on the most appropriate value of the star formation density threshold to use in cosmological simulations. The detailed study of the clustering properties of star formation, their dependence on model parameters (particularly on the star formation density threshold) and the implications for stellar feedback is outside the scope of the present paper and will be addressed in future work.}
\begin{figure*}
\includegraphics[width=0.325\textwidth]{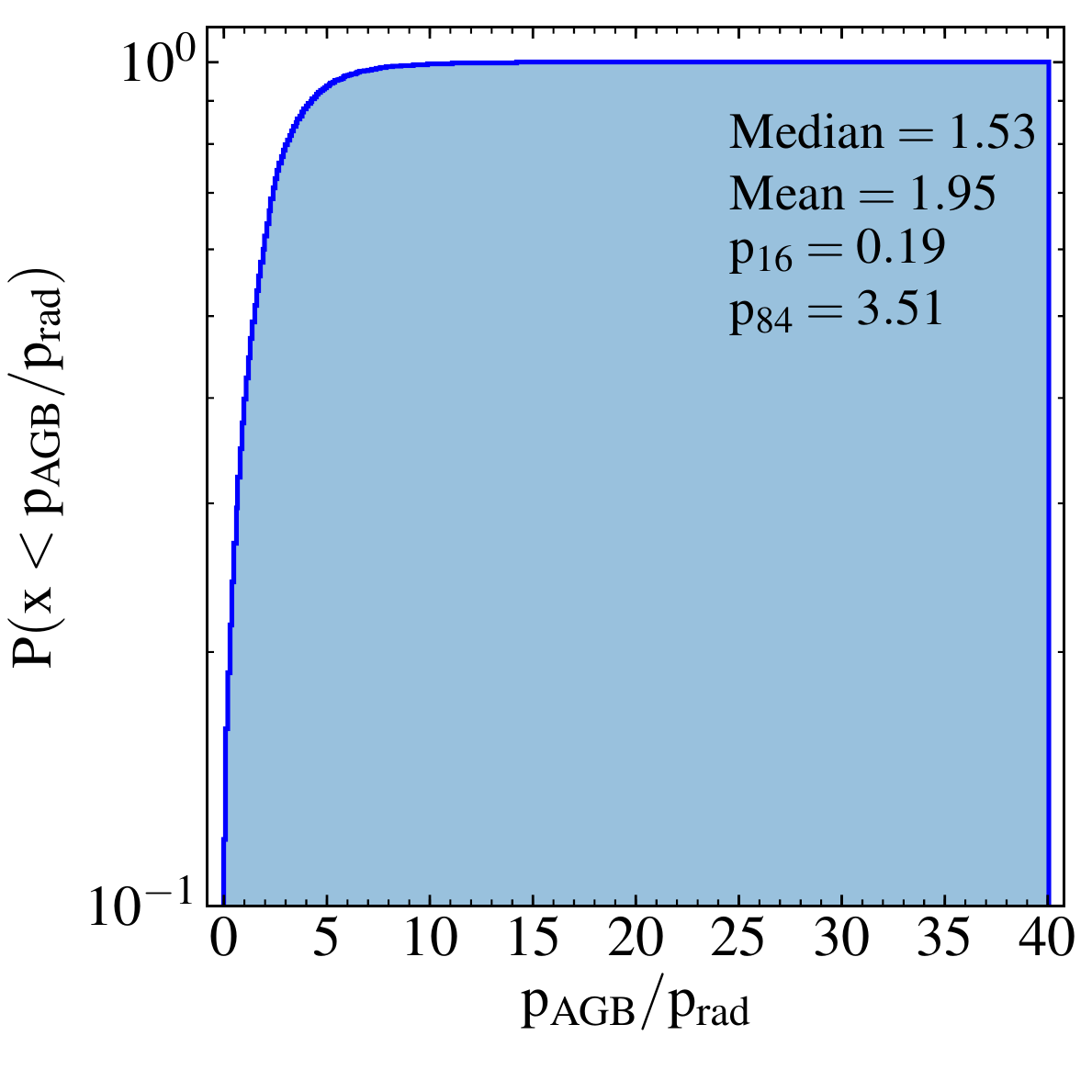}
\includegraphics[width=0.325\textwidth]{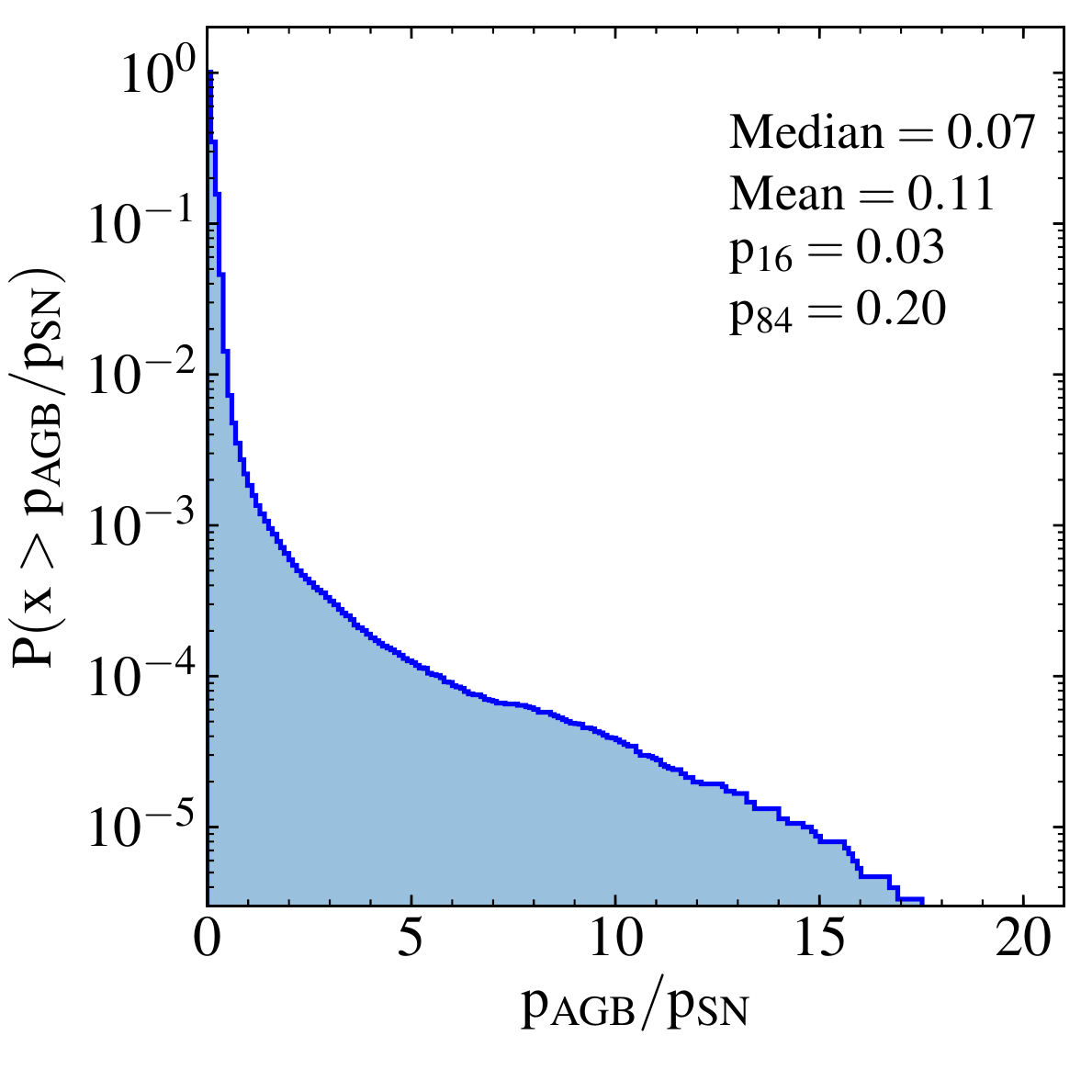}
\includegraphics[width=0.325\textwidth]{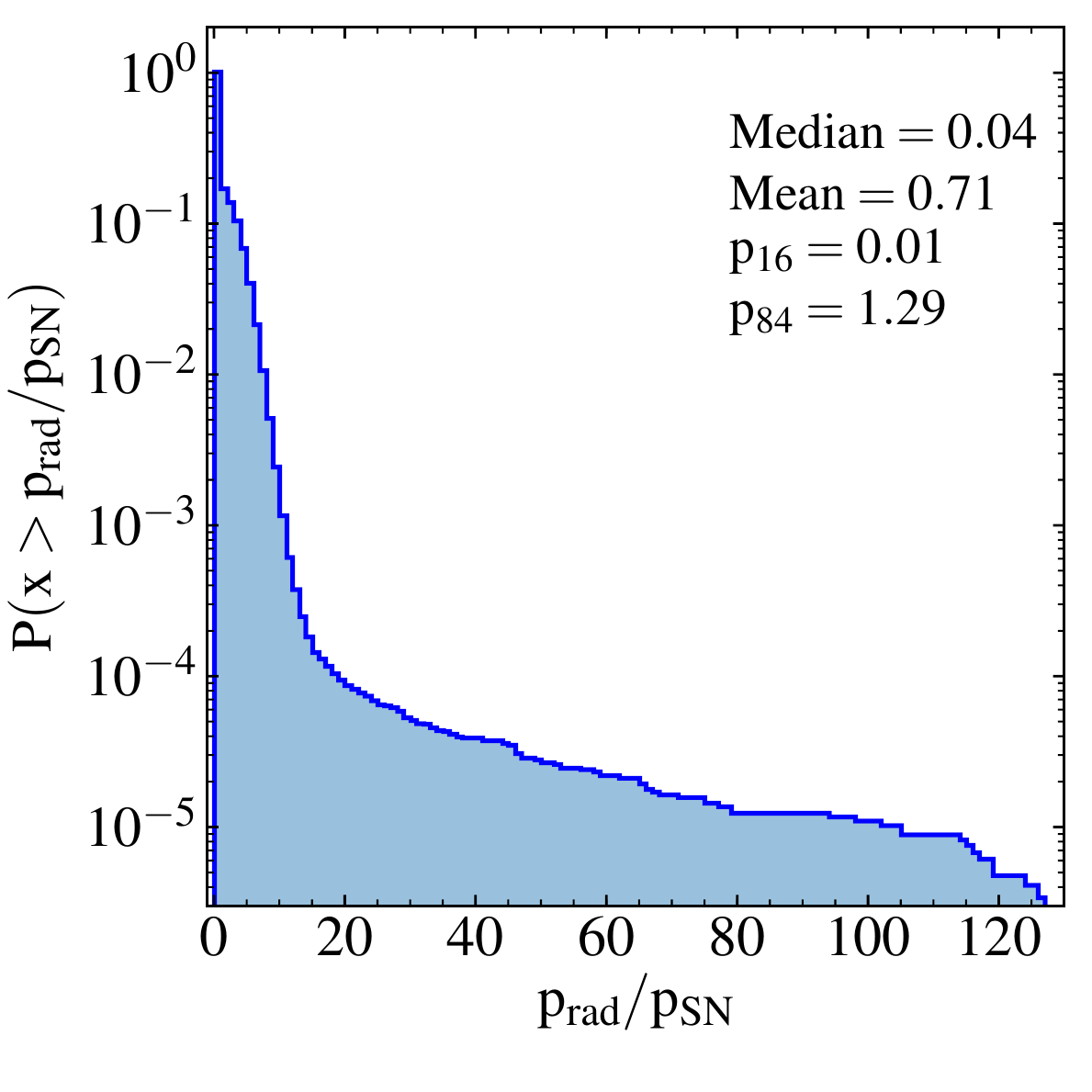}
\caption{Cumulative distribution of the ratio between OB/AGB injected momentum 
and radiation momentum (left-hand panel), OB/AGB and SN momentum (central panel), and 
radiation and SN momentum (right-hand panel) for the 
high resolution Milky Way simulation. All the momenta are given as cumulative 
quantities throughout the life time of each stellar particle. In all panels we report the median, mean and the 16-th and 84-th percentile of the 
resulting distributions to illustrate the relative contribution of each feedback channel considered by the present model. Stellar winds and radiation provide roughly the same amount of total momentum. This quantity 
is about $\sim 5-10$ pre cent of the one provided by SNe, which for Milky Way-like objects are therefore the dominant channel.}
\label{fig:pratios}
\end{figure*}

As stars are born and stellar evolution takes over they inject energy and momentum back to the surrounding media. 
We keep track of the cumulative momentum injected into the gas by all 
feedback channels (radiative, OB/AGB winds and SNe). This enables some 
interesting comparisons among the different channels and allows us to gauge their 
relative contributions to the star formation self regulation and outflows. 
Additionally, as in the case of radiative and SN feedback, one can 
evaluate their ``boost'' factors, i.e. the ratio between 
momentum actually injected into the gas compared to the one expected.

In Fig.~\ref{fig:pratios} we display the ratios between cumulative momenta 
injected into the gas by different feedback channels for our high 
resolution simulation. 
The different panels show the ratios between OB/AGB winds and 
radiative feedback (left-hand panel), OB/AGB winds and SNe (central panel), 
and radiative feedback and SNe (right-hand panel). The plots are 
computed as cumulative distribution of the ratio of the total momenta injected by 
each stellar particle in their life-time in each feedback channel (blue solid line). 
A variable number of constant-width bins has been used to compute the distribution 
of all the ratios. In particular, the bin width for ratios involving OB/AGB winds 
is 0.1, whereas it is fixed to 1 for the SN and radiative feedback case. For plots including SN momentum, only stars older than $5\,{\rm Myr}$ are considered. This is to exclude the nominal delay of $\sim 5\,{\rm Myr}$ between the creation of the star particle and the first SN event in order to not bias the analysis. We note, however, that because of our discrete sampling of the SN events there is a residual dependence of the tail of the distribution on the adopted age cut (see also the discussion below). 

Several interesting trends are noticeable. First, the left panel of Fig.~\ref{fig:pratios}
indicates that stellar winds (OB/AGB) and radiation feedback provide approximately 
the same amount of momentum to the gas within a $1\sigma$ dispersion\footnote{This is computed in analogy with a Gaussian distribution by considering half the range between $16-$th and $84-$th percentile 
of the distribution.} of $\simeq 1.66$. Second, as indicated by the middle panel, 
SN momentum clearly dominates over that from OB/AGB winds.
Indeed, the momentum imparted by stellar winds is on average 
about 10 per cent or less than the one delivered by SNe and the distribution 
around the mean is quite narrow ($1\sigma$ dispersion is $\simeq 0.09$). 
We note, however, that a small minority of stars have a tail 
extending to momentum ratios up to $p_{\rm AGB}/p_{\rm rad} \gsim 15$. 
We have checked that this is caused by young stars (age $\lsim 40\,\Myr$), 
for which there is a delay in the onset of the first SN explosions 
and the energy and momentum injected by stellar winds dominates for a short 
period of time from the creation of the star particle and before the first SN event. 

We now turn our attention to the ratio between the momentum injected by radiation
compared to SN, with the distribution shown in the rightmost panel of Fig.~\ref{fig:pratios}.
The behaviour is similar to the stellar winds analyzed before, 
with radiation being largely subdominant to the SN channel. 
Radiation can deliver on average a factor of $\sim 0.71$ times the momentum 
of SNe with a $1\sigma$ dispersion of $\simeq 0.64$. However, the estimate for the mean is biased by the long tail of stellar particles extending up to $p_{\rm SN}/p_{\rm rad} \gsim 100$. By using the median of the distribution, the momentum imparted by the radiation is merely $4$ per cent of the one injected by SNe. Once again the extended tail at large $p_{\rm rad}/p_{\rm SN}$ values is due to very young stars for which the momentum budget is dominated by radiation because there has not been enough time for SN to set in. Moreover, even by using a more conservative cut on stellar ages ($40\,{\rm Myr}$) leaves a tail of particles for which values of $p_{\rm rad}/p_{\rm SN} \sim 15$ can be reached. This signals that the momentum imparted by the radiation can locally play an important role, especially in high gas density regions. However, if one looks at the overall budget, this analysis suggests that SNe dominate the momentum feedback input in our simulations.

\rev{The results discussed above appear to be quite robust to numerical resolution at which the simulations are performed. Although the median values of the imparted momentum ratios between different feedback channels change with resolution, the conclusions on their relative contribution to the overall momentum output hold also at lower resolution levels. In particular, the median $p_{\rm AGB}/p_{\rm rad}$ ratio increases from 0.36 to 1.53 passing from low to high resolution, the median  $p_{\rm AGB}/p_{\rm SN}$ ratio is almost constant across levels (decreasing only from 0.1 to 0.07 from low to high resolution), and the median $p_{\rm rad}/p_{\rm SN}$ ration decreases from 0.23 at low resolution to 0.04 at high resolution. From these findings it is evident that, for increased resolution, the radiation pressure channel becomes on average less important for the overall momentum output, whereas the ratio between the momentum imparted to the gas by OB/AGB winds and supernovae remains approximately constant. The radiation pressure is the most sensitive feedback channel to resolution changes, because its effects are more strongly influenced by the detailed structure of the ISM given the explicit dependence of the imparted momentum on the gas column density (see eq.~[\ref{eq:radpress}]).}

\begin{figure}
\includegraphics[width=0.48\textwidth]{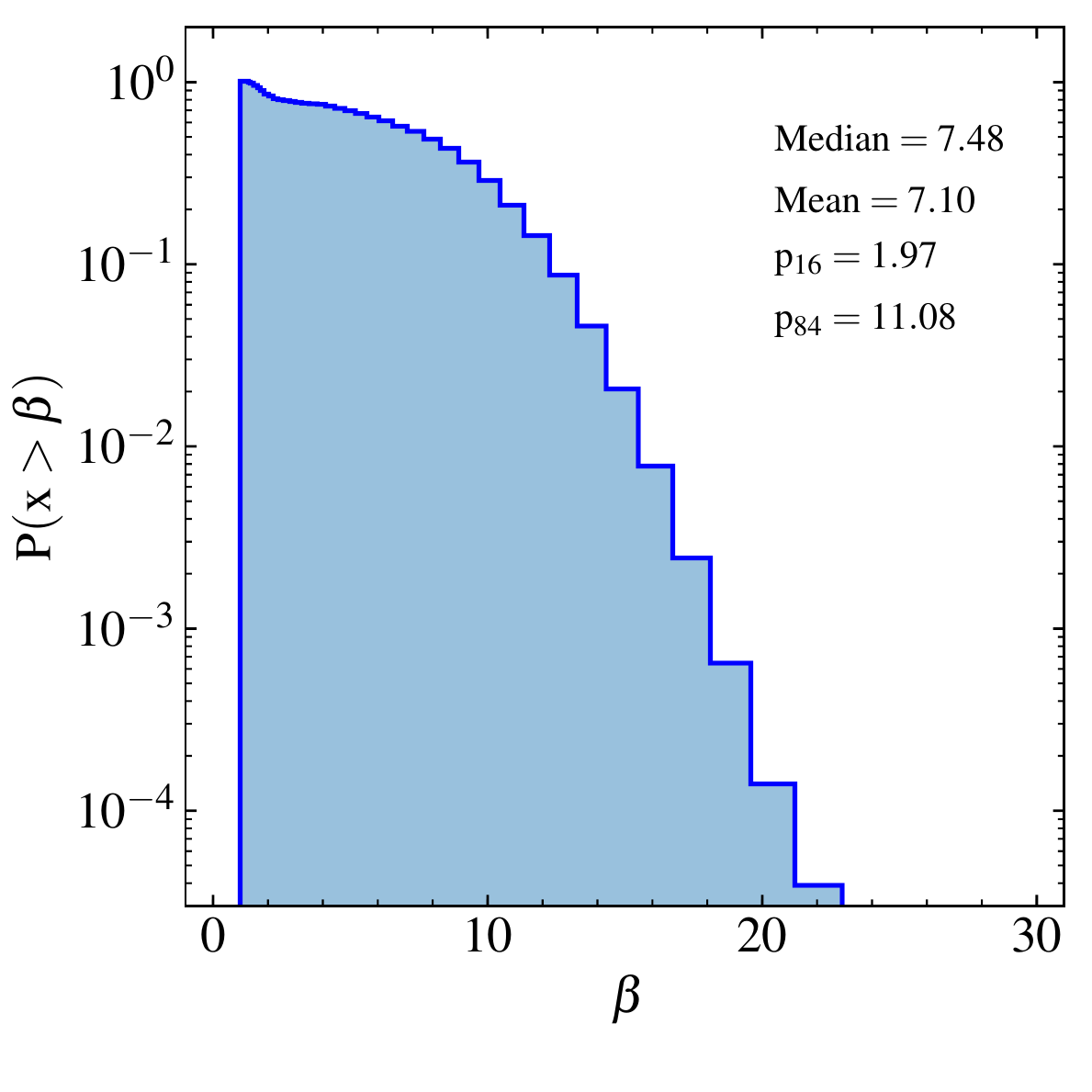}
\includegraphics[width=0.48\textwidth]{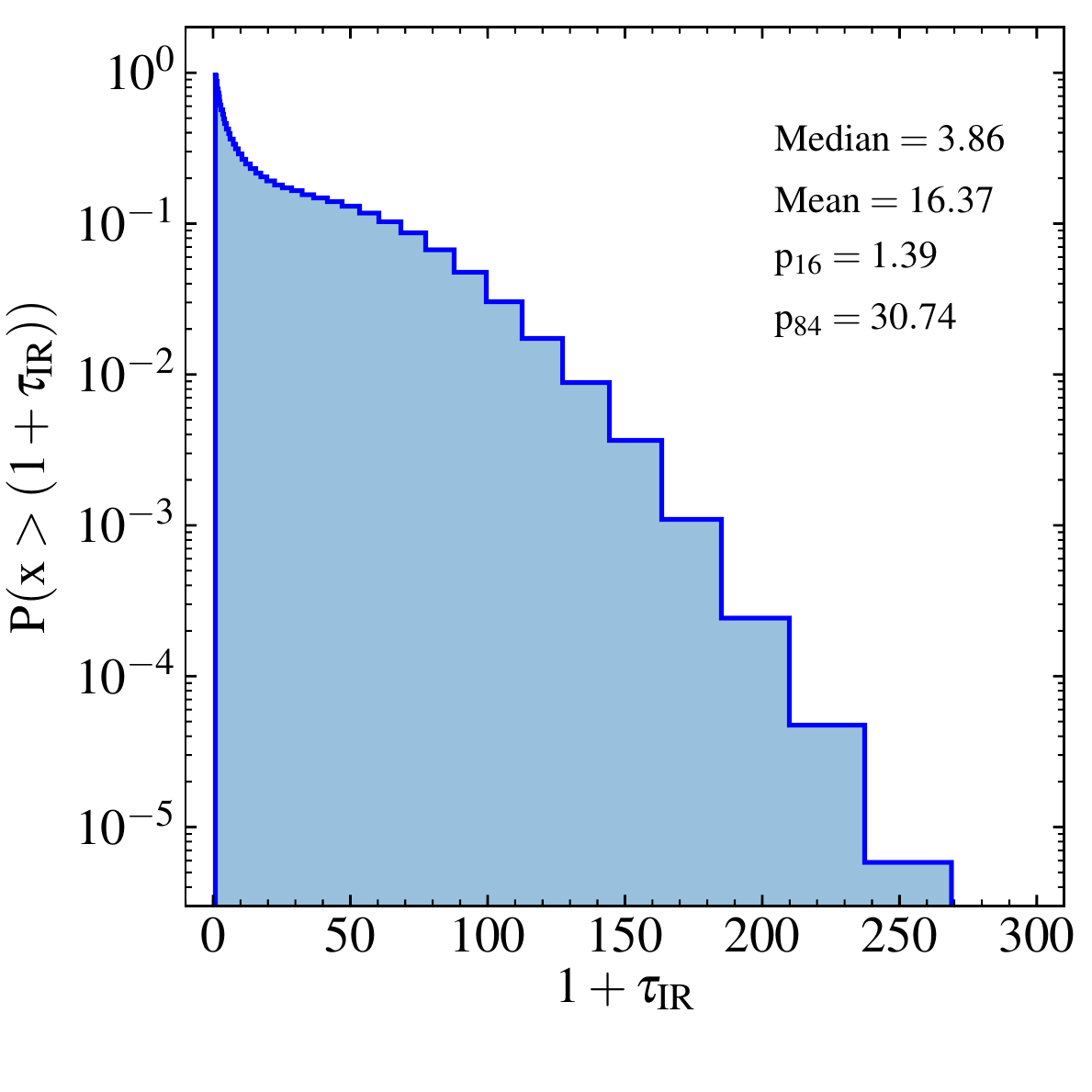}
\caption{Cumulative distribution of the boost factor for SN momentum (i.e. the ratio between the momentum actually imparted to the gas after the Sedov-Taylor phase and the initial momentum associated to the supernova event; top) and radiation momentum (that is the the ratio between the momentum actually imparted to the gas accounting for multiple infrared scattering the momentum associated to the stellar radiation field; bottom) for the high resolution Milky Way simulation. Both boost factors are computed as the ratio of the two cumulative momenta over the lifetime of the stellar particle. Each panel also indicates the median, mean and the 16-th and 84-th percentile of the resulting distributions. Supernovae usually show a mean increase of their initial momentum up to about a factor of $\sim 7.1$ after the Sedov-Taylor phase. The average boost factor for radiation is about $16.37$ and the distribution is broader, due to the $\tau_{\rm IR}$ strong dependence on the environmental conditions of the star's birthplace. 
}
\label{fig:boostfac}
\end{figure}

The coupling and propagation of feedback in the ISM can cause the effective momentum achieved in the gas to exceed the one originally injected, a phenomenon referred to as ``boost" factor. Figure~\ref{fig:boostfac} shows the cumulative distribution of these boost factors for SN (top) and radiation (bottom) momentum in the high resolution Milky Way simulation. As in Figure \ref{fig:pratios}, this is computed cumulatively with respect to the total momenta injected by each stellar particle during their entire life-time. Logarithmically-spaced bins are used to compute the histograms for both quantities.

In the SN case, the boost factor $\beta$ indicates the ratio between the momentum imparted to the gas after the Sedov-Taylor phase and the value injected at the explosion. We find on average a $\sim 7.1$ momentum enhancement for SNe. The dispersion around the mean value is $\simeq 4.6$, implying that larger 
values of the boost factor are possible, but that the probability of very large values is rather small -- only approximately less than $1$ per cent of stars have $\beta \gsim 15$ -- and that, therefore, $\beta$ has only a very weak dependence on the environmental conditions of the gas, i.e. density and metallicity (although metallicity is constant and equal to $Z_\odot$ in our simulations), surrounding the star particle. We would like to note that these values for the momentum boost factor are in line with previous high-resolution simulation studies \citep{Cioffi1988,Kim2015,Martizzi2015,Thornton1998,Walch2015}.

On the other hand, for the radiative feedback case, the boost factor accounts for multiple infrared scattering, which increases the amount of momentum transferred to the gas via radiation pressure by $(1 + \tau_{\rm IR})$. This quantity is proportional to the gas optical depth to infrared radiation, which is in turn depending quite strongly on the physical conditions (and in particular the column density) of the gas. Therefore, contrary to the SN case, the momentum boost is more affected and gets significantly larger in denser gas. On average, the momentum boost in our simulations is $\simeq 16.37$, with a median of $\simeq 3.86$, which are reasonable values based on previous work \citep{Agertz2013}. Compared to the SN case there is a larger scatter ($\simeq 16.14$) because of the (stronger) environmental dependence. 
For the environments analyzed here, namely a Milky-Way like disc at $z=0$, values for the $(1 + \tau_{\rm IR})$ boost factor can reach $> 100$ in very rare cases, but no more than $\sim 5$ per cent of the stars have $(1 + \tau_{\rm IR}) \gsim 100$. 

Based on this analysis, we conclude that all feedback channels included here play a role in regulating star formation, with boost factors of about $7.1$ and $16.37$ that seem to be common for SN and radiation pressure, respectively. However, the momentum input and generation of galactic-scale fountain flows for the scale of Milky Way galaxies analyzed here is largely dominated by the SNe channel only.  
We acknowledge, however, that according to the gas physical conditions, which may vary in galaxies of different mass and redshifts, the relative contribution to the 
global feedback energy and momentum input among these channels may vary, a topic that 
we plan to investigate in future work.

\section{Comparison to previous work}\label{sec:discussion}
The \namemodel\ ISM and feedback model is not the first attempt towards a more explicit description of the ISM physics in simulations of galaxy formation. Several of the processes included here (for example the parameterization for molecular cooling, heating process in the ISM and stellar winds) are in common with other models previously introduced in the literature. However, our adopted numerical implementation is different in several aspects. We will first focus on a comparison with the cases of the FIRE-2 \citep{Fire2} and the \citet{Agertz2013} ISM treatments, being among the closest match to our newly developed model. The main differences can be summarized as follows:

\begin{itemize}
    \item hydrodynamic solver: \arepo\ adopts a moving-mesh finite volume scheme versus mesh-less finite mass employed by {\sc gizmo} (FIRE-2) and the AMR technique in {\sc ramses} \citep{Agertz2013}. This makes a fundamental difference in the way feedback energy and momentum are distributed, especially for updating the fluid conserved variables in our model;
    \item feedback energy and momentum: one of the differing aspects of our model is the explicit justification of a maximum coupling radius for the SN feedback energy and momentum (see Sec.~2.3.4), which comes through superbubble evolution results. Instead, in \citep{Fire2} the maximum coupling radius is treated as a fixed parameter and set to $\simeq 1-2\,\kpc$. The distribution of feedback among the neighboring cells can also differ \rev{because of a different choice in the computations of weights or in the number of neighbours. For instance, in the computation of the weights $\tilde{w_i}$ (see equation~\ref{eq:weight2}), we do not take into account a vector/tensor correction term \citep[see e.g.][]{Hopkins2018, Smith2018} that would ensure linear momentum conservation to machine precision. This choice was made for reasons of computational efficiency/simplicity and it would be interesting to compare the outcome of the two approaches. Instead}, in \citet{Agertz2013} energy and momentum are spread over the 26 neighbours surrounding each stellar particle (basically the closest neighbours of the cell containing the stellar particle, which, however, is excluded from the feedback loop). In our model we choose an SPH-like kernel with 64 cells, but the weights may be adjusted based on the maximum coupling radius, as discussed in Sec.~2. We also note that in order to have an efficient feedback loop regulating star formation, \citet{Agertz2013} may also adopt a non-thermal energy reservoir approach \citep[based on the turbulent feedback model by][]{Teyssier2013} that in practice is comparable to a delayed cooling set-up. We do not include such energy source here;
    \item radiation feedback: our model adopts a stochastic photoionization scheme based on the calculation of the total number of ionizing photons that are then distributed to the neighbouring cells surrounding a star particle based on a solid angle weighting scheme (see equations~\ref{eq:weight1} and \ref{eq:weight2}). The probability of ionization is then computed as the ratio between the number of ionizing photons and that of the recombinations occurring within the cell. This is different from the FIRE-2 implementation in which cells are sorted by distance and become photoionized in such order until the total available numbers of ionizing photons is accounted for. Moreover, we do not consider the long-range radiation component in our model and we also use a prescribed mass-to-light ratio and a fixed time over which radiative feedback occurs. This is a different choice with respect to FIRE-2 in which these processes are included as the output of {\sc starburst99} calculations. \rev{Note that we follow photoionization and radiation pressure effectively as a single-band calculation and do not track separately the 5-band approximation (photo-ionizing, photo-electric, UV, optical, IR) as done in FIRE-2.} On the other hand, the model in \citet{Agertz2013} does not explicitly take into account the impact of photoionization, but includes radiation pressure. Their boost due to multiple scattering of the infrared photons is tracked only for the lifetime of the birth clouds of the stars ($3$ Myr, instead of the $5$ Myrs considered here based on the age of the OB stars) and the infrared opacity in \citet{Agertz2013} is half of our value ($\kappa_{\rm IR} = 5(Z/Z_\odot) {\rm cm}^2\, {\rm g}^{-1}$)\footnote{A multiplicative pre-factor of 2 is used in \citet{Agertz2013} for both direct and infrared momentum injection to account for grid smearing effects; see their equation~(5).}.
    \rev{\item Photo-electric heating: \namemodel\ does not include an explicit treatment of this processes, whereas FIRE-2 does. Comparison to radiative transfer calculations have shown that the inclusion of photo-electric heating effects does not significantly impact the dynamics and global properties of galaxies \citep{Hopkins2018_RT}, however, it may play a relevant role in the properties of the ISM \citep[e.g. ][]{Forbes2016}. We defer to future work a detail study of this in \namemodel.} 
    \item AGB winds: while we adopt the FIRE-2 parameterization for the wind speed, the mass loss due to the AGB phase is directly taken from our stellar evolutionary model (see below);
    \item stellar evolution: besides differing on the adopted metal production yields, the IMF adopted in \namemodel\ and in \citet{Agertz2013} follows a \citet{Chabrier2001} distribution\footnote{However, the metal enrichment outcome of the stellar evolution can differ, as only oxygen and iron are taken as proxy for gas metal enrichment in \citep{Agertz2013}.}, whereas FIRE-2 uses a \cite{Kroupa2001} IMF instead.  There are also subtle differences in the mass ranges for type II supernova progenitors. For instance, in our model we take $8\,{\rm M}_\odot$ as lower limit, whereas $6\,{\rm M}_\odot$ is used in FIRE-2. This has a direct influence on the number of SNII events per unit stellar mass formed, which is nearly twice as high in FIRE-2 compared to our model. Furthermore, this also translates in an average mass per SN event of \rev{$\sim 16\, {\rm M}_\odot$ in \namemodel\ } and $\sim 10 \, {\rm M}_\odot$ for FIRE-2. Finally, the sampling of type II supernova mass return is also treated differently. In particular, in FIRE-2 each supernova returns the same ejecta mass, which is equal to the global IMF average of the mass return of all supernova events. In our model, each supernova has a different ejecta mass depending on the age of the stellar particle. Indeed, the ejecta mass is computed as an IMF average over the local star particle time step only. Basically, this translates to a progressively lower ejecta mass as a function of the star particle age;
    \item star formation criteria: while for our model and in FIRE-2 this is a combination of local gas density and virial parameter conditions (see Sec. 2.2 and equation~\ref{eq:alpha}), \citet{Agertz2013} use a density threshold criterion with metallicity dependence to discriminate between star-forming and non-starforming cells. This follows the form $n_* \simeq 25 (Z/Z_\odot)^{-1}\, {\rm cm}^{-3}$ and represents the transition between the neutral and molecular dominated phase of the interstellar gas according to \citet{Gnedin2011}. Morevoer, only gas with $T < 10^{4}\,{\rm K}$ is eligible for star formation in their model.
\end{itemize}

We hasten to add that other efforts have been made to attempt to produce a resolved ISM medium in  \arepo\ \citep[see][]{Smith2018}. However, there are several major differences with respect to our new model. Most importantly, radiative feedback and OB/AGB winds are not taken into account in the \citet{Smith2018} treatment. Other differences include the scheme adopted to distribute SN feedback energy and momentum. \cite{Smith2018} identify the gas cell containing the star particle (which they denote as host cell) and they assign $5$ per cent of the energy budget to it. The remaining part is then given to the nearest neighbours of the host cell by employing a modified version of the solid angle weighting scheme that we present here. \citet{Smith2018} also adopts a \citet{Kroupa2001} IMF, in contrast with the \cite{Chabrier2001} in \namemodel.  Furthermore, each supernova event (sampled from a Poisson distribution) releases $10\,{\rm M}_\odot$ of which $2\,{\rm M}_\odot$ are metals or, in other words, a constant metallicity of $Z = 0.02$ is assumed for the supernova ejecta. No other metal enrichment channel is considered. Lastly, the star formation criterion demands only a gas density threshold, whereas our model imposes an extra check on the virialization state of the gas, and additionally, an explicit dependence on the molecular fraction (see equation~\ref{eq:H2}). 

Although the impact of these specific differences in the outcome of each model is unclear, it is a useful reminder of the many unconstrained numerical choices that are necessarily made in each attempt to follow the physics of the ISM at sub-kpc resolution level. A detailed evaluation of our results and their dependence on variations of some of these choices will be presented in future work. 

Finally, we note that important physical processes in the ISM are not taken into account in our model or are included with rather crude approximations. These processes include, for example, cosmic rays transport and physics (which are important as a source of pressure support and as an additional feedback channel to launch galactic-scale outflows), magnetic fields (which might provide additional pressure support to the gas and determine the transport properties of relativistic particles and thermal energy), dust production, destruction and evolution (important for both the molecular chemistry of the ISM and the reprocessing of radiation fields), thermal conduction (which operates at the transition layers between the supernova-heated hot gas and the cold phase). Also, while we include some form of feedback due to radiation, its transport and interaction with the gas are treated in a rather simplistic way. A more detailed implementation is available in \arepo\ for the majority of these physical processes \citep{Pakmor2013, Kannan2016b, Pfrommer2017, Kannan2018a, McKinnon2018}; their inclusion in our model is a line of research that we intend to pursue. For example, efforts are currently under way to combine the \namemodel\ model with a realistic treatment of radiation transport and dust physics (Kannan et al., in prep.).

\section{Summary and conclusions}\label{sec:conclusion}

Modelling the formation of galaxies and their evolution is a challenging task due to the large dynamic range and the variety of physical processes that concur in shaping galaxies. Numerical simulations are the ideal tool to tackle this non-linear and complex problem. Cosmological simulations have recently achieved important successes in modelling of Milky Way-like galaxies \citep{Fire, Grand2017, Wang2015} as well as in reproducing realistic populations of galaxies within the cosmological context~\citep{Vogelsberger2014a, Schaye2015, Springel2018}. However, these simulations have to rely on effective physical prescriptions, with different degrees of sophistication, to include important physical processes below their resolution scale -- the so-called sub-grid physics. For the galaxy formation problem, sub-grid models are particularly important for the treatment  of the interstellar medium (ISM) and stellar feedback processes. In this paper we haved presented the \namemodel\ model, a novel numerical implementation for the treatment of the ISM gas and stellar feedback in the moving mesh code \arepo. 

This new model does not employ an effective equation of state for the ISM, but instead aims at resolving the different gas phases in the ISM and self-consistently implements a local stellar feedback scheme that generates gaseous outflows. The energy and momentum feedback inputs, and the associated generation of outflows, are tied to the location and characteristics of newly created stars rather than by imposing a prescribed outflow mass loading set based on the more global properties of the simulated galaxy. Specifically, our model includes the following physical processes:  low-temperature molecular cooling; cosmic ray and photo-electric heating; star formation based on a high density threshold, virial parameter and molecular gas criteria; mechanical feedback from supernovae taking into account the momentum boost due to $P{\rm d}V$ work during the (unresolved) Sedov-Taylor expansion phase; radiative feedback from young, massive stars both in the form of photoionization and radiation pressure accounting for multiple infrared scattering; and the energy and momentum injection from OB and AGB stellar winds.

We have explored this novel implementation with simulations of an isolated non-cosmological Milky Way-like disc. Our model yields a star formation rate of $\sim 3\,{\rm M_\odot\, yr^{-1}}$, consistent with observational data and reproduces the observed Kennicutt-Schmidt relation. The regulation of star formation increases the gas depletion time-scale from a few hundred $\Myr$ in the no feedback case to a few $\Gyr$ once feedback processes are included, which is in line with observations. These results are largely independent of resolution. As a consequence of feedback, gaseous outflows across the star-forming disc are generated with mass loading factors of order unity. The gas outflow rates are quite bursty on a time-scale of $\simeq 100\,\Myr$, which is reflected in the behaviour of the mass loading factor. Contrary to the outflow rate, however, the mass loading factor remains approximately constant. The majority of the ejected gas cycles back onto the disc, as the similarity between gas outflow and inflow rates demonstrates. 

Most importantly, the model is also able to produce a distinct multiphase ISM in the galactic disc with three gas phases coexisting. There is still a residual dependence in the amount of the hot gas present in the simulation and the numerical resolution. In particular, higher resolution corresponds to larger amounts of hot gas. This is likely due to the fact that the cooling of the gas is overestimated with coarser numerical resolution. The creation of the multiphase medium in the simulated galaxies is due to the interplay between gas cooling, star formation and the different feedback channels implemented in the model. Although all feedback processes considered here are important towards regulation of star formation, in the Milky Way-type galaxy simulated in this work, the dominant channel is represented by supernova feedback. Indeed, compared to radiative feedback and stellar winds, supernovae inject into the ISM about a factor of $10$ times more momentum. We note, however, that these ratios might change in more dense environments or at high redshifts.  

Our new model is intended to pave the way towards the next generation of cosmological large  volumes simulations together with associated zoom-ins of individual objects. While the future of galaxy
formation simulations will require the direct treatment of additional physics such as radiative transfer, effects of cosmic rays, etc., the computational power needed to achieve such goals on large cosmological scales are beyond reach in the near future. 
 Our model aims to bridge the gap between detailed physical modelling of the ISM requiring pc-resolution scale \citep[e.g. ][]{Rosdahl2015,Simpson2016, Raskutti2016, Kannan2018b, Krumholz2018, Emerick2018,Haid2018} and the current cosmological volume simulations. The application of our model to such volume simulations will increase the predictive power of these simulations, in particular for galactic structure, outflows and circum-galactic gas predictions. Such theoretical forecasts over a large number of galaxies will be directly contrasted with observations of galaxy population offering further constraints to galaxy formation models within the $\Lambda$CDM scenario.

\section*{Acknowledgments}
The authors wish to thank Rahul Kannan, Eve Ostriker, Andrey Kravtsov, Lars Hernquist and Philip Hopkins for stimulating discussions about this work. FM acknowledges support through the Program ``Rita Levi Montalcini'' of the Italian MIUR. LVS is thankful for financial support from the Hellman Foundation as well as NSF and NASA grants, AST-1817233, HST-AR-14552 and HST-AR-14553. MV acknowledges support through an MIT RSC award, a Kavli Research Investment Fund, NASA ATP grant NNX17AG29G, and NSF grants AST-1814053 and AST-1814259.
The simulations were performed on the joint Harvard-MIT computing cluster, supported by the Faculty of Arts and Sciences of Harvard University and the MIT Kavli Institute for Astrophysics and Space Research, and the HiPerGator computing cluster at the University of Florida. Preparatory work was made possible by the award of the XSEDE grants AST-140082, AST-140078, AST-150007 and AST-160006. All the figures in this work were created with the {\sc matplotlib} graphics environment \citep{Matplotlib}.

\bibliographystyle{mnras}
\bibliography{paper}

\begin{thebibliography}{}
\makeatletter
\relax
\def\mn@urlcharsother{\let\do\@makeother \do\$\do\&\do\#\do\^\do\_\do\%\do\~}
\def\mn@doi{\begingroup\mn@urlcharsother \@ifnextchar [ {\mn@doi@}
  {\mn@doi@[]}}
\def\mn@doi@[#1]#2{\def\@tempa{#1}\ifx\@tempa\@empty \href
  {http://dx.doi.org/#2} {doi:#2}\else \href {http://dx.doi.org/#2} {#1}\fi
  \endgroup}
\def\mn@eprint#1#2{\mn@eprint@#1:#2::\@nil}
\def\mn@eprint@arXiv#1{\href {http://arxiv.org/abs/#1} {{\tt arXiv:#1}}}
\def\mn@eprint@dblp#1{\href {http://dblp.uni-trier.de/rec/bibtex/#1.xml}
  {dblp:#1}}
\def\mn@eprint@#1:#2:#3:#4\@nil{\def\@tempa {#1}\def\@tempb {#2}\def\@tempc
  {#3}\ifx \@tempc \@empty \let \@tempc \@tempb \let \@tempb \@tempa \fi \ifx
  \@tempb \@empty \def\@tempb {arXiv}\fi \@ifundefined
  {mn@eprint@\@tempb}{\@tempb:\@tempc}{\expandafter \expandafter \csname
  mn@eprint@\@tempb\endcsname \expandafter{\@tempc}}}

\bibitem[\protect\citeauthoryear{{Agertz}, {Teyssier}  \& {Moore}}{{Agertz}
  et~al.}{2011}]{Agertz2011}
{Agertz} O.,  {Teyssier} R.,   {Moore} B.,  2011, \mn@doi [\mnras]
  {10.1111/j.1365-2966.2010.17530.x}, \href
  {http://adsabs.harvard.edu/abs/2011MNRAS.410.1391A} {410, 1391}

\bibitem[\protect\citeauthoryear{{Agertz}, {Kravtsov}, {Leitner}  \&
  {Gnedin}}{{Agertz} et~al.}{2013}]{Agertz2013}
{Agertz} O.,  {Kravtsov} A.~V.,  {Leitner} S.~N.,   {Gnedin} N.~Y.,  2013,
  \mn@doi [\apj] {10.1088/0004-637X/770/1/25}, \href
  {http://adsabs.harvard.edu/abs/2013ApJ...770...25A} {770, 25}

\bibitem[\protect\citeauthoryear{{Angl{\'e}s-Alc{\'a}zar},
  {Faucher-Gigu{\`e}re}, {Kere{\v s}}, {Hopkins}, {Quataert}  \&
  {Murray}}{{Angl{\'e}s-Alc{\'a}zar} et~al.}{2017}]{AngelesAlcazar2017}
{Angl{\'e}s-Alc{\'a}zar} D.,  {Faucher-Gigu{\`e}re} C.-A.,  {Kere{\v s}} D.,
  {Hopkins} P.~F.,  {Quataert} E.,   {Murray} N.,  2017, \mn@doi [\mnras]
  {10.1093/mnras/stx1517}, \href
  {http://adsabs.harvard.edu/abs/2017MNRAS.470.4698A} {470, 4698}

\bibitem[\protect\citeauthoryear{{Asplund}, {Grevesse}, {Sauval}  \&
  {Scott}}{{Asplund} et~al.}{2009}]{Asplund2009}
{Asplund} M.,  {Grevesse} N.,  {Sauval} A.~J.,   {Scott} P.,  2009, \mn@doi
  [\araa] {10.1146/annurev.astro.46.060407.145222}, \href
  {http://adsabs.harvard.edu/abs/2009ARA%26A..47..481A} {47, 481}

\bibitem[\protect\citeauthoryear{{Aumer}, {White}, {Naab}  \&
  {Scannapieco}}{{Aumer} et~al.}{2013}]{Aumer2013b}
{Aumer} M.,  {White} S.~D.~M.,  {Naab} T.,   {Scannapieco} C.,  2013, \mn@doi
  [\mnras] {10.1093/mnras/stt1230}, \href
  {http://adsabs.harvard.edu/abs/2013MNRAS.434.3142A} {434, 3142}

\bibitem[\protect\citeauthoryear{{Barnes} \& {Hut}}{{Barnes} \&
  {Hut}}{1986}]{Barnes1986}
{Barnes} J.,  {Hut} P.,  1986, \mn@doi [\nat] {10.1038/324446a0}, \href
  {https://ui.adsabs.harvard.edu/#abs/1986Natur.324..446B} {324, 446}

\bibitem[\protect\citeauthoryear{{Behroozi}, {Wechsler}, {Hearin}  \&
  {Conroy}}{{Behroozi} et~al.}{2019}]{Behroozi2018}
{Behroozi} P.,  {Wechsler} R.~H.,  {Hearin} A.~P.,   {Conroy} C.,  2019,
  \mn@doi [\mnras] {10.1093/mnras/stz1182}, \href
  {https://ui.adsabs.harvard.edu/abs/2019MNRAS.tmp.1134B} {p.~1134}

\bibitem[\protect\citeauthoryear{{Ben{\'{\i}}tez-Llambay}, {Navarro}, {Frenk}
  \& {Ludlow}}{{Ben{\'{\i}}tez-Llambay} et~al.}{2018}]{Benitez-Llambay2018}
{Ben{\'{\i}}tez-Llambay} A.,  {Navarro} J.~F.,  {Frenk} C.~S.,   {Ludlow}
  A.~D.,  2018, \mn@doi [\mnras] {10.1093/mnras/stx2420}, \href
  {http://adsabs.harvard.edu/abs/2018MNRAS.473.1019B} {473, 1019}

\bibitem[\protect\citeauthoryear{{Benson}, {Lacey}, {Baugh}, {Cole}  \&
  {Frenk}}{{Benson} et~al.}{2002}]{Benson2002}
{Benson} A.~J.,  {Lacey} C.~G.,  {Baugh} C.~M.,  {Cole} S.,   {Frenk} C.~S.,
  2002, \mn@doi [\mnras] {10.1046/j.1365-8711.2002.05387.x}, \href
  {http://adsabs.harvard.edu/abs/2002MNRAS.333..156B} {333, 156}

\bibitem[\protect\citeauthoryear{{Bigiel}, {Leroy}, {Walter}, {Brinks}, {de
  Blok}, {Madore}  \& {Thornley}}{{Bigiel} et~al.}{2008}]{Bigiel2008}
{Bigiel} F.,  {Leroy} A.,  {Walter} F.,  {Brinks} E.,  {de Blok} W.~J.~G.,
  {Madore} B.,   {Thornley} M.~D.,  2008, \mn@doi [\aj]
  {10.1088/0004-6256/136/6/2846}, \href
  {http://adsabs.harvard.edu/abs/2008AJ....136.2846B} {136, 2846}

\bibitem[\protect\citeauthoryear{{Bigiel} et~al.,}{{Bigiel}
  et~al.}{2011}]{Bigiel2011}
{Bigiel} F.,  et~al., 2011, \mn@doi [\apjl] {10.1088/2041-8205/730/2/L13},
  \href {http://adsabs.harvard.edu/abs/2011ApJ...730L..13B} {730, L13}

\bibitem[\protect\citeauthoryear{{Bregman}}{{Bregman}}{1980}]{Bregman1980}
{Bregman} J.~N.,  1980, \mn@doi [\apj] {10.1086/157776}, \href
  {http://adsabs.harvard.edu/abs/1980ApJ...236..577B} {236, 577}

\bibitem[\protect\citeauthoryear{{Bregman}}{{Bregman}}{2007}]{Bregman2007}
{Bregman} J.~N.,  2007, \mn@doi [Annual Review of Astronomy and Astrophysics]
  {10.1146/annurev.astro.45.051806.110619}, \href
  {https://ui.adsabs.harvard.edu/#abs/2007ARA&A..45..221B} {45, 221}

\bibitem[\protect\citeauthoryear{{Buck}, {Dutton}  \& {Macci{\`o}}}{{Buck}
  et~al.}{2019}]{Buck2019}
{Buck} T.,  {Dutton} A.~A.,   {Macci{\`o}} A.~V.,  2019, \mn@doi [\mnras]
  {10.1093/mnras/stz969}, \href
  {https://ui.adsabs.harvard.edu/abs/2019MNRAS.486.1481B} {486, 1481}

\bibitem[\protect\citeauthoryear{{Chabrier}}{{Chabrier}}{2001}]{Chabrier2001}
{Chabrier} G.,  2001, \mn@doi [\apj] {10.1086/321401}, \href
  {http://adsabs.harvard.edu/abs/2001ApJ...554.1274C} {554, 1274}

\bibitem[\protect\citeauthoryear{{Christensen}, {Dav{\'e}}, {Governato},
  {Pontzen}, {Brooks}, {Munshi}, {Quinn}  \& {Wadsley}}{{Christensen}
  et~al.}{2016}]{Christensen2016}
{Christensen} C.~R.,  {Dav{\'e}} R.,  {Governato} F.,  {Pontzen} A.,  {Brooks}
  A.,  {Munshi} F.,  {Quinn} T.,   {Wadsley} J.,  2016, \mn@doi [\apj]
  {10.3847/0004-637X/824/1/57}, \href
  {http://adsabs.harvard.edu/abs/2016ApJ...824...57C} {824, 57}

\bibitem[\protect\citeauthoryear{{Cioffi}, {McKee}  \& {Bertschinger}}{{Cioffi}
  et~al.}{1988}]{Cioffi1988}
{Cioffi} D.~F.,  {McKee} C.~F.,   {Bertschinger} E.,  1988, \mn@doi [\apj]
  {10.1086/166834}, \href {http://adsabs.harvard.edu/abs/1988ApJ...334..252C}
  {334, 252}

\bibitem[\protect\citeauthoryear{{Conroy} \& {Wechsler}}{{Conroy} \&
  {Wechsler}}{2009}]{Conroy2009}
{Conroy} C.,  {Wechsler} R.~H.,  2009, \mn@doi [\apj]
  {10.1088/0004-637X/696/1/620}, \href
  {http://adsabs.harvard.edu/abs/2009ApJ...696..620C} {696, 620}

\bibitem[\protect\citeauthoryear{{Dale}}{{Dale}}{2017}]{Dale2017}
{Dale} J.~E.,  2017, \mn@doi [\mnras] {10.1093/mnras/stx028}, \href
  {https://ui.adsabs.harvard.edu/#abs/2017MNRAS.467.1067D} {467, 1067}

\bibitem[\protect\citeauthoryear{{Dalla Vecchia} \& {Schaye}}{{Dalla Vecchia}
  \& {Schaye}}{2012}]{DallaVecchia2012}
{Dalla Vecchia} C.,  {Schaye} J.,  2012, \mn@doi [\mnras]
  {10.1111/j.1365-2966.2012.21704.x}, \href
  {https://ui.adsabs.harvard.edu/#abs/2012MNRAS.426..140D} {426, 140}

\bibitem[\protect\citeauthoryear{{Dav{\'e}}, {Rafieferantsoa}, {Thompson}  \&
  {Hopkins}}{{Dav{\'e}} et~al.}{2017}]{Dave2017}
{Dav{\'e}} R.,  {Rafieferantsoa} M.~H.,  {Thompson} R.~J.,   {Hopkins} P.~F.,
  2017, \mn@doi [\mnras] {10.1093/mnras/stx108}, \href
  {http://adsabs.harvard.edu/abs/2017MNRAS.467..115D} {467, 115}

\bibitem[\protect\citeauthoryear{{Emerick}, {Bryan}  \& {Mac Low}}{{Emerick}
  et~al.}{2018}]{Emerick2018}
{Emerick} A.,  {Bryan} G.~L.,   {Mac Low} M.-M.,  2018, \mn@doi [\apjl]
  {10.3847/2041-8213/aae315}, \href
  {http://adsabs.harvard.edu/abs/2018ApJ...865L..22E} {865, L22}

\bibitem[\protect\citeauthoryear{{Faucher-Gigu{\`e}re}, {Lidz}, {Zaldarriaga}
  \& {Hernquist}}{{Faucher-Gigu{\`e}re} et~al.}{2009}]{FaucherGiguere2009}
{Faucher-Gigu{\`e}re} C.-A.,  {Lidz} A.,  {Zaldarriaga} M.,   {Hernquist} L.,
  2009, \mn@doi [\apj] {10.1088/0004-637X/703/2/1416}, \href
  {http://adsabs.harvard.edu/abs/2009ApJ...703.1416F} {703, 1416}

\bibitem[\protect\citeauthoryear{{Ferland}, {Korista}, {Verner}, {Ferguson},
  {Kingdon}  \& {Verner}}{{Ferland} et~al.}{1998}]{Ferland1998}
{Ferland} G.~J.,  {Korista} K.~T.,  {Verner} D.~A.,  {Ferguson} J.~W.,
  {Kingdon} J.~B.,   {Verner} E.~M.,  1998, \mn@doi [\pasp] {10.1086/316190},
  \href {http://adsabs.harvard.edu/abs/1998PASP..110..761F} {110, 761}

\bibitem[\protect\citeauthoryear{{Ferri{\`e}re}}{{Ferri{\`e}re}}{2001}]{Ferriere2001}
{Ferri{\`e}re} K.~M.,  2001, \mn@doi [Reviews of Modern Physics]
  {10.1103/RevModPhys.73.1031}, \href
  {http://adsabs.harvard.edu/abs/2001RvMP...73.1031F} {73, 1031}

\bibitem[\protect\citeauthoryear{{Field}, {Goldsmith}  \& {Habing}}{{Field}
  et~al.}{1969}]{Field1969}
{Field} G.~B.,  {Goldsmith} D.~W.,   {Habing} H.~J.,  1969, \mn@doi [\apjl]
  {10.1086/180324}, \href {http://adsabs.harvard.edu/abs/1969ApJ...155L.149F}
  {155, L149}

\bibitem[\protect\citeauthoryear{{Fitts} et~al.,}{{Fitts}
  et~al.}{2017}]{Fitts2017}
{Fitts} A.,  et~al., 2017, \mn@doi [\mnras] {10.1093/mnras/stx1757}, \href
  {http://adsabs.harvard.edu/abs/2017MNRAS.471.3547F} {471, 3547}

\bibitem[\protect\citeauthoryear{{Forbes}, {Krumholz}, {Goldbaum}  \&
  {Dekel}}{{Forbes} et~al.}{2016}]{Forbes2016}
{Forbes} J.~C.,  {Krumholz} M.~R.,  {Goldbaum} N.~J.,   {Dekel} A.,  2016,
  \mn@doi [\nat] {10.1038/nature18292}, \href
  {https://ui.adsabs.harvard.edu/abs/2016Natur.535..523F} {535, 523}

\bibitem[\protect\citeauthoryear{{Frank} et~al.,}{{Frank}
  et~al.}{2014}]{Frank2014}
{Frank} A.,  et~al., 2014, \mn@doi [Protostars and Planets VI]
  {10.2458/azu_uapress_9780816531240-ch020}, \href
  {http://adsabs.harvard.edu/abs/2014prpl.conf..451F} {pp 451--474}

\bibitem[\protect\citeauthoryear{{Gatto} et~al.,}{{Gatto}
  et~al.}{2015}]{Gatto2015}
{Gatto} A.,  et~al., 2015, \mn@doi [\mnras] {10.1093/mnras/stv324}, \href
  {http://adsabs.harvard.edu/abs/2015MNRAS.449.1057G} {449, 1057}

\bibitem[\protect\citeauthoryear{{Gnedin} \& {Kravtsov}}{{Gnedin} \&
  {Kravtsov}}{2011}]{Gnedin2011}
{Gnedin} N.~Y.,  {Kravtsov} A.~V.,  2011, \mn@doi [\apj]
  {10.1088/0004-637X/728/2/88}, \href
  {http://adsabs.harvard.edu/abs/2011ApJ...728...88G} {728, 88}

\bibitem[\protect\citeauthoryear{{Grand} et~al.,}{{Grand}
  et~al.}{2017}]{Grand2017}
{Grand} R.~J.~J.,  et~al., 2017, \mn@doi [\mnras] {10.1093/mnras/stx071}, \href
  {http://adsabs.harvard.edu/abs/2017MNRAS.467..179G} {467, 179}

\bibitem[\protect\citeauthoryear{{Greggio}}{{Greggio}}{2005}]{Greggio2005}
{Greggio} L.,  2005, \mn@doi [\aap] {10.1051/0004-6361:20052926}, \href
  {https://ui.adsabs.harvard.edu/#abs/2005A&A...441.1055G} {441, 1055}

\bibitem[\protect\citeauthoryear{{Guedes}, {Callegari}, {Madau}  \&
  {Mayer}}{{Guedes} et~al.}{2011}]{Guedes2011}
{Guedes} J.,  {Callegari} S.,  {Madau} P.,   {Mayer} L.,  2011, \mn@doi [\apj]
  {10.1088/0004-637X/742/2/76}, \href
  {http://adsabs.harvard.edu/abs/2011ApJ...742...76G} {742, 76}

\bibitem[\protect\citeauthoryear{{Guo} \& {Oh}}{{Guo} \& {Oh}}{2008}]{Guo2008}
{Guo} F.,  {Oh} S.~P.,  2008, \mn@doi [\mnras]
  {10.1111/j.1365-2966.2007.12692.x}, \href
  {http://adsabs.harvard.edu/abs/2008MNRAS.384..251G} {384, 251}

\bibitem[\protect\citeauthoryear{{Gupta}, {Nath}, {Sharma}  \&
  {Shchekinov}}{{Gupta} et~al.}{2016}]{Gupta2016}
{Gupta} S.,  {Nath} B.~B.,  {Sharma} P.,   {Shchekinov} Y.,  2016, \mn@doi
  [\mnras] {10.1093/mnras/stw1920}, \href
  {https://ui.adsabs.harvard.edu/#abs/2016MNRAS.462.4532G} {462, 4532}

\bibitem[\protect\citeauthoryear{{Haid}, {Walch}, {Seifried}, {W{\"u}nsch},
  {Dinnbier}  \& {Naab}}{{Haid} et~al.}{2018}]{Haid2018}
{Haid} S.,  {Walch} S.,  {Seifried} D.,  {W{\"u}nsch} R.,  {Dinnbier} F.,
  {Naab} T.,  2018, \mn@doi [\mnras] {10.1093/mnras/sty1315}, \href
  {https://ui.adsabs.harvard.edu/#abs/2018MNRAS.478.4799H} {478, 4799}

\bibitem[\protect\citeauthoryear{{Heckman}, {Lehnert}, {Strickland}  \&
  {Armus}}{{Heckman} et~al.}{2000}]{Heckman2000}
{Heckman} T.~M.,  {Lehnert} M.~D.,  {Strickland} D.~K.,   {Armus} L.,  2000,
  \mn@doi [The Astrophysical Journal Supplement Series] {10.1086/313421}, \href
  {https://ui.adsabs.harvard.edu/#abs/2000ApJS..129..493H} {129, 493}

\bibitem[\protect\citeauthoryear{{Hernquist}}{{Hernquist}}{1990}]{Hernquist1990}
{Hernquist} L.,  1990, \mn@doi [\apj] {10.1086/168845}, \href
  {http://adsabs.harvard.edu/abs/1990ApJ...356..359H} {356, 359}

\bibitem[\protect\citeauthoryear{{Hernquist}}{{Hernquist}}{1993}]{Hernquist1993}
{Hernquist} L.,  1993, \mn@doi [\apjs] {10.1086/191784}, \href
  {http://adsabs.harvard.edu/abs/1993ApJS...86..389H} {86, 389}

\bibitem[\protect\citeauthoryear{{Hopkins}, {Quataert}  \& {Murray}}{{Hopkins}
  et~al.}{2011}]{Hopkins2011}
{Hopkins} P.~F.,  {Quataert} E.,   {Murray} N.,  2011, \mn@doi [\mnras]
  {10.1111/j.1365-2966.2011.19306.x}, \href
  {http://adsabs.harvard.edu/abs/2011MNRAS.417..950H} {417, 950}

\bibitem[\protect\citeauthoryear{{Hopkins}, {Quataert}  \& {Murray}}{{Hopkins}
  et~al.}{2012}]{Hopkins2012}
{Hopkins} P.~F.,  {Quataert} E.,   {Murray} N.,  2012, \mn@doi [\mnras]
  {10.1111/j.1365-2966.2012.20578.x}, \href
  {http://adsabs.harvard.edu/abs/2012MNRAS.421.3488H} {421, 3488}

\bibitem[\protect\citeauthoryear{{Hopkins}, {Kere{\v s}}, {O{\~n}orbe},
  {Faucher-Gigu{\`e}re}, {Quataert}, {Murray}  \& {Bullock}}{{Hopkins}
  et~al.}{2014a}]{Fire}
{Hopkins} P.~F.,  {Kere{\v s}} D.,  {O{\~n}orbe} J.,  {Faucher-Gigu{\`e}re}
  C.-A.,  {Quataert} E.,  {Murray} N.,   {Bullock} J.~S.,  2014a, \mn@doi
  [\mnras] {10.1093/mnras/stu1738}, \href
  {http://adsabs.harvard.edu/abs/2014MNRAS.445..581H} {445, 581}

\bibitem[\protect\citeauthoryear{{Hopkins}, {Kere{\v s}}, {O{\~n}orbe},
  {Faucher-Gigu{\`e}re}, {Quataert}, {Murray}  \& {Bullock}}{{Hopkins}
  et~al.}{2014b}]{Hopkins2014}
{Hopkins} P.~F.,  {Kere{\v s}} D.,  {O{\~n}orbe} J.,  {Faucher-Gigu{\`e}re}
  C.-A.,  {Quataert} E.,  {Murray} N.,   {Bullock} J.~S.,  2014b, \mn@doi
  [\mnras] {10.1093/mnras/stu1738}, \href
  {http://adsabs.harvard.edu/abs/2014MNRAS.445..581H} {445, 581}

\bibitem[\protect\citeauthoryear{{Hopkins}, {Grudic}, {Wetzel}, {Keres},
  {Gaucher-Giguere}, {Ma}, {Murray}  \& {Butcher}}{{Hopkins}
  et~al.}{2018a}]{Hopkins2018_RT}
{Hopkins} P.~F.,  {Grudic} M.~Y.,  {Wetzel} A.~R.,  {Keres} D.,
  {Gaucher-Giguere} C.-A.,  {Ma} X.,  {Murray} N.,   {Butcher} N.,  2018a,
  arXiv e-prints, \href {https://ui.adsabs.harvard.edu/abs/2018arXiv181112462H}
  {p. arXiv:1811.12462}

\bibitem[\protect\citeauthoryear{{Hopkins} et~al.,}{{Hopkins}
  et~al.}{2018b}]{Hopkins2018}
{Hopkins} P.~F.,  et~al., 2018b, \mn@doi [\mnras] {10.1093/mnras/sty674}, \href
  {https://ui.adsabs.harvard.edu/#abs/2018MNRAS.477.1578H} {477, 1578}

\bibitem[\protect\citeauthoryear{{Hopkins} et~al.,}{{Hopkins}
  et~al.}{2018c}]{Fire2}
{Hopkins} P.~F.,  et~al., 2018c, \mn@doi [\mnras] {10.1093/mnras/sty1690},
  \href {http://adsabs.harvard.edu/abs/2018MNRAS.480..800H} {480, 800}

\bibitem[\protect\citeauthoryear{{Hunter}}{{Hunter}}{2007}]{Matplotlib}
{Hunter} J.~D.,  2007, \mn@doi [Computing In Science \& Engineering]
  {10.1109/MCSE.2007.55}, 9, 90

\bibitem[\protect\citeauthoryear{{Ikeuchi} \& {Ostriker}}{{Ikeuchi} \&
  {Ostriker}}{1986}]{Ikeuchi1986}
{Ikeuchi} S.,  {Ostriker} J.~P.,  1986, \mn@doi [\apj] {10.1086/163921}, \href
  {http://adsabs.harvard.edu/abs/1986ApJ...301..522I} {301, 522}

\bibitem[\protect\citeauthoryear{{Jacob}, {Pakmor}, {Simpson}, {Springel}  \&
  {Pfrommer}}{{Jacob} et~al.}{2018}]{Jacob2018}
{Jacob} S.,  {Pakmor} R.,  {Simpson} C.~M.,  {Springel} V.,   {Pfrommer} C.,
  2018, \mn@doi [\mnras] {10.1093/mnras/stx3221}, \href
  {https://ui.adsabs.harvard.edu/#abs/2018MNRAS.475..570J} {475, 570}

\bibitem[\protect\citeauthoryear{{Jenkins} \& {Tripp}}{{Jenkins} \&
  {Tripp}}{2001}]{Jenkins2001}
{Jenkins} E.~B.,  {Tripp} T.~M.,  2001, \mn@doi [\apjs] {10.1086/323326}, \href
  {http://adsabs.harvard.edu/abs/2001ApJS..137..297J} {137, 297}

\bibitem[\protect\citeauthoryear{{Kannan}, {Springel}, {Pakmor}, {Marinacci}
  \& {Vogelsberger}}{{Kannan} et~al.}{2016a}]{Kannan2016b}
{Kannan} R.,  {Springel} V.,  {Pakmor} R.,  {Marinacci} F.,   {Vogelsberger}
  M.,  2016a, \mn@doi [\mnras] {10.1093/mnras/stw294}, \href
  {http://adsabs.harvard.edu/abs/2016MNRAS.458..410K} {458, 410}

\bibitem[\protect\citeauthoryear{{Kannan}, {Vogelsberger}, {Stinson},
  {Hennawi}, {Marinacci}, {Springel}  \& {Macci{\`o}}}{{Kannan}
  et~al.}{2016b}]{Kannan2016}
{Kannan} R.,  {Vogelsberger} M.,  {Stinson} G.~S.,  {Hennawi} J.~F.,
  {Marinacci} F.,  {Springel} V.,   {Macci{\`o}} A.~V.,  2016b, \mn@doi
  [\mnras] {10.1093/mnras/stw463}, \href
  {https://ui.adsabs.harvard.edu/#abs/2016MNRAS.458.2516K} {458, 2516}

\bibitem[\protect\citeauthoryear{{Kannan}, {Marinacci}, {Simpson}, {Glover}  \&
  {Hernquist}}{{Kannan} et~al.}{2018}]{Kannan2018b}
{Kannan} R.,  {Marinacci} F.,  {Simpson} C.~M.,  {Glover} S.~C.~O.,
  {Hernquist} L.,  2018, arXiv e-prints, \href
  {http://adsabs.harvard.edu/abs/2018arXiv181201614K} {p. arXiv:1812.01614}

\bibitem[\protect\citeauthoryear{{Kannan}, {Vogelsberger}, {Marinacci},
  {McKinnon}, {Pakmor}  \& {Springel}}{{Kannan} et~al.}{2019}]{Kannan2018a}
{Kannan} R.,  {Vogelsberger} M.,  {Marinacci} F.,  {McKinnon} R.,  {Pakmor} R.,
    {Springel} V.,  2019, \mn@doi [\mnras] {10.1093/mnras/stz287}, \href
  {https://ui.adsabs.harvard.edu/\#abs/2019MNRAS.485..117K} {485, 117}

\bibitem[\protect\citeauthoryear{{Katz}, {Weinberg}  \& {Hernquist}}{{Katz}
  et~al.}{1996}]{Katz1996}
{Katz} N.,  {Weinberg} D.~H.,   {Hernquist} L.,  1996, \mn@doi [\apjs]
  {10.1086/192305}, \href {http://adsabs.harvard.edu/abs/1996ApJS..105...19K}
  {105, 19}

\bibitem[\protect\citeauthoryear{{Kennicutt}}{{Kennicutt}}{1998}]{Kennicutt1998}
{Kennicutt} Jr. R.~C.,  1998, \mn@doi [\apj] {10.1086/305588}, \href
  {http://cdsads.u-strasbg.fr/abs/1998ApJ...498..541K} {498, 541}

\bibitem[\protect\citeauthoryear{{Kennicutt} \& {Evans}}{{Kennicutt} \&
  {Evans}}{2012}]{Kennicutt2012}
{Kennicutt} R.~C.,  {Evans} N.~J.,  2012, \mn@doi [\araa]
  {10.1146/annurev-astro-081811-125610}, \href
  {http://adsabs.harvard.edu/abs/2012ARA%26A..50..531K} {50, 531}

\bibitem[\protect\citeauthoryear{{Kim} \& {Ostriker}}{{Kim} \&
  {Ostriker}}{2015}]{Kim2015}
{Kim} C.-G.,  {Ostriker} E.~C.,  2015, \mn@doi [\apj]
  {10.1088/0004-637X/802/2/99}, \href
  {http://adsabs.harvard.edu/abs/2015ApJ...802...99K} {802, 99}

\bibitem[\protect\citeauthoryear{{Kim} \& {Ostriker}}{{Kim} \&
  {Ostriker}}{2017}]{Kim2017}
{Kim} C.-G.,  {Ostriker} E.~C.,  2017, \mn@doi [\apj]
  {10.3847/1538-4357/aa8599}, \href
  {https://ui.adsabs.harvard.edu/#abs/2017ApJ...846..133K} {846, 133}

\bibitem[\protect\citeauthoryear{{Kim} \& {Ostriker}}{{Kim} \&
  {Ostriker}}{2018}]{Kim2018}
{Kim} C.-G.,  {Ostriker} E.~C.,  2018, \mn@doi [\apj]
  {10.3847/1538-4357/aaa5ff}, \href
  {https://ui.adsabs.harvard.edu/abs/2018ApJ...853..173K} {853, 173}

\bibitem[\protect\citeauthoryear{{Kim}, {Kim}  \& {Ostriker}}{{Kim}
  et~al.}{2018}]{Kim2018b}
{Kim} J.-G.,  {Kim} W.-T.,   {Ostriker} E.~C.,  2018, \mn@doi [\apj]
  {10.3847/1538-4357/aabe27}, \href
  {https://ui.adsabs.harvard.edu/abs/2018ApJ...859...68K} {859, 68}

\bibitem[\protect\citeauthoryear{{Kimm} \& {Cen}}{{Kimm} \&
  {Cen}}{2014}]{Kimm2014}
{Kimm} T.,  {Cen} R.,  2014, \mn@doi [\apj] {10.1088/0004-637X/788/2/121},
  \href {http://adsabs.harvard.edu/abs/2014ApJ...788..121K} {788, 121}

\bibitem[\protect\citeauthoryear{{Kimm}, {Cen}, {Devriendt}, {Dubois}  \&
  {Slyz}}{{Kimm} et~al.}{2015}]{Kimm2015}
{Kimm} T.,  {Cen} R.,  {Devriendt} J.,  {Dubois} Y.,   {Slyz} A.,  2015,
  \mn@doi [\mnras] {10.1093/mnras/stv1211}, \href
  {http://adsabs.harvard.edu/abs/2015MNRAS.451.2900K} {451, 2900}

\bibitem[\protect\citeauthoryear{{Kroupa}}{{Kroupa}}{2001}]{Kroupa2001}
{Kroupa} P.,  2001, in {Deiters} S.,  {Fuchs} B.,  {Just} A.,  {Spurzem} R.,
  {Wielen} R.,  eds,  Astronomical Society of the Pacific Conference Series
  Vol. 228, Dynamics of Star Clusters and the Milky Way. p.~187 (\mn@eprint {}
  {astro-ph/0011328})

\bibitem[\protect\citeauthoryear{{Krumholz}}{{Krumholz}}{2018}]{Krumholz2018}
{Krumholz} M.~R.,  2018, \mn@doi [\mnras] {10.1093/mnras/sty2105}, \href
  {http://adsabs.harvard.edu/abs/2018MNRAS.480.3468K} {480, 3468}

\bibitem[\protect\citeauthoryear{{Krumholz} \& {Gnedin}}{{Krumholz} \&
  {Gnedin}}{2011}]{Krumholz2011}
{Krumholz} M.~R.,  {Gnedin} N.~Y.,  2011, \mn@doi [\apj]
  {10.1088/0004-637X/729/1/36}, \href
  {http://adsabs.harvard.edu/abs/2011ApJ...729...36K} {729, 36}

\bibitem[\protect\citeauthoryear{{Krumholz} \& {Matzner}}{{Krumholz} \&
  {Matzner}}{2009}]{Krumholz2009}
{Krumholz} M.~R.,  {Matzner} C.~D.,  2009, \mn@doi [\apj]
  {10.1088/0004-637X/703/2/1352}, \href
  {http://adsabs.harvard.edu/abs/2009ApJ...703.1352K} {703, 1352}

\bibitem[\protect\citeauthoryear{{Krumholz} \& {Tan}}{{Krumholz} \&
  {Tan}}{2007}]{Krumholz2007}
{Krumholz} M.~R.,  {Tan} J.~C.,  2007, \mn@doi [\apj] {10.1086/509101}, \href
  {https://ui.adsabs.harvard.edu/#abs/2007ApJ...654..304K} {654, 304}

\bibitem[\protect\citeauthoryear{{Krumholz} \& {Thompson}}{{Krumholz} \&
  {Thompson}}{2012}]{Krumholz2012}
{Krumholz} M.~R.,  {Thompson} T.~A.,  2012, \mn@doi [\apj]
  {10.1088/0004-637X/760/2/155}, \href
  {https://ui.adsabs.harvard.edu/#abs/2012ApJ...760..155K} {760, 155}

\bibitem[\protect\citeauthoryear{{Krumholz} \& {Thompson}}{{Krumholz} \&
  {Thompson}}{2013}]{Krumholz2013}
{Krumholz} M.~R.,  {Thompson} T.~A.,  2013, \mn@doi [\mnras]
  {10.1093/mnras/stt1174}, \href
  {https://ui.adsabs.harvard.edu/#abs/2013MNRAS.434.2329K} {434, 2329}

\bibitem[\protect\citeauthoryear{{Leroy} et~al.,}{{Leroy}
  et~al.}{2013}]{Leroy2013}
{Leroy} A.~K.,  et~al., 2013, \mn@doi [\aj] {10.1088/0004-6256/146/2/19}, \href
  {http://adsabs.harvard.edu/abs/2013AJ....146...19L} {146, 19}

\bibitem[\protect\citeauthoryear{{Li}, {Bryan}  \& {Ostriker}}{{Li}
  et~al.}{2017}]{Li2017}
{Li} M.,  {Bryan} G.~L.,   {Ostriker} J.~P.,  2017, \mn@doi [\apj]
  {10.3847/1538-4357/aa7263}, \href
  {https://ui.adsabs.harvard.edu/#abs/2017ApJ...841..101L} {841, 101}

\bibitem[\protect\citeauthoryear{{Lopez}, {Krumholz}, {Bolatto}, {Prochaska}
  \& {Ramirez-Ruiz}}{{Lopez} et~al.}{2011}]{Lopez2011}
{Lopez} L.~A.,  {Krumholz} M.~R.,  {Bolatto} A.~D.,  {Prochaska} J.~X.,
  {Ramirez-Ruiz} E.,  2011, \mn@doi [\apj] {10.1088/0004-637X/731/2/91}, \href
  {http://adsabs.harvard.edu/abs/2011ApJ...731...91L} {731, 91}

\bibitem[\protect\citeauthoryear{{Mac Low} \& {McCray}}{{Mac Low} \&
  {McCray}}{1988}]{MacLow1988}
{Mac Low} M.-M.,  {McCray} R.,  1988, \mn@doi [\apj] {10.1086/165936}, \href
  {http://adsabs.harvard.edu/abs/1988ApJ...324..776M} {324, 776}

\bibitem[\protect\citeauthoryear{{Maoz}, {Mannucci}  \& {Brandt}}{{Maoz}
  et~al.}{2012}]{Maoz2012}
{Maoz} D.,  {Mannucci} F.,   {Brandt} T.~D.,  2012, \mn@doi [\mnras]
  {10.1111/j.1365-2966.2012.21871.x}, \href
  {http://adsabs.harvard.edu/abs/2012MNRAS.426.3282M} {426, 3282}

\bibitem[\protect\citeauthoryear{{Marinacci}, {Binney}, {Fraternali}, {Nipoti},
  {Ciotti}  \& {Londrillo}}{{Marinacci} et~al.}{2010}]{Marinacci2010}
{Marinacci} F.,  {Binney} J.,  {Fraternali} F.,  {Nipoti} C.,  {Ciotti} L.,
  {Londrillo} P.,  2010, \mn@doi [\mnras] {10.1111/j.1365-2966.2010.16352.x},
  \href {http://adsabs.harvard.edu/abs/2010MNRAS.404.1464M} {404, 1464}

\bibitem[\protect\citeauthoryear{{Marinacci}, {Pakmor}  \&
  {Springel}}{{Marinacci} et~al.}{2014}]{Marinacci2014}
{Marinacci} F.,  {Pakmor} R.,   {Springel} V.,  2014, \mn@doi [\mnras]
  {10.1093/mnras/stt2003}, \href
  {http://adsabs.harvard.edu/abs/2014MNRAS.437.1750M} {437, 1750}

\bibitem[\protect\citeauthoryear{{Martizzi}, {Faucher-Gigu{\`e}re}  \&
  {Quataert}}{{Martizzi} et~al.}{2015}]{Martizzi2015}
{Martizzi} D.,  {Faucher-Gigu{\`e}re} C.-A.,   {Quataert} E.,  2015, \mn@doi
  [\mnras] {10.1093/mnras/stv562}, \href
  {https://ui.adsabs.harvard.edu/#abs/2015MNRAS.450..504M} {450, 504}

\bibitem[\protect\citeauthoryear{{Martizzi}, {Fielding}, {Faucher-Gigu{\`e}re}
  \& {Quataert}}{{Martizzi} et~al.}{2016}]{Martizzi2016}
{Martizzi} D.,  {Fielding} D.,  {Faucher-Gigu{\`e}re} C.-A.,   {Quataert} E.,
  2016, \mn@doi [\mnras] {10.1093/mnras/stw745}, \href
  {https://ui.adsabs.harvard.edu/#abs/2016MNRAS.459.2311M} {459, 2311}

\bibitem[\protect\citeauthoryear{{Matzner}}{{Matzner}}{2002}]{Matzner2002}
{Matzner} C.~D.,  2002, \mn@doi [\apj] {10.1086/338030}, \href
  {http://adsabs.harvard.edu/abs/2002ApJ...566..302M} {566, 302}

\bibitem[\protect\citeauthoryear{{McKee} \& {Krumholz}}{{McKee} \&
  {Krumholz}}{2010}]{McKee2010}
{McKee} C.~F.,  {Krumholz} M.~R.,  2010, \mn@doi [\apj]
  {10.1088/0004-637X/709/1/308}, \href
  {http://adsabs.harvard.edu/abs/2010ApJ...709..308M} {709, 308}

\bibitem[\protect\citeauthoryear{{McKee} \& {Ostriker}}{{McKee} \&
  {Ostriker}}{1977}]{McKee1977}
{McKee} C.~F.,  {Ostriker} J.~P.,  1977, \mn@doi [\apj] {10.1086/155667}, \href
  {http://cdsads.u-strasbg.fr/abs/1977ApJ...218..148M} {218, 148}

\bibitem[\protect\citeauthoryear{{McKinnon}, {Vogelsberger}, {Torrey},
  {Marinacci}  \& {Kannan}}{{McKinnon} et~al.}{2018}]{McKinnon2018}
{McKinnon} R.,  {Vogelsberger} M.,  {Torrey} P.,  {Marinacci} F.,   {Kannan}
  R.,  2018, \mn@doi [\mnras] {10.1093/mnras/sty1248}, \href
  {http://adsabs.harvard.edu/abs/2018MNRAS.478.2851M} {478, 2851}

\bibitem[\protect\citeauthoryear{{Monaghan} \& {Lattanzio}}{{Monaghan} \&
  {Lattanzio}}{1985}]{Monaghan1985}
{Monaghan} J.~J.,  {Lattanzio} J.~C.,  1985, \aap, \href
  {http://adsabs.harvard.edu/abs/1985A%26A...149..135M} {149, 135}

\bibitem[\protect\citeauthoryear{{Moster}, {Naab}  \& {White}}{{Moster}
  et~al.}{2013}]{Moster2013}
{Moster} B.~P.,  {Naab} T.,   {White} S. D.~M.,  2013, \mn@doi [\mnras]
  {10.1093/mnras/sts261}, \href
  {https://ui.adsabs.harvard.edu/#abs/2013MNRAS.428.3121M} {428, 3121}

\bibitem[\protect\citeauthoryear{{Murante}, {Monaco}, {Giovalli}, {Borgani}  \&
  {Diaferio}}{{Murante} et~al.}{2010}]{Murante2010}
{Murante} G.,  {Monaco} P.,  {Giovalli} M.,  {Borgani} S.,   {Diaferio} A.,
  2010, \mn@doi [\mnras] {10.1111/j.1365-2966.2010.16567.x}, \href
  {https://ui.adsabs.harvard.edu/abs/2010MNRAS.405.1491M} {405, 1491}

\bibitem[\protect\citeauthoryear{{Murante}, {Monaco}, {Borgani}, {Tornatore},
  {Dolag}  \& {Goz}}{{Murante} et~al.}{2015}]{Murante2015}
{Murante} G.,  {Monaco} P.,  {Borgani} S.,  {Tornatore} L.,  {Dolag} K.,
  {Goz} D.,  2015, \mn@doi [\mnras] {10.1093/mnras/stu2400}, \href
  {https://ui.adsabs.harvard.edu/abs/2015MNRAS.447..178M} {447, 178}

\bibitem[\protect\citeauthoryear{{Muratov}, {Kere{\v s}},
  {Faucher-Gigu{\`e}re}, {Hopkins}, {Quataert}  \& {Murray}}{{Muratov}
  et~al.}{2015}]{Muratov2015}
{Muratov} A.~L.,  {Kere{\v s}} D.,  {Faucher-Gigu{\`e}re} C.-A.,  {Hopkins}
  P.~F.,  {Quataert} E.,   {Murray} N.,  2015, \mn@doi [\mnras]
  {10.1093/mnras/stv2126}, \href
  {http://adsabs.harvard.edu/abs/2015MNRAS.454.2691M} {454, 2691}

\bibitem[\protect\citeauthoryear{{Murray}, {Quataert}  \& {Thompson}}{{Murray}
  et~al.}{2005}]{Murray2005}
{Murray} N.,  {Quataert} E.,   {Thompson} T.~A.,  2005, \mn@doi [\apj]
  {10.1086/426067}, \href {http://adsabs.harvard.edu/abs/2005ApJ...618..569M}
  {618, 569}

\bibitem[\protect\citeauthoryear{{Murray}, {Quataert}  \& {Thompson}}{{Murray}
  et~al.}{2010}]{Murray2010}
{Murray} N.,  {Quataert} E.,   {Thompson} T.~A.,  2010, \mn@doi [\apj]
  {10.1088/0004-637X/709/1/191}, \href
  {http://adsabs.harvard.edu/abs/2010ApJ...709..191M} {709, 191}

\bibitem[\protect\citeauthoryear{{Nelson} et~al.,}{{Nelson}
  et~al.}{2019}]{Nelson2019}
{Nelson} D.,  et~al., 2019, arXiv e-prints, \href
  {https://ui.adsabs.harvard.edu/\#abs/2019arXiv190205554N} {p.
  arXiv:1902.05554}

\bibitem[\protect\citeauthoryear{{O{\~n}orbe}, {Boylan-Kolchin}, {Bullock},
  {Hopkins}, {Kere{\v{s}}}, {Faucher-Gigu{\`e}re}, {Quataert}  \&
  {Murray}}{{O{\~n}orbe} et~al.}{2015}]{Onorbe2015}
{O{\~n}orbe} J.,  {Boylan-Kolchin} M.,  {Bullock} J.~S.,  {Hopkins} P.~F.,
  {Kere{\v{s}}} D.,  {Faucher-Gigu{\`e}re} C.-A.,  {Quataert} E.,   {Murray}
  N.,  2015, \mn@doi [\mnras] {10.1093/mnras/stv2072}, \href
  {https://ui.adsabs.harvard.edu/#abs/2015MNRAS.454.2092O} {454, 2092}

\bibitem[\protect\citeauthoryear{{Ochsendorf}, {Verdolini}, {Cox}, {Bern{\'e}},
  {Kaper}  \& {Tielens}}{{Ochsendorf} et~al.}{2014}]{Ochsendorf2014}
{Ochsendorf} B.~B.,  {Verdolini} S.,  {Cox} N.~L.~J.,  {Bern{\'e}} O.,  {Kaper}
  L.,   {Tielens} A.~G.~G.~M.,  2014, \mn@doi [\aap]
  {10.1051/0004-6361/201423545}, \href
  {https://ui.adsabs.harvard.edu/#abs/2014A&A...566A..75O} {566, A75}

\bibitem[\protect\citeauthoryear{{Oppenheimer}, {Dav{\'e}}, {Kere{\v s}},
  {Fardal}, {Katz}, {Kollmeier}  \& {Weinberg}}{{Oppenheimer}
  et~al.}{2010}]{oppenheimer2010}
{Oppenheimer} B.~D.,  {Dav{\'e}} R.,  {Kere{\v s}} D.,  {Fardal} M.,  {Katz}
  N.,  {Kollmeier} J.~A.,   {Weinberg} D.~H.,  2010, \mn@doi [\mnras]
  {10.1111/j.1365-2966.2010.16872.x}, \href
  {http://adsabs.harvard.edu/abs/2010MNRAS.406.2325O} {406, 2325}

\bibitem[\protect\citeauthoryear{{Ostriker} \& {Shetty}}{{Ostriker} \&
  {Shetty}}{2011}]{Ostriker2011}
{Ostriker} E.~C.,  {Shetty} R.,  2011, \mn@doi [\apj]
  {10.1088/0004-637X/731/1/41}, \href
  {https://ui.adsabs.harvard.edu/#abs/2011ApJ...731...41O} {731, 41}

\bibitem[\protect\citeauthoryear{{Pais}, {Pfrommer}, {Ehlert}  \&
  {Pakmor}}{{Pais} et~al.}{2018}]{Pais2018}
{Pais} M.,  {Pfrommer} C.,  {Ehlert} K.,   {Pakmor} R.,  2018, \mn@doi [\mnras]
  {10.1093/mnras/sty1410}, \href
  {https://ui.adsabs.harvard.edu/#abs/2018MNRAS.478.5278P} {478, 5278}

\bibitem[\protect\citeauthoryear{{Pakmor} \& {Springel}}{{Pakmor} \&
  {Springel}}{2013}]{Pakmor2013}
{Pakmor} R.,  {Springel} V.,  2013, \mn@doi [\mnras] {10.1093/mnras/stt428},
  \href {http://adsabs.harvard.edu/abs/2013MNRAS.432..176P} {432, 176}

\bibitem[\protect\citeauthoryear{{Pakmor}, {Springel}, {Bauer}, {Mocz},
  {Munoz}, {Ohlmann}, {Schaal}  \& {Zhu}}{{Pakmor} et~al.}{2016}]{Pakmor2016}
{Pakmor} R.,  {Springel} V.,  {Bauer} A.,  {Mocz} P.,  {Munoz} D.~J.,
  {Ohlmann} S.~T.,  {Schaal} K.,   {Zhu} C.,  2016, \mn@doi [\mnras]
  {10.1093/mnras/stv2380}, \href
  {https://ui.adsabs.harvard.edu/#abs/2016MNRAS.455.1134P} {455, 1134}

\bibitem[\protect\citeauthoryear{{Pettini}, {Shapley}, {Steidel}, {Cuby},
  {Dickinson}, {Moorwood}, {Adelberger}  \& {Giavalisco}}{{Pettini}
  et~al.}{2001}]{Pettini2001}
{Pettini} M.,  {Shapley} A.~E.,  {Steidel} C.~C.,  {Cuby} J.-G.,  {Dickinson}
  M.,  {Moorwood} A. F.~M.,  {Adelberger} K.~L.,   {Giavalisco} M.,  2001,
  \mn@doi [\apj] {10.1086/321403}, \href
  {https://ui.adsabs.harvard.edu/#abs/2001ApJ...554..981P} {554, 981}

\bibitem[\protect\citeauthoryear{{Pfrommer}, {Pakmor}, {Schaal}, {Simpson}  \&
  {Springel}}{{Pfrommer} et~al.}{2017}]{Pfrommer2017}
{Pfrommer} C.,  {Pakmor} R.,  {Schaal} K.,  {Simpson} C.~M.,   {Springel} V.,
  2017, \mn@doi [\mnras] {10.1093/mnras/stw2941}, \href
  {http://adsabs.harvard.edu/abs/2017MNRAS.465.4500P} {465, 4500}

\bibitem[\protect\citeauthoryear{{Pillepich} et~al.,}{{Pillepich}
  et~al.}{2018}]{Pillepich2018b}
{Pillepich} A.,  et~al., 2018, \mn@doi [\mnras] {10.1093/mnras/stx2656}, \href
  {https://ui.adsabs.harvard.edu/#abs/2018MNRAS.473.4077P} {473, 4077}

\bibitem[\protect\citeauthoryear{{Pillepich} et~al.,}{{Pillepich}
  et~al.}{2019}]{Pillepich2019}
{Pillepich} A.,  et~al., 2019, arXiv e-prints, \href
  {https://ui.adsabs.harvard.edu/\#abs/2019arXiv190205553P} {p.
  arXiv:1902.05553}

\bibitem[\protect\citeauthoryear{{Planck Collaboration} et~al.,}{{Planck
  Collaboration} et~al.}{2016}]{Planck2016}
{Planck Collaboration} et~al., 2016, \mn@doi [\aap]
  {10.1051/0004-6361/201525830}, \href
  {http://adsabs.harvard.edu/abs/2016A%26A...594A..13P} {594, A13}

\bibitem[\protect\citeauthoryear{{Portinari}, {Chiosi}  \&
  {Bressan}}{{Portinari} et~al.}{1998}]{Portinari1998}
{Portinari} L.,  {Chiosi} C.,   {Bressan} A.,  1998, \aap, \href
  {http://adsabs.harvard.edu/abs/1998A%26A...334..505P} {334, 505}

\bibitem[\protect\citeauthoryear{{Putman}}{{Putman}}{2017}]{Putman2017}
{Putman} M.~E.,  2017, in {Fox} A.,  {Dav{\'e}} R.,  eds,  Astrophysics and
  Space Science Library Vol. 430, Gas Accretion onto Galaxies. p.~1 (\mn@eprint
  {arXiv} {1612.00461}), \mn@doi{10.1007/978-3-319-52512-9_1}

\bibitem[\protect\citeauthoryear{{Putman}, {Peek}  \& {Joung}}{{Putman}
  et~al.}{2012}]{Putman2012}
{Putman} M.~E.,  {Peek} J.~E.~G.,   {Joung} M.~R.,  2012, \mn@doi [Annual
  Review of Astronomy and Astrophysics] {10.1146/annurev-astro-081811-125612},
  \href {https://ui.adsabs.harvard.edu/#abs/2012ARA&A..50..491P} {50, 491}

\bibitem[\protect\citeauthoryear{{Rahmati}, {Pawlik}, {Rai{\v c}evi{\'c}}  \&
  {Schaye}}{{Rahmati} et~al.}{2013}]{Rahmati2013}
{Rahmati} A.,  {Pawlik} A.~H.,  {Rai{\v c}evi{\'c}} M.,   {Schaye} J.,  2013,
  \mn@doi [\mnras] {10.1093/mnras/stt066}, \href
  {http://adsabs.harvard.edu/abs/2013MNRAS.430.2427R} {430, 2427}

\bibitem[\protect\citeauthoryear{{Raskutti}, {Ostriker}  \&
  {Skinner}}{{Raskutti} et~al.}{2016}]{Raskutti2016}
{Raskutti} S.,  {Ostriker} E.~C.,   {Skinner} M.~A.,  2016, \mn@doi [\apj]
  {10.3847/0004-637X/829/2/130}, \href
  {http://adsabs.harvard.edu/abs/2016ApJ...829..130R} {829, 130}

\bibitem[\protect\citeauthoryear{{Recchia}, {Blasi}  \& {Morlino}}{{Recchia}
  et~al.}{2016}]{Recchia2016}
{Recchia} S.,  {Blasi} P.,   {Morlino} G.,  2016, \mn@doi [\mnras]
  {10.1093/mnras/stw1966}, \href
  {https://ui.adsabs.harvard.edu/#abs/2016MNRAS.462.4227R} {462, 4227}

\bibitem[\protect\citeauthoryear{{Rosdahl}, {Schaye}, {Teyssier}  \&
  {Agertz}}{{Rosdahl} et~al.}{2015}]{Rosdahl2015}
{Rosdahl} J.,  {Schaye} J.,  {Teyssier} R.,   {Agertz} O.,  2015, \mn@doi
  [\mnras] {10.1093/mnras/stv937}, \href
  {http://adsabs.harvard.edu/abs/2015MNRAS.451...34R} {451, 34}

\bibitem[\protect\citeauthoryear{{Rosdahl}, {Schaye}, {Dubois}, {Kimm}  \&
  {Teyssier}}{{Rosdahl} et~al.}{2017}]{Rosdahl2017}
{Rosdahl} J.,  {Schaye} J.,  {Dubois} Y.,  {Kimm} T.,   {Teyssier} R.,  2017,
  \mn@doi [\mnras] {10.1093/mnras/stw3034}, \href
  {http://adsabs.harvard.edu/abs/2017MNRAS.466...11R} {466, 11}

\bibitem[\protect\citeauthoryear{{Rybicki} \& {Lightman}}{{Rybicki} \&
  {Lightman}}{1986}]{Rybicki_Lightman}
{Rybicki} G.~B.,  {Lightman} A.~P.,  1986, {Radiative Processes in
  Astrophysics}.
Wiley-VCH, Weinheim, Germany

\bibitem[\protect\citeauthoryear{{Sales}, {Navarro}, {Schaye}, {Dalla Vecchia},
  {Springel}  \& {Booth}}{{Sales} et~al.}{2010}]{Sales2010}
{Sales} L.~V.,  {Navarro} J.~F.,  {Schaye} J.,  {Dalla Vecchia} C.,  {Springel}
  V.,   {Booth} C.~M.,  2010, \mn@doi [\mnras]
  {10.1111/j.1365-2966.2010.17391.x}, \href
  {https://ui.adsabs.harvard.edu/#abs/2010MNRAS.409.1541S} {409, 1541}

\bibitem[\protect\citeauthoryear{{Sales}, {Marinacci}, {Springel}  \&
  {Petkova}}{{Sales} et~al.}{2014}]{Sales2014}
{Sales} L.~V.,  {Marinacci} F.,  {Springel} V.,   {Petkova} M.,  2014, \mn@doi
  [\mnras] {10.1093/mnras/stu155}, \href
  {https://ui.adsabs.harvard.edu/#abs/2014MNRAS.439.2990S} {439, 2990}

\bibitem[\protect\citeauthoryear{{Sancisi}, {Fraternali}, {Oosterloo}  \& {van
  der Hulst}}{{Sancisi} et~al.}{2008}]{Sancisi2008}
{Sancisi} R.,  {Fraternali} F.,  {Oosterloo} T.,   {van der Hulst} T.,  2008,
  \mn@doi [Astronomy and Astrophysics Review] {10.1007/s00159-008-0010-0},
  \href {https://ui.adsabs.harvard.edu/#abs/2008A&ARv..15..189S} {15, 189}

\bibitem[\protect\citeauthoryear{{Scannapieco} et~al.}{{Scannapieco}
  et~al.}{2012}]{Scannapieco2012}
{Scannapieco} C.,  et~al., 2012, \mn@doi [\mnras]
  {10.1111/j.1365-2966.2012.20993.x}, \href
  {http://esoads.eso.org/abs/2012MNRAS.423.1726S} {423, 1726}

\bibitem[\protect\citeauthoryear{{Schaye} et~al.,}{{Schaye}
  et~al.}{2015}]{Schaye2015}
{Schaye} J.,  et~al., 2015, \mn@doi [\mnras] {10.1093/mnras/stu2058}, \href
  {http://adsabs.harvard.edu/abs/2015MNRAS.446..521S} {446, 521}

\bibitem[\protect\citeauthoryear{{Semenov}, {Kravtsov}  \& {Gnedin}}{{Semenov}
  et~al.}{2017}]{Semenov2017}
{Semenov} V.~A.,  {Kravtsov} A.~V.,   {Gnedin} N.~Y.,  2017, \mn@doi [\apj]
  {10.3847/1538-4357/aa8096}, \href
  {https://ui.adsabs.harvard.edu/#abs/2017ApJ...845..133S} {845, 133}

\bibitem[\protect\citeauthoryear{{Simpson}, {Bryan}, {Johnston}, {Smith}, {Mac
  Low}, {Sharma}  \& {Tumlinson}}{{Simpson} et~al.}{2013}]{Simpson2013}
{Simpson} C.~M.,  {Bryan} G.~L.,  {Johnston} K.~V.,  {Smith} B.~D.,  {Mac Low}
  M.-M.,  {Sharma} S.,   {Tumlinson} J.,  2013, \mn@doi [\mnras]
  {10.1093/mnras/stt474}, \href
  {https://ui.adsabs.harvard.edu/#abs/2013MNRAS.432.1989S} {432, 1989}

\bibitem[\protect\citeauthoryear{{Simpson}, {Pakmor}, {Marinacci}, {Pfrommer},
  {Springel}, {Glover}, {Clark}  \& {Smith}}{{Simpson}
  et~al.}{2016}]{Simpson2016}
{Simpson} C.~M.,  {Pakmor} R.,  {Marinacci} F.,  {Pfrommer} C.,  {Springel} V.,
   {Glover} S.~C.~O.,  {Clark} P.~C.,   {Smith} R.~J.,  2016, \mn@doi [\apjl]
  {10.3847/2041-8205/827/2/L29}, \href
  {http://adsabs.harvard.edu/abs/2016ApJ...827L..29S} {827, L29}

\bibitem[\protect\citeauthoryear{{Smith}, {Sijacki}  \& {Shen}}{{Smith}
  et~al.}{2018}]{Smith2018}
{Smith} M.~C.,  {Sijacki} D.,   {Shen} S.,  2018, \mn@doi [\mnras]
  {10.1093/mnras/sty994}, \href
  {http://adsabs.harvard.edu/abs/2018MNRAS.478..302S} {478, 302}

\bibitem[\protect\citeauthoryear{{Springel}}{{Springel}}{2000}]{Springel2000}
{Springel} V.,  2000, \mn@doi [\mnras] {10.1046/j.1365-8711.2000.03187.x},
  \href {http://adsabs.harvard.edu/abs/2000MNRAS.312..859S} {312, 859}

\bibitem[\protect\citeauthoryear{{Springel}}{{Springel}}{2010}]{Springel2010}
{Springel} V.,  2010, \mn@doi [\mnras] {10.1111/j.1365-2966.2009.15715.x},
  \href {http://cdsads.u-strasbg.fr/abs/2010MNRAS.401..791S} {401, 791}

\bibitem[\protect\citeauthoryear{{Springel} \& {Hernquist}}{{Springel} \&
  {Hernquist}}{2003}]{Springel2003}
{Springel} V.,  {Hernquist} L.,  2003, \mn@doi [\mnras]
  {10.1046/j.1365-8711.2003.06206.x}, \href
  {http://cdsads.u-strasbg.fr/abs/2003MNRAS.339..289S} {339, 289}

\bibitem[\protect\citeauthoryear{{Springel}, {Di Matteo}  \&
  {Hernquist}}{{Springel} et~al.}{2005}]{Springel2005b}
{Springel} V.,  {Di Matteo} T.,   {Hernquist} L.,  2005, \mn@doi [\mnras]
  {10.1111/j.1365-2966.2005.09238.x}, \href
  {http://adsabs.harvard.edu/abs/2005MNRAS.361..776S} {361, 776}

\bibitem[\protect\citeauthoryear{{Springel} et~al.,}{{Springel}
  et~al.}{2018}]{Springel2018}
{Springel} V.,  et~al., 2018, \mn@doi [\mnras] {10.1093/mnras/stx3304}, \href
  {http://adsabs.harvard.edu/abs/2018MNRAS.475..676S} {475, 676}

\bibitem[\protect\citeauthoryear{{Stinson}, {Seth}, {Katz}, {Wadsley},
  {Governato}  \& {Quinn}}{{Stinson} et~al.}{2006}]{Stinson2006}
{Stinson} G.,  {Seth} A.,  {Katz} N.,  {Wadsley} J.,  {Governato} F.,   {Quinn}
  T.,  2006, \mn@doi [\mnras] {10.1111/j.1365-2966.2006.11097.x}, \href
  {http://cdsads.u-strasbg.fr/abs/2006MNRAS.373.1074S} {373, 1074}

\bibitem[\protect\citeauthoryear{{Stinson}, {Brook}, {Macci{\`o}}, {Wadsley},
  {Quinn}  \& {Couchman}}{{Stinson} et~al.}{2013}]{Stinson2013}
{Stinson} G.~S.,  {Brook} C.,  {Macci{\`o}} A.~V.,  {Wadsley} J.,  {Quinn}
  T.~R.,   {Couchman} H.~M.~P.,  2013, \mn@doi [\mnras] {10.1093/mnras/sts028},
  \href {http://adsabs.harvard.edu/abs/2013MNRAS.428..129S} {428, 129}

\bibitem[\protect\citeauthoryear{{Teyssier}, {Pontzen}, {Dubois}  \&
  {Read}}{{Teyssier} et~al.}{2013}]{Teyssier2013}
{Teyssier} R.,  {Pontzen} A.,  {Dubois} Y.,   {Read} J.~I.,  2013, \mn@doi
  [\mnras] {10.1093/mnras/sts563}, \href
  {http://adsabs.harvard.edu/abs/2013MNRAS.429.3068T} {429, 3068}

\bibitem[\protect\citeauthoryear{{Thielemann} et~al.,}{{Thielemann}
  et~al.}{2003}]{Thielemann2003}
{Thielemann} F.-K.,  et~al., 2003, in {Hillebrandt} W.,  {Leibundgut} B.,  eds,
  From Twilight to Highlight: The Physics of Supernovae. p.~331,
  \mn@doi{10.1007/10828549_46}

\bibitem[\protect\citeauthoryear{{Thornton}, {Gaudlitz}, {Janka}  \&
  {Steinmetz}}{{Thornton} et~al.}{1998}]{Thornton1998}
{Thornton} K.,  {Gaudlitz} M.,  {Janka} H.-T.,   {Steinmetz} M.,  1998, \mn@doi
  [\apj] {10.1086/305704}, \href
  {http://adsabs.harvard.edu/abs/1998ApJ...500...95T} {500, 95}

\bibitem[\protect\citeauthoryear{{Tielens}}{{Tielens}}{2010}]{Tielens2010}
{Tielens} A.~G.~G.~M.,  2010, {The Physics and Chemistry of the Interstellar
  Medium}.
Cambridge University Press, Cambridge UK

\bibitem[\protect\citeauthoryear{{Tollet}, {Cattaneo}, {Macci{\`o}}, {Dutton}
  \& {Kang}}{{Tollet} et~al.}{2019}]{Tollet2019}
{Tollet} {\'E}.,  {Cattaneo} A.,  {Macci{\`o}} A.~V.,  {Dutton} A.~A.,   {Kang}
  X.,  2019, \mn@doi [\mnras] {10.1093/mnras/stz545}, \href
  {http://adsabs.harvard.edu/abs/2019MNRAS.485.2511T} {485, 2511}

\bibitem[\protect\citeauthoryear{{{\"U}bler}, {Naab}, {Oser}, {Aumer}, {Sales}
  \& {White}}{{{\"U}bler} et~al.}{2014}]{Ubler2014}
{{\"U}bler} H.,  {Naab} T.,  {Oser} L.,  {Aumer} M.,  {Sales} L.~V.,   {White}
  S.~D.~M.,  2014, \mn@doi [\mnras] {10.1093/mnras/stu1275}, \href
  {http://adsabs.harvard.edu/abs/2014MNRAS.443.2092U} {443, 2092}

\bibitem[\protect\citeauthoryear{{Valentini}, {Murante}, {Borgani}, {Monaco},
  {Bressan}  \& {Beck}}{{Valentini} et~al.}{2017}]{Valentini2017}
{Valentini} M.,  {Murante} G.,  {Borgani} S.,  {Monaco} P.,  {Bressan} A.,
  {Beck} A.~M.,  2017, \mn@doi [\mnras] {10.1093/mnras/stx1352}, \href
  {https://ui.adsabs.harvard.edu/abs/2017MNRAS.470.3167V} {470, 3167}

\bibitem[\protect\citeauthoryear{{Veilleux}, {Cecil}  \&
  {Bland-Hawthorn}}{{Veilleux} et~al.}{2005}]{Veilleux2005}
{Veilleux} S.,  {Cecil} G.,   {Bland-Hawthorn} J.,  2005, \mn@doi [Annual
  Review of Astronomy and Astrophysics]
  {10.1146/annurev.astro.43.072103.150610}, \href
  {https://ui.adsabs.harvard.edu/#abs/2005ARA&A..43..769V} {43, 769}

\bibitem[\protect\citeauthoryear{{Vogelsberger}, {Sijacki}, {Kere{\v{s}}},
  {Springel}  \& {Hernquist}}{{Vogelsberger} et~al.}{2012}]{Vogelsberger2012}
{Vogelsberger} M.,  {Sijacki} D.,  {Kere{\v{s}}} D.,  {Springel} V.,
  {Hernquist} L.,  2012, \mn@doi [\mnras] {10.1111/j.1365-2966.2012.21590.x},
  \href {https://ui.adsabs.harvard.edu/#abs/2012MNRAS.425.3024V} {425, 3024}

\bibitem[\protect\citeauthoryear{{Vogelsberger}, {Genel}, {Sijacki}, {Torrey},
  {Springel}  \& {Hernquist}}{{Vogelsberger} et~al.}{2013}]{Vogelsberger2013}
{Vogelsberger} M.,  {Genel} S.,  {Sijacki} D.,  {Torrey} P.,  {Springel} V.,
  {Hernquist} L.,  2013, \mn@doi [\mnras] {10.1093/mnras/stt1789}, \href
  {http://adsabs.harvard.edu/abs/2013MNRAS.436.3031V} {436, 3031}

\bibitem[\protect\citeauthoryear{{Vogelsberger} et~al.,}{{Vogelsberger}
  et~al.}{2014}]{Vogelsberger2014a}
{Vogelsberger} M.,  et~al., 2014, \mn@doi [\mnras] {10.1093/mnras/stu1536},
  \href {http://adsabs.harvard.edu/abs/2014MNRAS.444.1518V} {444, 1518}

\bibitem[\protect\citeauthoryear{{Walch} \& {Naab}}{{Walch} \&
  {Naab}}{2015}]{Walch2015}
{Walch} S.,  {Naab} T.,  2015, \mn@doi [\mnras] {10.1093/mnras/stv1155}, \href
  {http://adsabs.harvard.edu/abs/2015MNRAS.451.2757W} {451, 2757}

\bibitem[\protect\citeauthoryear{{Walch}, {Whitworth}, {Bisbas}, {W{\"u}nsch}
  \& {Hubber}}{{Walch} et~al.}{2012}]{Walch2012}
{Walch} S.~K.,  {Whitworth} A.~P.,  {Bisbas} T.,  {W{\"u}nsch} R.,   {Hubber}
  D.,  2012, \mn@doi [\mnras] {10.1111/j.1365-2966.2012.21767.x}, \href
  {https://ui.adsabs.harvard.edu/#abs/2012MNRAS.427..625W} {427, 625}

\bibitem[\protect\citeauthoryear{{Wang}, {Dutton}, {Stinson}, {Macci{\`o}},
  {Penzo}, {Kang}, {Keller}  \& {Wadsley}}{{Wang} et~al.}{2015}]{Wang2015}
{Wang} L.,  {Dutton} A.~A.,  {Stinson} G.~S.,  {Macci{\`o}} A.~V.,  {Penzo} C.,
   {Kang} X.,  {Keller} B.~W.,   {Wadsley} J.,  2015, \mn@doi [\mnras]
  {10.1093/mnras/stv1937}, \href
  {https://ui.adsabs.harvard.edu/#abs/2015MNRAS.454...83W} {454, 83}

\bibitem[\protect\citeauthoryear{{Weaver}, {McCray}, {Castor}, {Shapiro}  \&
  {Moore}}{{Weaver} et~al.}{1977}]{Weaver1977}
{Weaver} R.,  {McCray} R.,  {Castor} J.,  {Shapiro} P.,   {Moore} R.,  1977,
  \mn@doi [\apj] {10.1086/155692}, \href
  {http://adsabs.harvard.edu/abs/1977ApJ...218..377W} {218, 377}

\bibitem[\protect\citeauthoryear{{Wetzel}, {Hopkins}, {Kim},
  {Faucher-Gigu{\`e}re}, {Kere{\v{s}}}  \& {Quataert}}{{Wetzel}
  et~al.}{2016}]{Wetzel2016}
{Wetzel} A.~R.,  {Hopkins} P.~F.,  {Kim} J.-h.,  {Faucher-Gigu{\`e}re} C.-A.,
  {Kere{\v{s}}} D.,   {Quataert} E.,  2016, \mn@doi [\apj]
  {10.3847/2041-8205/827/2/L23}, \href
  {https://ui.adsabs.harvard.edu/#abs/2016ApJ...827L..23W} {827, L23}

\bibitem[\protect\citeauthoryear{{White} \& {Rees}}{{White} \&
  {Rees}}{1978}]{White1978}
{White} S.~D.~M.,  {Rees} M.~J.,  1978, \mn@doi [\mnras]
  {10.1093/mnras/183.3.341}, \href
  {http://adsabs.harvard.edu/abs/1978MNRAS.183..341W} {183, 341}

\bibitem[\protect\citeauthoryear{{Wolfire}, {Hollenbach}, {McKee}, {Tielens}
  \& {Bakes}}{{Wolfire} et~al.}{1995}]{Wolfire1995}
{Wolfire} M.~G.,  {Hollenbach} D.,  {McKee} C.~F.,  {Tielens} A.~G.~G.~M.,
  {Bakes} E.~L.~O.,  1995, \mn@doi [\apj] {10.1086/175510}, \href
  {http://adsabs.harvard.edu/abs/1995ApJ...443..152W} {443, 152}

\bibitem[\protect\citeauthoryear{{Wolfire}, {McKee}, {Hollenbach}  \&
  {Tielens}}{{Wolfire} et~al.}{2003}]{Wolfire2003}
{Wolfire} M.~G.,  {McKee} C.~F.,  {Hollenbach} D.,   {Tielens} A.~G.~G.~M.,
  2003, \mn@doi [\apj] {10.1086/368016}, \href
  {http://adsabs.harvard.edu/abs/2003ApJ...587..278W} {587, 278}

\makeatother
\end{thebibliography}

\appendix

\section{Discrete supernova rates}\label{sec:appA}

Supernovae are discrete and rare\footnote{In our stellar evolutionary model, 
each stellar particle produces $\simeq 3\times 10^{-4}\,{\rm 
SN\,M_{\odot}^{-1}\, Myr^{-1}}$ of type II supernovae for a main sequence lifetime of an $8\,{\rm M_{\odot}}$ star of $40\,\Myr$. Type Ia supernovae are even rarer with a maximum 
rate of $\simeq 7.8\times 10^{-8}\,{\rm SN\,M_{\odot}^{-1}\,Myr^{-1}}$ at a 
stellar age of $40\,\Myr$ (see equation~\ref{eq:SNIDTD}).} events, as such, they are 
sampled in our model as originating from a Poisson distribution 
\begin{equation}
 p(n; \lambda) = \frac{\lambda^ne^{-\lambda}}{n!},
\end{equation}
where $p$ is the probability of having $n$ supernova events given an expected 
value $\lambda$ over the time step $\Delta t$. 

\begin{figure}
\includegraphics[width=0.47\textwidth]{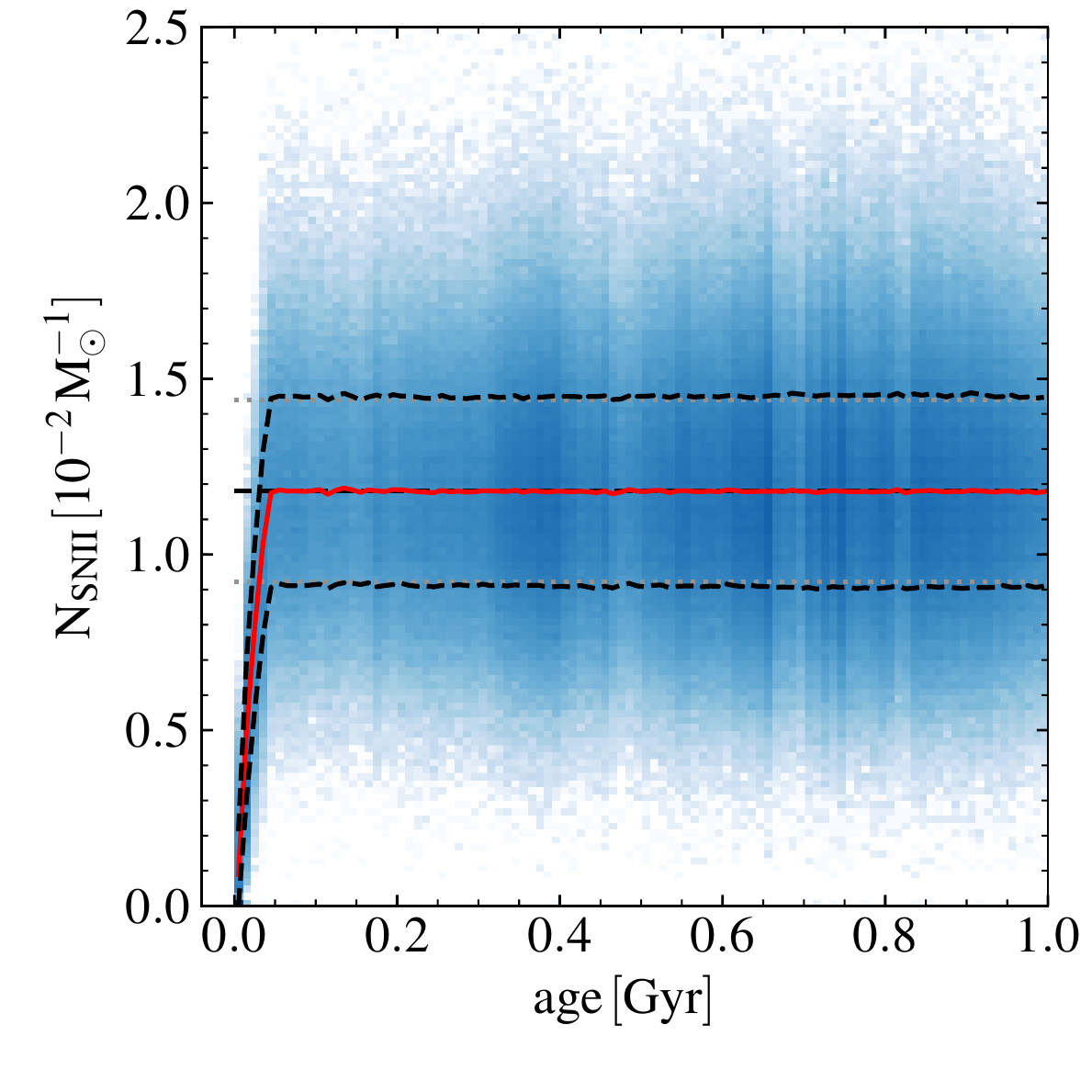}
\includegraphics[width=0.47\textwidth]{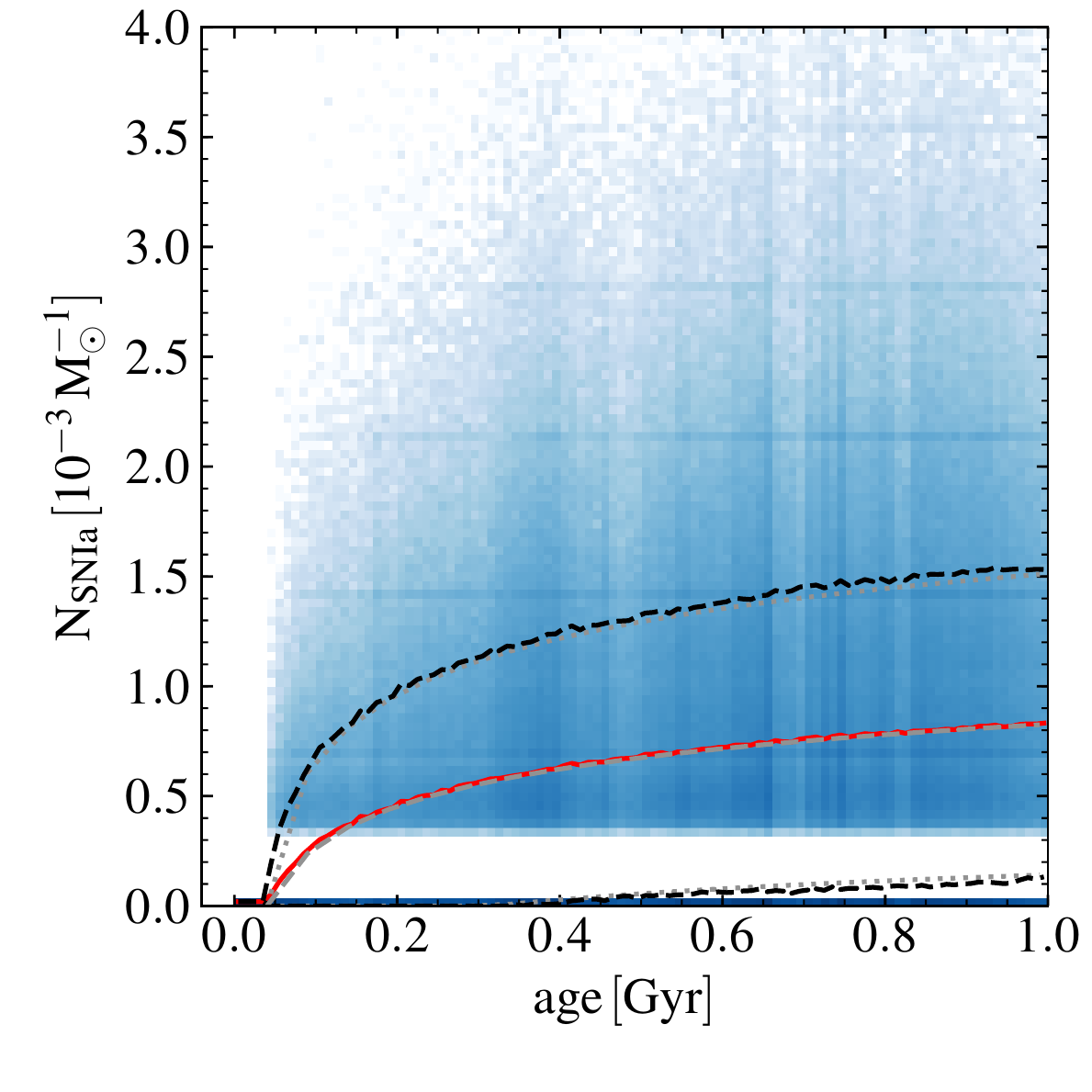}
\caption{Cumulative number of supernova type II events (top) and type Ia (bottom) per unit formed stellar mass as a function of stellar age for the high resolution Milky Way simulation. The two-dimensional histogram (blue shades) shows the number of star particles falling into each 2D-bin having a width of $10\,\Myr$ in age and $2\times 10^{-4}$ and $4\times10^{-5}$ supernova events per unit mass for supernova type II and type Ia, respectively. The red solid line represents the average supernova events as a function of age, whereas the black dashed lines show the standard deviation around the mean value. The grey dashed lines indicate the expected number of supernova events per unit mass computed from the \citet{Chabrier2001} IMF for core-collapse SN and equation~(\ref{eq:SNIanum}) for SNIa. Please note that our Poisson sampling of the distribution recovers on average the expected number of events and their scatter around the mean value.}
\label{fig:snrate}
\end{figure}

Figure \ref{fig:snrate} presents the cumulative number of SNII (top) and SNIa (bottom) as a function of stellar age to illustrate the accuracy of our discrete sampling. 
Each plot shows a two-dimensional histogram (blue shades) counting the number 
of stellar particles of a given age yielding 
a cumulative number of SN \textit{per unit} stellar mass formed (see 
figure caption for details on binning). The red solid line is the average number of 
cumulative SN events per age bin with black dashed lines showing the standard 
deviation around this mean value. The grey dashed lines are the number of SN 
events per unit mass expected from the stellar evolution model adopted in this 
work, with the associated standard deviation represented by the grey dotted 
lines.

It is immediately obvious that the two supernova types have two different 
distributions as a function of the stellar age. In particular, type II supernovae 
display a very steep rise in the first $40\,\Myr$ after which they plateau. This 
is a direct consequence of the fact that we set the minimum mass for a star to 
explode as a type II supernova to $8\,{\rm M_\odot}$, that in turn implies a 
stellar lifetime of $\sim 40\,\Myr$ (with a very slight dependence on 
metallicity). In other words after about that age, all supernova II events have 
exploded and no further events can occur. Type Ia supernova instead have a 
different trend: events start only after $40\,\Myr$ and then follow the trend 
described in equation~(\ref{eq:SNIanum}). Please note that our sampling method is 
able to recover both the average number of supernova events as a function of age 
and their scatter around the mean value. Given the lower number of type Ia events, the signatures of discrete supernova explosions are clearly visible in the bottom panel as the horizontal stripes in the histogram.

It is worth noting that the level of scatter around the mean value, i.e. its 
amplitude, is expected to increase with resolution. This is because 
progressively less massive stellar particles (with increasing resolution) tend, 
on average, to produce less supernova events per particle, thus increasing the 
scatter. This can be easily shown as follows. Let us define 
\begin{equation}
 N_{\sigma_\pm} \equiv N \pm \sigma = N \pm \sqrt{N},
\end{equation}
as the expected number of supernova events at $1\sigma$ above/below the average --
the last equality holds because for a Poisson distribution with expected value 
$\lambda$, $\sigma^2 = \lambda$. 
The amplitude of the scatter will be 
\begin{equation}
\Delta N_{\sigma_\pm} = 2 \sqrt{N},
\end{equation}
or per unit mass 
\begin{equation}
\frac{\Delta N_{\sigma_\pm}}{m_\star} = 2 \frac{\sqrt{N}}{m_\star} = 2 \sqrt{\frac{n}{m_\star}},
\end{equation}
in which $n = N / m_\star$ and $m_{\star}$ is the average mass of a stellar 
particle. The number of supernova events (of both types) per unit mass $n$ is 
independent of resolution and determined only by the stellar evolutionary model 
adopted. However, for increasing resolution $m_\star$ decreases and the 
amplitude of the scatter around the expected number of supernova events per unit mass increases accordingly. 

\label{lastpage}

\end{document}